\documentclass[vecphys]{svmult}

\usepackage{makeidx}         
\usepackage{graphicx}        
\usepackage{multicol}        
\usepackage{xspace}
\usepackage{amsmath}
\usepackage{amssymb}
\usepackage{citesort}

\makeindex             

\DeclareMathAlphabet{\bi}{OML}{cmm}{b}{it}
\newfont{\tensy}{cmsy10}

\newcommand{\bm}[1]{\boldsymbol{#1}}

\renewcommand{\Im}[0]{{\rm Im}\ }

\newcommand{\etal}[0]{{\it et al.}\@\xspace}
\newcommand{\ie}[0]{i.e.\@\xspace}
\newcommand{\eg}[0]{e.g.\@\xspace}

\newcommand{\UP}[0]{\uparrow}
\newcommand{\DO}[0]{\downarrow}

\newcommand{\om}[0]{\omega}

\newcommand{\si}[0]{\sigma}
\newcommand{\las}[0]{\langle}
\newcommand{\ras}[0]{\rangle}
\newcommand{\la}[0]{\left\las}

\newcommand{\ra}[0]{\right\ras}

\newcommand{\ol}[1]{\overline{#1}}

\newcommand{\bk}[0]{\vec{k}}

\newcommand{\sign}{\las\text{sign}\ras}
\newcommand{\Ek}{\ol{E}_\text{kin}}
\newcommand{\Ekin}{E_\text{kin}}
\newcommand{\Ep}{E_\text{P}}
\newcommand{\Ub}[0]{U/t}

\newcommand{\rme}{\mathrm{e}}
\newcommand{\rmi}{\mathrm{i}}
\newcommand{\rmd}{\mathrm{d}}

\newcommand{\ket}[1]{\left|#1\ra}
\newcommand{\bra}[1]{\la#1\right|}
\newcommand{\sket}[1]{|#1\ras}  
\newcommand{\sbra}[1]{\las#1|} 

\newcommand{\com}[2]{[#1,#2]}

\newcommand{\Z}[0]{\mathcal{Z}}
\renewcommand{\D}[0]{\mathcal{D}}

\newcommand{\tr}[0]{\text{Tr}\,}
\newcommand{\dtau}{\Delta\tau}

\newcommand{\Rs}{\mathbb{R}}
\newcommand{\op}{\hat{p}}
\newcommand{\ox}{\hat{x}}
\newcommand{\on}{\hat{n}}
\newcommand{\wb}{w_\text{b}}
\newcommand{\wf}{w_\text{f}}
\newcommand{\U}{\mathcal{U}}
\newcommand{\rD}{\text{D}}
\newcommand{\oB}[0]{\hat{B}}
\newcommand{\V}{\mathcal{V}}

\renewcommand{\text}[1]{\mathrm{#1}}
\renewcommand{\tilde}[1]{\widetilde{#1}}
\renewcommand{\hat}[1]{\widehat{#1}}

\newcommand{\nag}{\phantom{\dag}}

\renewcommand{\H}{\mathcal{H}}

\begin{document}

\title*{Lang-Firsov approaches to polaron physics: From variational methods to
unbiased quantum Monte Carlo simulations}
\titlerunning{Lang-Firsov approaches to polaron physics}
\author{Martin Hohenadler \and Wolfgang von der Linden}
\institute{%
Institute for Theoretical and Computational Physics, TU Graz, Austria\\
\texttt{hohenadler@itp.tu-graz.ac.at}, \texttt{wvl@itp.tu-graz.ac.at}}
%
%
\maketitle

\section{Introduction} %
\label{sec:intro}

In the last decades, there has been substantial interest in simple models for
electron-phonon (el-ph) interaction in condensed matter.  Despite intensive
theoretical efforts, it was not before the advent of numerical methods in the
1980's that a thorough understanding on the basis of exact, unbiased results
was achieved. Although at the present our knowledge of the rather simple
cases of a single carrier (the {\it polaron problem}) or two carriers (the
{\it bipolaron problem}) in Holstein and Fr\"ohlich models is fairly
complete, this is not true for arbitrary band fillings.  There is
still a major desire to develop more efficient simulation techniques to
tackle strongly correlated many-polaron models, which are expected to
describe several aspects of real materials currently under
investigation, such as quantum dots and quantum wires, high-temperature
superconductors or colossal-magnetoresistance manganites.

One of the principle problems in computer simulations of microscopic models
is the limitation in both system size and parameter values. Whereas the former can
be overcome for the polaron and the bipolaron problem in some cases, it is
very difficult to obtain results of similar quality in the many-electron
case. Moreover, many
approaches still suffer from severe restrictions concerning the parameter
regions accessible. For example, interesting materials such as the
cuprates and manganites are characterized by small but finite phonon
frequencies---as compared to the electronic hopping integral---and
intermediate to strong el-ph interaction. Unfortunately,
simulations turn out to be most difficult exactly for such parameters, and
it is therefore highly desirable to improve existing simulation methods.

In this chapter, we shall mainly review different versions of a recently
developed quantum Monte Carlo (QMC) method applicable to Holstein-type models with
one, two or many electrons. The appealing advantages of QMC over other
numerical methods include the accessibility of rather large systems, the
exact treatment of bosonic degrees of freedom (\ie, no truncation is
necessary), and the possibility to consider finite temperatures to study phase
transitions. The important new aspect here is the use of canonically
transformed Hamiltonians, which permits the introduction of exact sampling for
the phonon degrees of freedom, enabling us to carry out accurate simulations
in practically all interesting parameter regimes.

Additionally, based on a generalization of the Lang-Firsov transformation,
we shall present a simple variational approach to the
polaron and the bipolaron problem which yields surprisingly accurate results.

The chapter is organized as follows. In section~\ref{sec:model}, we present
the general model Hamiltonian. Section~\ref{sec:lf} is devoted to a
discussion of the Lang-Firsov transformation, and section~\ref{sec:vpa}
contains the derivation of the variational approach. The QMC method is
introduced in section~\ref{sec:qmc}. Section~\ref{sec:results} gives a
selection of results for the cases of one, two and many electrons.  Finally,
we summarize in section~\ref{sec:summary}.

\section{Model} %
\label{sec:model}

In this paper we focus on the {\it extended Holstein-Hubbard model} defined by
\begin{eqnarray}\label{eq:H}\nonumber
  H
  =
  &-&
  t\sum_{\las ij\ras\si}c^\dag_{i\si}c_{j\si}
  +
  U\sum_i \on_{i\UP} \on_{i\DO}
  +
  V\sum_{\las ij\ras} \on_i \on_j
  \\
  &+&
  \frac{\om_0}{2}\sum_i(\op^2_i+\ox^2_i)
  -
  g'\sum_i \on_i \ox_i
  \,.
\end{eqnarray}
Here $c^\dag_{i\si}$ creates an electron with spin $\si$ at site $i$, and
$\on_i=\sum_\si\on_{i\si}$ with $\on_{i\si}=c^\dag_{i\si}c^{\nag}_{i\si}$.
The phonon degrees of freedom at site $i$ are described by the momentum
$\op_i$ and coordinate (displacement) $\ox_i$ of a harmonic oscillator. The
microscopic parameters are the nearest-neighbour (denoted by $\las\ras$)
hopping amplitude $t$, the on-site (Hubbard--) repulsion $U$, the
nearest-neighbour Coulomb repulsion $V$, the Einstein phonon frequency $\om_0$
and the el-ph coupling $g'$.

This model neglects both long-range Coulomb and el-ph interaction,
which is often a suitable approximation for metallic systems due to
screening. Two simple limiting cases of the Hamiltonian~(\ref{eq:H}) are the
Holstein model ($U=V=0$) and the Hubbard model ($g'=V=0$). In general,
the physics of the model~(\ref{eq:H}) is determined by the competition of the
various interactions. Depending on the choice of parameters and band filling,
it describes fascinating phenomena such as (bi-)polaron formation, Mott-- and
Peierls quantum phase transitions or superconductivity. As we shall see
below, the {\it adiabaticity ratio}
\begin{equation}
\alpha=\om_0/t
\end{equation}
permits us to
distinguish two physically different regimes, namely the {\it adiabatic
  regime} $\alpha<1$ and the {\it non-adiabatic regime} $\alpha>1$.

We further define the dimensionless el-ph coupling parameter $\lambda=g'^2/(\om_0 W)$, where
$W=4t\mathrm{D}$ is the bare bandwidth in D dimensions. Alternatively,
$\lambda$ may also be written as $\lambda=2\Ep/W$, \ie, the ratio of the
polaron binding energy in the atomic limit $t=0$, $\Ep=g'^2/2\om_0$, and
half the bare bandwidth. A useful constant in the non-adiabatic regime is
$g^2=\Ep/\om_0$. We exclusively consider hypercubic
lattices with linear size $N$ and volume $N^\mathrm{D}$, and assume periodic
boundary conditions in real space.

\section{Lang-Firsov transformation} %
\label{sec:lf}

The cornerstone of the methods presented here is the canonical {\it (extended)
Lang-Firsov transformation} of the Hamiltonian~(\ref{eq:H}). The original
Lang-Firsov (LF) transformation \cite{LangFirsov} has
been used extensively to study Holstein-type models. A well-known, early
approximation is due to Holstein \cite{Ho59a}, who replaced the hopping term
by its expectation value in a zero-phonon state, neglecting emission and
absorption of phonons during electron transfer. However, this approach yields
reliable results only in the non-adiabatic strong-coupling (SC) limit.
For $\lambda=\infty$ (or $t=0$), the LF transformation provides an exact
solution of the single-site problem \cite{Ma90}.

Whereas transformed Hamiltonians have been treated numerically before
\cite{dMeRa97,Robin97,FeLoWe00}, the first QMC method making use of the LF
transformation has been proposed in \cite{HoEvvdL03}.

We introduce the extended LF transformation by defining the unitary operator
\begin{equation}\label{eq:polaron:U}
  \hat{\Phi}
  =
  \rme^S
  \,,\quad
  S = \rmi\sum_{ij}\gamma_{ij} \on_i \op_j
\end{equation}
with real parameters $\gamma_{ij}$,  $i,j=1,\dots,N^\mathrm{D}$. $\hat{\Phi}$ as defined in
equation~(\ref{eq:polaron:U}) has the form of a translation operator,
and fulfills $\hat{\Phi}^\dag=\hat{\Phi}^{-1}$. Given an electron at site $i$,
$\hat{\Phi}$ mediates displacements $\gamma_{ij}$ of the harmonic oscillators
at all sites $j$. Hence, the extended transformation is capable of
describing an extended phonon cloud, important in the large-polaron or
bipolaron regime. We shall use this transformation for the
variational approach. However, the standard
(local) LF transformation will be expedient as a basis for unbiased QMC
simulations, in which the transformed Hamiltonian is treated exactly.

Operators have to be transformed according to $\tilde{\hat{A}}=
\hat{\Phi}\hat{A}\hat{\Phi}^\dag$. Defining the function $f(\eta) =
\rme^{\eta\hat{S}} \hat{A} \rme^{-\eta\hat{S}}$ we obtain
\begin{equation}\label{eq:polaron:fprime}
  f'(\eta)
  =
  \rme^{\eta\hat{S}}\com{\hat{S}}{\hat{A}}\rme^{-\eta\hat{S}}
  \,,
\end{equation}
where $f'\equiv\partial f/\partial \eta$. A simple calculation gives
\begin{equation}
  \com{\hat{S}}{c_{i\si}}
  =
  -\rmi\sum_l\gamma_{il}\,\op_l\,c_{i\si}
  \,,\quad
  \com{\hat{S}}{c^\dag_{i\si}}
  =
  \rmi\sum_l\gamma_{il}\,\op_l\,c^\dag_{i\si}
  \,.
\end{equation}
Substitution in equation~(\ref{eq:polaron:fprime}), integration with
respect to $\eta$ and setting $\eta=1$ results in
\begin{equation}
 \tilde{c}_{i\si}^\dag
  =
  c_{i\si}^\dag\, \rme^{ \rmi\sum_j\gamma_{ij}\op_j}
  \,,\qquad
  \tilde{c}_{i\si}
  =
  c_{i\si}\, \rme^{-\rmi\sum_j\gamma_{ij}\op_j}
  \,.
\end{equation}
For phonon operators, the relation
\begin{equation}
  \tilde{\hat{A}}
  =
  \rme^{\hat{S}} \hat{A} \rme^{-\hat{S}}
  =
  \hat{A} + \com{\hat{S}}{\hat{A}} + \frac{1}{2!}\com{\hat{S}}{\com{\hat{S}}{\hat{A}}} + \cdots
  \,,
\end{equation}
yields
\begin{equation}\label{eq:polaron:trop}
  \tilde{\ox}_i
  =
  \ox_i + \sum_j \gamma_{ij} \on_j
  \,,\qquad
  \tilde{\op}_i\;
  =
  {\op}_i
  \,.
\end{equation}
Collecting these results, the transformation of the Hamiltonian~(\ref{eq:H}) leads to
\begin{eqnarray}\label{eq:LF-H-extended}\nonumber
  \hspace{-0.75em}
  \tilde{H}
  &=&
  \underbrace{%
  -t\sum_{\las ij\ras\si}
  c^\dag_{i\si}c^{\nag}_{j\si} \rme^{\rmi\sum_l(\gamma_{il}-\gamma_{jl})\op_l}}
  _{\tilde{H}_\text{kin}}
  +
  \underbrace{%
    \frac{\om_0}{2}\sum_{\phantom{\las\ras}i\phantom{\las\ras}}(\op^2_i+\ox^2_i)}_{\tilde H_\text{ph}\equiv \tilde H_\text{ph}^p+
    \tilde H_\text{ph}^x}
  +
  \underbrace{%
  \sum_{ij} \on_j
  \ox_i(\om_0\gamma_{ij}-g'\delta_{ij})}
  _{\tilde{H}_\text{ep}}
  \\
  &&
  +
  \underbrace{%
  \sum_{ij} \on_i \on_j\hspace{-0.25em}
  \left(
    \frac{\om_0}{2}\sum_l\gamma_{lj}\gamma_{li}-g'\gamma_{ij} +
    \frac{U}{2}\delta_{ij} +
    V\delta_{\las ij\ras}
    \right)
    - \mbox{\small$\frac{1}{2}$}U\sum_i \on_i}_{\tilde{H}_\text{ee}}
  .
\end{eqnarray}
Here the term $\tilde{H}_\text{ep}$ describes the coupling between electrons
and phonons, whereas $\tilde{H}_\text{ee}$ represents an effective
el-el interaction. Hamiltonian~(\ref{eq:LF-H-extended}) will be the starting point
for the variational approach in section~\ref{sec:vpa}.

For QMC simulations, it is more suitable to require that
the el-ph terms in $\tilde{H}_\mathrm{ep}$ cancel. This can be
achieved by setting $\gamma_{ij} = \gamma\delta_{ij}$ with
\begin{equation}\label{eq:polaron:gamma}
  \gamma
  =
  \sqrt{\frac{\lambda W}{\om_0}}
  \,.
\end{equation}
The parameter $\gamma$ corresponds to the distortion which minimizes the potential energy
of the shifted harmonic oscillator $E_\text{pot}=\frac{\omega_0}{2} x^2 - g'x$.
This leads us to the standard LF transformation
\begin{equation}\label{eq:polaron:LFop}
  \hat{\Phi}_0
  =
  \rme^{S_0}
  \,,\quad
  S_0=\rmi\gamma\sum_i \on_i \op_i
  \,,
\end{equation}
and the familiar results for the transformed operators
\begin{equation}
   \tilde{c}^\dag_{i\si}
   =
   c^\dag_{i\si}\rme^{\rmi\gamma \op_i}
   \,,\qquad
   \tilde{c}_{i\si}
   =
   c_{i\si}\rme^{-\rmi\gamma \op_i}
\end{equation}
and
\begin{equation}\label{eq:polaron:LFvariables}
  \tilde{\ox}_i
  =
  \ox_i + \gamma \on_i
  \,,\qquad
  \tilde{\op}_i
  =
  \op_i
  \,.
\end{equation}

In contrast to the non-local transformation~(\ref{eq:polaron:U}),
only the oscillator at the site of the electron is displaced. The transformed
Hamiltonian reads
\begin{eqnarray}\label{eq:LF-H-local}\nonumber
  \tilde{H}
  &=&
  \underbrace{%
  -t\sum_{\las ij\ras\si}c^\dag_{i\si}c^{\nag}_{j\si}
  \rme^{\rmi\gamma(\op_i-\op_j)}}
  _{\tilde{H}_\text{kin}}
  +
  \underbrace{%
    \frac{\om_0}{2}\sum_{\phantom{\las\ras}i\phantom{\las\ras}}(\op^2_i+\ox^2_i)}_{\tilde H_\text{ph}}
  \\
  &&
  +
  \underbrace{%
    (\mbox{\small$\frac{1}{2}$}U-\Ep)
    \sum_i\on_i^2
    +
    V \sum_{\las ij\ras}\on_i \on_j
    -
    \mbox{\small$\frac{1}{2}$}U
    \sum_i \on_i
  }_{\tilde{H}_\text{ee}}
  \,.
\end{eqnarray}

As we shall discuss in detail in section~\ref{sec:qmc}, the difficulties
encountered in QMC simulations of the original Hamiltonian~(\ref{eq:H}) are
to a certain extent related to (bi-)polaron effects, \ie, to the dynamic
formation of spatially rather localized lattice distortions which
surround the charge carriers and follow their motion in the lattice.

For a single electron, the aforementioned {\it Holstein-Lang-Firsov} (HLF)
approximation \cite{Ho59a} becomes exact in the non-adiabatic SC or
small-polaron limit, and agrees qualitatively with exact results also in the
intermediate-coupling (IC) regime \cite{ZhJeWh99}. Although it overestimates
the shift $\gamma$ of the equilibrium position of the oscillator in the
presence of an electron, and does not reproduce the retardation effects when
the electron hops onto a previously unoccupied site, the approximation
mediates the crucial impact of el-ph interaction on the lattice.
Consequently, the transformed Hamiltonian~(\ref{eq:LF-H-local}) can be
expected to be a good starting point for QMC simulations, which then
merely need to account for the rather small fluctuations around the (shifted
or unshifted) equilibrium positions. In principle, it would also be possible
to develop a QMC algorithm based on the
Hamiltonian~(\ref{eq:LF-H-extended})---the basis of our variational
approach---with the parameters $\gamma_{ij}$ determined variationally, but
the local LF transformation proves to be sufficient.

The Hamiltonian~(\ref{eq:LF-H-local}) does no longer contain a term coupling
the electron density $\on$ and the lattice displacement $\ox$.  By contrast,
the extended transformation does not eliminate the interaction term
completely. On top of that, the hopping term involves all phonon momenta
$\op_i$ as well as the parameters $\gamma_{ij}$, and the el-el interaction
becomes long ranged [cf equation~(\ref{eq:LF-H-extended})].

For spin dependent carriers with $\on_i^2\neq\on_i$, the interaction term
$\tilde{H}_\text{ee}$ contains a Hubbard-like attractive interaction. Whereas the
latter can be treated exactly in the case of two electrons
(section~\ref{sec:partition-function-two-electrons}), the many-electron case
requires the introduction of auxiliary fields which complicate simulations.
However, no such difficulties arise for the spinless Holstein model
considered in section~\ref{sec:results}.

\section{Variational approach}
\label{sec:vpa}

For simplicity, we shall restrict the following derivation to one dimension;
an extension to $\mathrm{D}>1$ is straight forward.  Furthermore, we only
consider finite clusters with periodic boundary conditions, although infinite
systems may also be treated. The results of this section have originally been
presented in \cite{HoEvvdL03,HovdL05}.

\subsection{One electron}

As noted before, the simple variational method presented here is based on the
extended transformation~(\ref{eq:polaron:U}), leading to the
Hamiltonian~(\ref{eq:LF-H-extended}).  We treat the $\gamma_{ij}$ as
variational parameters which are determined by minimizing the ground-state
energy in a zero-phonon basis in which $\las \tilde{H}_\mathrm{ep}\ras=0$.

For systems with translation invariance the {\em displacement fields} satisfy the
condition $\gamma_{ij}=\gamma_{|i-j|}$. Together with $\sum_i\on_i=1$ for a
single electron we get
$\tilde{H}_\text{ee}=\frac{\om_0}{2}\sum_l\gamma^2_l-g'\gamma_0$.

The eigenvalue problem of the transformed
Hamiltonian~(\ref{eq:LF-H-extended}) is solved by making the following ansatz
for the one-electron basis states
\begin{equation}\label{eq:polaron:basis_vpa}
  \left\{
  \ket{l}
  = c^\dag_{l\si}\ket{0}
  \otimes
  \prod_{\nu=1}^N \sket{\phi_0^{(\nu)}}
  \,,\quad
  l=1,\dots,N
  \right\}
  \,,
\end{equation}
where $\sket{\phi_0^{(\nu)}}$ denotes the ground state of the harmonic
oscillator at site $\nu$. The non-zero matrix elements of the hopping term are
\begin{eqnarray}\label{eq:polaron:K}\nonumber
   \bra{l}\tilde{H}_\text{kin}\ket{l'}
   &=&
   -t\delta_{\la ll'\ra} \prod_\nu \sbra{\phi^{(\nu)}_0}
   \rme^{\rmi(\gamma_{l\nu}-\gamma_{l'\nu})\op_\nu}
   \sket{\phi^{(\nu)}_0}
   \\\nonumber
   &=&
   -t\delta_{\la ll'\ra}\;\prod_\nu\;\int\,\rmd x\,
   \phi(x+\gamma_{l\nu})\phi(x+\gamma_{l'\nu})
   \\
   &=&
   -t\delta_{\la ll'\ra} \rme^{-\frac{1}{4}\sum_\nu(\gamma_{\nu}-\gamma_{\nu+l-l'})^2}
   \,,
\end{eqnarray}
where $\phi(x)$ denotes the real-space wavefunction of the harmonic-oscillator
ground state. The Kronecker symbol $\delta_{\la ll'\ra}$ forces $l$ and $l'$ to represent
nearest-neighbor sites. A simple calculation gives for
the other terms in equation~(\ref{eq:LF-H-extended})
\begin{equation}\label{eq:polaron:matelem}
  \bra{l}\tilde{H}_\text{ph}\ket{l'}
  =
  \delta_{ll'}\frac{\om_0}{2}
  \,,\,\,\,
  \bra{l}\tilde{H}_\text{ep}\ket{l'}
  =
  0
  \,,\,\,\,
  \bra{l}\tilde{H}_\text{ee}\ket{l'}
  =
  \delta_{ll'}\left(
    \frac{\om_0}{2}\sum_l\gamma_l^2
    -g'\gamma_0
  \right)
  .
\end{equation}
In the zero-phonon subspace spanned by the basis
states~(\ref{eq:polaron:basis_vpa}), the eigenstates of
Hamiltonian~(\ref{eq:LF-H-extended}) with momentum $k$ are
\begin{equation}\label{eq:polaron:eigen}
  \sket{\psi_k}
  =
  c^\dag_{k\si}\ket{0}\otimes\prod_\nu\sket{\phi_0^{(\nu)}}
\end{equation}
with eigenvalues
\begin{equation}\label{eq:polaron:E(k)}
  E(k)
  =
  \Ekin
  + \frac{N\om_0}{2} + \frac{\om_0}{2}\sum_l\gamma_l^2-g'\gamma_0
\end{equation}
and the kinetic energy
\begin{equation}\label{eq:polaron-vpa:Ekin}
  \Ekin
  =
  -t\sum_{\delta=\pm1} \rme^{\rmi k\delta}
  \rme^{-\frac{1}{4}\sum_\nu(\gamma_\nu-\gamma_{\nu+\delta})^2}
  \,.
\end{equation}
Defining the Fourier transform
\begin{equation}\label{eq:polaron:FTgamma}
  \overline{\gamma}_q
  =
  \frac{1}{\sqrt{N}}\sum_l \rme^{\rmi ql}\gamma_l
\end{equation}
and using ($\gamma_l \in \Rs$)
\begin{equation}
  \sum_\nu \gamma_\nu \gamma_{\nu+\delta}
  =
  \sum_q\overline{\gamma}_q\overline{\gamma}_{-q} \rme^{\rmi q\delta}
  =
  \sum_q\overline{\gamma}_q^2\cos q\delta
  \,,
\end{equation}
we may write
\begin{equation}
  \Ekin
  =
  -t\sum_\delta \rme^{\rmi k\delta} \rme^{-\frac{1}{2}
  \sum_q(1-\cos q\delta)\overline{\gamma}_q^2}
  =
  \varepsilon_0(k) \rme^{-\frac{1}{2} \sum_q(1-\cos q)\overline{\gamma}_q^2}
  =
  \varepsilon(k)
\end{equation}
with the tight-binding dispersion $\varepsilon_0(k)=-2t\cos k$. Hence
the ground-state energy becomes
\begin{equation}\label{eq:polaron:evs}
  E(k)
  =
  \varepsilon(k) + \frac{N\om_0}{2}
  +\frac{\om_0}{2}\sum_q\overline{\gamma}_q^2 -
  \frac{g'}{\sqrt{N}}\sum_q\overline{\gamma}_q
  \,.
\end{equation}
The variational parameters $\overline{\gamma}_p$ are determined by requiring
\begin{equation}\label{eq:polaron:deriv}
  \frac{\partial E}{\partial \overline{\gamma}_p}
  =
  -\overline{\gamma}_p \varepsilon(k)
  (1-\cos p) + \om_0 \overline{\gamma}_p -
  \frac{g'}{\sqrt{N}}
  \overset{!}{=}0
  \,,
\end{equation}
so that the optimal values $\overline{\gamma}_p$ can be obtained from
\begin{equation}\label{eq:polaron:gammap}
  \overline{\gamma}_p
  =
  \frac{g'}{\sqrt{N}}
  \frac{1}{\om_0 +
  \varepsilon(k)(1-\cos p)}
  \,.
\end{equation}
Since $\varepsilon(k)$ depends implicitly on the $\overline{\gamma}_p$,
equation~(\ref{eq:polaron:gammap}) has to be solved self-consistently. It has
the typical form of the random-phase approximation since a variational ansatz
for the untransformed Hamiltonian may be written as
\begin{equation}\label{eq:polaron:rpa_type}
 \hat{\Phi}^\dag\sket{\psi_k}
 =
 \frac{1}{\sqrt{N}} \sum_j
 \rme^{\rmi k j} \,c^\dag_{j\si} \,
 \rme^{-\rmi \sum_l \gamma_{jl} \hat p_l}\,
 \sket{0}\otimes\prod_\nu \sket{\phi_0^{(\nu)}}
 \,,
\end{equation}
with $\hat{\Phi}$ as defined in equation~(\ref{eq:polaron:U}).

We shall also calculate the quasiparticle spectral weight for momentum $k=0$,
defined as
\begin{equation}\label{eq:polaron:def_z0}
  \sqrt{z_0}
  =
  \bra{0} \tilde{c}_{k=0,\si}\ket{\psi_0}
  \,.
\end{equation}
Here $\ket{\psi_0}$ denotes the ground state with one electron of momentum
$p=0$ and the oscillators in the ground state $\ket{\phi_0}$. Fourier
transformation and the same manipulations as in equation~(\ref{eq:polaron:K})
lead to
\begin{eqnarray}\label{eq:polaron:z0}\nonumber
  \sqrt{z_0}
  &=&
  \frac{1}{N}\sum_{ij}\bra{\phi_0}\bra{0}
  \tilde{c}^{\phantom{\dag}}_{i\si}c^\dag_{j\si}\ket{0}\ket{\phi_0}
  \\\nonumber
  &=&
  \frac{1}{N}\sum_i \bra{\phi_0} \rme^{-\rmi\sum_k\gamma_{ik} \op_k}\ket{\phi_0}
  \\
  &=&
  \rme^{-\frac{1}{4}\sum_q\tilde{\gamma}_q^2}
  \,.
\end{eqnarray}

Just as the HLF approximation, the present variational method becomes exact in the
non-interacting limit ($\lambda=0$) and in the non-adiabatic SC
limit. Furthermore, it yields the correct results both for $\alpha=0$
(classical phonons) and $\alpha=\infty$, and also gives accurate results for
large $\alpha$ and finite $\lambda$, since the displacements of the
oscillators---only local and generally overestimated in the HLF
approximation---are determined variationally.

\subsection{Two electrons}

As in the one-electron case, the use of a zero-phonon basis leads to $\las\tilde{H}_\text{ep}\ras=0$
and, neglecting the ground-state energy of the oscillators, we also have
$\las \tilde{H}_\text{ph}\ras=0$. Hence,
$\tilde{H}=\tilde{H}_\text{kin}+\tilde{H}_\text{ee}$ with the transformed hopping term
\begin{equation}
  \tilde{H}_\text{kin}
  =
  -t_\text{eff}\sum_{\las ij\ras\si}
  c^\dag_{i\si}c^{\nag}_{j\si}
  =
  \sum_{k\si} \varepsilon(k)\; c^\dag_{k\si}c^{\nag }_{k\si}
\end{equation}
and $\varepsilon(k)=- 2\; t_\text{eff} \cos(k)$.
Here the effective hopping
\begin{equation}\label{eq:bipolaron:teff}
  t_\text{eff}
  =
  \frac{1}{2}\sum_{\delta=\pm1}
  \rme^{-\frac{1}{4}\sum_l(\gamma_{l-\delta}-\gamma_{l})^2}
  \,,
\end{equation}
where rotational invariance has been exploited. For two electrons of opposite
spin (\ie, $\on_{i\sigma}\on_{j\sigma}=0$ for $i\ne j$) and $V=0$,
$\tilde{H}_\mathrm{ee}$ in equation~(\ref{eq:LF-H-extended}) reduces to
\begin{equation}\label{eq:bipolaron:Iee_vij}
  \tilde{H}_\text{ee}
  =
  2 v_0 - U + 2 \sum_{ij} v_{ij} \on_{i\UP} \on_{j\DO}
  \,,\quad
  v_{ij}
  =
  \frac{\om_0}{2}
  \sum_l\gamma_{lj}\gamma_{li}-g'\gamma_{ij}
  + \mbox{\small$\frac{1}{2}$}\delta_{ij} U
  \,.
\end{equation}
The eigenstates of the two-electron problem have the form
\begin{equation}\label{eq:bipolaron:states_k}
  \ket{\psi_k}
  =
  \sum_p  \overline{d}^{\nag}_p c^\dag_{k-p\DO} c^\dag_{p\UP}\ket{0}
  \,,
\end{equation}
suppressing the phonon component
[cf equation~(\ref{eq:polaron:eigen})], and may be written as
\begin{equation}\label{eq:bipolaron:psi_rs}
  \ket{\psi_k}
  =
  \frac{1}{\sqrt{N}}\sum_i \rme^{\rmi k x_i}
  \sum_l d^{\nag}_l c^\dag_{i\DO} c^\dag_{i+l\UP}\ket{0}
  \,,
\end{equation}
with the Fourier transform
\begin{equation}\label{eq:bipolaron:FT}
  \bm{d}
  =
  F \overline{\bm{d}}
  \,,\quad
  (F)_{lp} = \rme^{\rmi  x_l p} / \sqrt{N}
  \,.
\end{equation}
The normalization of equation~(\ref{eq:bipolaron:states_k}) reads
\begin{equation}
  \bra{\psi_k}\psi_k\ras
  =
  \sum_p |d_p|^2\,.
\end{equation}
Using equation~(\ref{eq:bipolaron:states_k}), we find
for the expectation value of $\tilde{H}_\text{kin}$
\begin{eqnarray}\nonumber
  \bra{\psi_k}\tilde{H}_\text{kin}\ket{\psi_k}
   &=&
   \sum_{pp'}
   \overline{d}_p^* \overline{d}_{p'}^{\phantom{*}}
   \sum_{q} \varepsilon(q)
   \\\nonumber
   &&
   \times
   \bigg(
   \underbrace{%
     \bra{0}
     c^{\nag }_{p\UP} c^{\nag}_{k-p\DO}
     \on^{\nag }_{q\UP}
     c^{\dag }_{k-p'\DO} c^\dag_{p'\UP}\ket{0}
     }
     _{\delta_{p,p'}\delta_{q,p}}
   +
   \underbrace{%
     \bra{0}
     c^{\nag }_{p\UP} c^{\nag}_{k-p\DO}
     \on^{\nag }_{q\DO}
     c^{\dag }_{k-p'\DO} c^\dag_{p'\UP}\ket{0}
     }
     _{\delta_{p,p'}\delta_{q,k-p}}
   \bigg)
   \\\nonumber
   &=&
   \sum_p |\overline{d}_p|^2 \left[\varepsilon(p) + \varepsilon(k-p)\right]
   \\
   &=&
   -4\;t_\text{eff}\,\bm{d}^\dag T_k \bm{d}
   \,.
\end{eqnarray}
In the last step we have introduced $T_k=\frac{1}{2}F
\,\text{diag}[\cos (p) + \cos (k-p)\big]\,F^\dag$ and made use of
equation~(\ref{eq:bipolaron:FT}).

The expectation value of the interaction term, best computed in the
real-space representation~(\ref{eq:bipolaron:psi_rs}), takes the form
\begin{eqnarray}\nonumber
  \bra{\psi_k}\tilde{H}_\text{ee}\ket{\psi_k}
  &=&
  (2v_0-U)\sum_l |d_l|^2
  + \frac{2}{N} \sum_{ij} v_{ij}
  \sum_{j'j''}\sum_{ll'}
  d_l^* d^{\phantom{*}}_{l'}
   \rme^{\rmi k (x_{l\phantom{'}}-x_{l'})}
  \\\nonumber
  &&
   \qquad\qquad\qquad\qquad\qquad\quad
   \times
  \underbrace{
    \bra{0}
    c^{\nag}_{j' + l\UP}  c^{\nag}_{j'\DO}
    \on^{\nag}_{i\UP} \on^{\nag}_{j\DO}
    c^\dag_{j''\DO} c^\dag_{j'' + l'\UP}
    \ket{0}}_{
    \delta_{jj'}\delta_{jj''}\delta_{i,j+l}\delta_{l,l'}
    }
  \\\nonumber
  &=&
  (2v_0-U) \sum_l |d_l|^2
  + \frac{2}{N} \sum_{jl} v_{j+l,j} |d_{l}|^2
  \\
  &=&
  (2v_0-U) \bm{d}^\dag \bm{d} + 2\bm{d}^\dag V \bm{d}
  \,,
\end{eqnarray}
with the diagonal matrix $V_{ij}=\delta_{ij} v_{i}$.

The minimization of the total energy with respect to $\bm{d}$ yields the
eigenvalue problem
\begin{equation}\label{eq:bipolaron:minimize}
  (-4t_\text{eff} \;T_k + 2 V)\,\bm{d}
  =
  (E_0 - 2v_0 + U)\,\bm{d}
  \,.
\end{equation}
The vector of coefficients $\bm{d}$ and thereby the ground state are found
by minimizing the ground-state energy $E_0$ through variation of the
displacement fields $\gamma_{ij}$. Similar to the one-electron case, this
procedure takes into account displacements of the oscillators not only at the
same but also at surrounding sites of the two electrons, and is therefore
capable of describing extended bipolaron states (see
section~\ref{sec:res-bipolaron}). Note that the two-electron problem is
diagonalized exactly without phonons (\ie, for $\lambda=0$).

\section{Quantum Monte Carlo}
\label{sec:qmc}

In this section, we present an overview of our recently developed QMC
algorithms for Holstein-type models
\cite{HoEvvdL03,HoEvvdL05,HovdL05,HoNevdLWeLoFe04}.

As mentioned before, in contrast to the variational approach, the QMC
approaches discussed here, based on the local LF
transformation~(\ref{eq:LF-H-local}) which does not contain any free
parameters, are unbiased.
They yield exact results with only statistical errors
that can in principle be made arbitrarily small.

The motivation for the development of improved QMC schemes for Holstein
models stems from the fact that calculations with existing methods often
suffer from strong autocorrelations, \ie, non-negligible statistical
correlations between successive MC configurations
\cite{wvl1992,HoEvvdL03}. In fact, autocorrelations may render accurate
simulations impossible within reasonable computing time. As discussed
in \cite{HoEvvdL03}, the problem becomes particularly noticeable for small
phonon frequencies and low temperatures.

Whereas autocorrelations can be avoided to a large extent for one or two
electrons by integrating out the phonons analytically, no efficient general schemes
exist for finite charge-carrier densities (see discussion in
\cite{HoEvvdL03}).

In the sequel, we present a general (\ie, applicable for
all densities) solution for this problem in several steps.
First, the effects due to el-ph interaction are separated from the free
lattice dynamics by means of the LF transformation~(\ref{eq:LF-H-local}).
Since the latter contains the crucial impact of the electronic degrees of
freedom on the lattice, simulations may be based only on the purely phononic
part of the resulting action. The fermionic degrees of freedom can then be
taken into account exactly by {\it reweighting} of the probability
distribution.  Consequently, we may completely ignore the electronic weights
in the updating process, and thereby dramatically reduce the computational
effort. The {\it principal component representation} of the phonon
coordinates allows exact sampling and avoids any autocorrelations.

\subsection{Partition function}\label{sec:partition-function}

We begin by deriving the partition function for the case of a single
electron. Then we discuss the differences occurring in the cases of two or
more carriers.

\subsubsection{One electron}\label{sec:partition-function-one-electron}

The partition function is defined as
\begin{equation}
  \Z=\tr \rme^{-\beta\tilde{H}}
\end{equation}
with $\tilde{H}$ given by equation~(\ref{eq:LF-H-local}) and the inverse
temperature $\beta=(k_\text{B}T)^{-1}$. For a single
electron, $\tilde{H}_\text{ee}=-\Ep$ becomes a constant which needs only to be considered in
calculating the total energy.

Using the Suzuki-Trotter decomposition \cite{wvl1992}, we obtain
\begin{equation}\label{eq:polaron:suzuki-trotter}
  \rme^{-\beta\tilde{H}}
  \approx
  (\rme^{-\dtau\tilde{H}_\text{kin}} \rme^{-\dtau \tilde{H}_\text{ph}^p} \rme^{-\dtau \tilde{H}_\text{ph}^x})^L
  \equiv
  \U^L
  \,,
\end{equation}
where $\dtau=\beta/L\ll1$. Splitting up the trace
into a bosonic and a fermionic part and inserting $L-1$ complete sets of oscillator
momentum eigenstates we find the approximation
\begin{equation}
  \Z_L
  =
  \tr_\text{f}\int\,\rmd p_1\rmd p_2\cdots\rmd p_L
  \bra{p_1}\U\ket{p_2}\cdots\bra{p_L}\U\ket{p_1}
\end{equation}
with $\rmd p_\tau\equiv\prod_i \rmd p_{i,\tau}$. Each matrix element can be
evaluated by inserting a complete set of phonon coordinate eigenstates
$\int\rmd x \sket{x}\sbra{x}$, since all $x$-integrals are of
Gaussian form and can easily be carried out. The result is
\begin{equation}
  \bra{p_\tau} \rme^{-\dtau \tilde{H}_\text{ph}^x}\ket{p_{\tau+1}}
  =
  C^{N^\rD}
  \rme^{-\frac{1}{2\om_0\dtau}\sum_i\left(p_{i,\tau}-p_{i,\tau+1}\right)^2}
  \,,\quad
  C
  =
  \sqrt{\frac{2\pi}{\om_0\dtau}}
  \,.
\end{equation}
The normalization factor in front of the exponential has to be taken into
account in the calculation of the total energy, but cancels when we measure
other observables. With the abbreviation $\mathcal{D}p=\rmd p_1\rmd
p_2\cdots\rmd p_L$ the partition function finally becomes
\begin{equation}\label{eq:polaron:Z}
  \Z_L
   =
   C^{N^\rD L} \int\,\mathcal{D}p\,\, \wb \,\wf
   \,,
\end{equation}
where
\begin{equation}\label{eq:polaron:omega}
   \wb
   =
   \rme^{-\dtau S_\text{b}}
   , ~~
   \wf
   =
   \tr_\text{f}\,\Omega
   ,~~
   \Omega
   =
   \prod_{\tau=1}^L \rme^{-\dtau\tilde{H}_\text{kin}^{(\tau)}}
   \,.
\end{equation}
Here $\tilde{H}_\text{kin}^{(\tau)}$ corresponds to $\tilde{H}_\text{kin}$ with the phonon operators $\op_i$,
$\op_j$ replaced by the momenta
$p_{i,\tau}$, $p_{j,\tau}$ on the $\tau$th Trotter slice, and its exponential
may be written as
\begin{equation}\label{eq:polaron:matrices}
  \rme^{-\dtau\tilde{H}_\text{kin}^{(\tau)}}
  =
  D_\tau \kappa D_\tau^\dag
  \,,\quad
  \kappa_{jj'}
  =
  \left(\rme^{\dtau t\,h^\text{tb}}\right)_{jj'}
  \,,\quad
  (D_\tau)_{jj'}
  =
  \delta_{jj'}\rme^{\rmi\gamma p_{j,\tau}}
  \,,
\end{equation}
where $h^\text{tb}$ is the $N^{\rD}\times N^{\rD}$ tight-binding hopping
matrix. To save some
computer time, we employ the checkerboard breakup \cite{LoGu92}
\begin{equation}\label{eq:many-electrons:checker}
  \rme^{\dtau t\sum_{\las ij\ras} c^\dag_i c^{\phantom{\dag}}_j}
  \approx
  \prod_{\las ij\ras}
  \rme^{\dtau t c^\dag_i c^{\phantom{\dag}}_j}
  \,.
\end{equation}
Using equation~(\ref{eq:many-electrons:checker}), the numerical effort scales as
$N^{2\rD}$ instead of $N^{3\rD}$ (see also section~\ref{sec:qmc_comparison}),
but the error due to this additional
approximation is of the same order $\dtau^2$ as the Trotter error in
equation~(\ref{eq:polaron:suzuki-trotter}).

According to equation~(\ref{eq:polaron:matrices}), we have the same matrix
$\kappa$ for every time slice, which is transformed by the diagonal unitary
matrices $D_\tau$.  The matrix $\Omega$ can be calculated in an efficient way
by noting that the transformation matrices $D^\dag_\tau$ and $D_{\tau+1}$ at
time slice $\tau$ may be combined to a diagonal matrix
\begin{equation}\label{eq:polaron:trickforM}
  (D_{\tau,\tau+1})_{ij}
  =
  \delta_{ij}
  \rme^{\rmi\gamma(p_{i,\tau+1}-p_{i,\tau})}
  \,.
\end{equation}
Due to the cyclic invariance of the fermionic trace, $D_1$ can be shifted to
the end of the product, where it combines with $D^\dag_L$ to
$D_{L,1}$. Hence we can write
\begin{equation}\label{eq:polaron:omega2}
  \Omega
  =
  \prod_{\tau=1}^L
  \kappa \,D_{\tau,\tau+1}
  \,,
\end{equation}
with periodic boundary conditions in imaginary time. In the one-electron
case, the fermionic weight $\wf=\sum_n \bra{n}\Omega\ket{n}$ is given by the
sum over the diagonal elements of the matrix representation of $\Omega$ in
the basis of one-electron states (dropping unnecessary spin indices)
\begin{equation}\label{eq:polaron:oneelbasis}
  \ket{n}
  =
  c^\dag_n\ket{0}
  \,.
\end{equation}
The bosonic action in equation~(\ref{eq:polaron:omega}) contains only
classical variables:
\begin{equation}\label{eq:polaron:action}
  S_\text{b}
  =
  \frac{\om_0}{2}\sum_{i,\tau}p_{i,\tau}^2 +
  \frac{1}{2\om_0\dtau^2}\sum_{i,\tau}
  \left(p_{i,\tau} - p_{i,\tau+1}\right)^2
  \,,
\end{equation}
where the indices $i=1,\dots,N^\rD$ and $\tau=1,\dots,L$ run over all lattice
sites and time slices, respectively, and
$p_{i,L+1}=p_{i,1}$. It may also be written as
\begin{equation}\label{eq:polaron:action-w-matrix}
  S_\text{b}
  =
  \sum_i \bm{p}_i^\text{T} A \bm{p}_i
\end{equation}
with $\bm{p}_i=(p_{i,1},\dots,p_{i,L})$ and a {\it periodic}, tridiagonal $L\times
L$ matrix $A$ with non-zero elements
\begin{equation}\label{eq:polaron:matrixA}
  (A)_{l,l}
  =
  \frac{\om_0}{2}+\frac{1}{\om_0\dtau^2}
  \,,\quad
  (A)_{l,l\pm1}
  =
  -\frac{1}{2\om_0\dtau^2}
  \,.
\end{equation}
Since $\Z_L$ is a trace, it follows that $(A)_{1,L}=(A)_{L,1}=-(2\om_0\dtau^2)^{-1}$.

\subsubsection{Two electrons}\label{sec:partition-function-two-electrons}

In contrast to \cite{HovdL05}, here we also take into account
nearest-neighbour Coulomb repulsion $V$. For two electrons, the
Hamiltonian~(\ref{eq:LF-H-local}) simplifies to
\begin{equation}\label{eq:LF-H-QMC-bipolaron}
  \tilde{H}
  =
  \tilde{H}_\text{kin}
  +
  \tilde{H}_\text{ph}
  +
  \tilde{H}_\text{ee}
  -2\Ep
  \,,\quad
  \tilde{H}_\text{ee}=
  (U-2\Ep)\sum_i\on_{i\UP}\on_{i\DO}
  +
  V\sum_{\las ij\ras} \on_i\on_j
  \,.
\end{equation}
Again, the constant shift can be neglected in the QMC simulation,
but in contrast to the single-electron case, we have a non-trivial
interaction term. The Suzuki-Trotter decomposition yields
\begin{equation}\label{eq:bipolaron:suzuki-trotter}
  \rme^{-\beta\tilde{H}}
  \approx
  \left(
    \rme^{-\dtau\tilde{H}_\text{kin}}
    \rme^{-\dtau \tilde{H}_\text{ph}^p}
    \rme^{-\dtau \tilde{H}_\text{ph}^x}
    \rme^{-\dtau \tilde{H}_\text{ee}}
  \right)^L \equiv \U^L\,.
\end{equation}
Using the same steps as above we obtain
\begin{equation}\label{eq:bipolaron:omega}
   \wb
   =
   \rme^{-\dtau S_\text{b}}
   \,,\quad
   \wf
   =
   \tr_\text{f}\,\Omega
   \,,
   \quad
   \Omega
   =
   \prod_{\tau=1}^L
   \rme^{-\dtau\tilde{H}_\text{kin}^{(\tau)}}
   \rme^{-\dtau\tilde{H}_\text{ee}}
   \,,
\end{equation}
with $S_\text{b}$ given by equation~(\ref{eq:polaron:action}).

As pointed out in \cite{HovdL05}, the numerical effort for two electrons
increases substantially in higher dimensions. Therefore, we restrict
ourselves to $\rD=1$.

Previously, we only considered the case of two electrons of opposite spin
(forming a singlet) \cite{HovdL05}. Here we shall also present results for
the triplet state.

\paragraph*{Singlet}
In the singlet case we choose the two-electron basis states
\begin{equation}\label{eq:bipolaron:basis}
 \left\{
   \ket{l}\equiv
   \ket{i,j}
   \equiv
  c^\dag_{i\UP}c^\dag_{j\DO}\ket{0}
  \,,\quad
  i,j = 1,\dots,N
  \right\}
  \,,
\end{equation}
where we have used a combined index $l=1,\dots,N^2$. The tight-binding hopping matrix, denoted as $\kappa$, has
dimension $N^2\times N^2$, and the corresponding exponential in
equation~(\ref{eq:bipolaron:omega}) can again be written as
$\rme^{-\dtau
\tilde{H}_\text{kin}^{(\tau)}
}=D_\tau \kappa D_\tau^\dag$
[cf equation~(\ref{eq:polaron:omega})], where
\begin{equation}
  (D_\tau)_{ll'}
  =
  \delta_{ll'}
  \rme^{\rmi\gamma(p_{i,\tau}+p_{j,\tau})}
\end{equation}
is diagonal in the basis~(\ref{eq:bipolaron:basis}).

The remaining contribution to $\Omega$ comes
from the effective el-el interaction term
$\tilde{H}_\text{ee}$ in terms of the sparse matrix
\begin{equation}
  (\V)_{ll'}
  =
  \sum_k
  (\delta_{lk}\;\rme^{-\dtau(U-2\Ep )\delta_{ij}})_{lk}
  (\rme^{-\dtau V\delta_{\las ij\ras}})_{kl'}
  \,.
\end{equation}
The momenta $\bm{p}$ merely enter the diagonal matrix $D$;
the $N^2\times N^2$ matrices $\V$ and $\kappa$ are fixed
throughout the entire MC simulation. Finally, we have
\begin{equation}\label{eq:bipolaron:matproduct}
  \Omega
  =
  \prod_\tau D_\tau \kappa D^\dag_\tau \V
  \,,
\end{equation}
and the fermionic trace can be calculated as the sum over the diagonal
elements of the matrix $\Omega$ in the basis~(\ref{eq:bipolaron:basis}), \ie,
\begin{equation}
  \tr_\text{f}\,\Omega
  =
  \sum_{ij} \bra{i,j} \Omega \ket{i,j}
  \,.
\end{equation}

\paragraph*{Triplet}

For two electrons with parallel spin we use the basis states
\begin{equation}\label{eq:bipolaron:basis_triplet}
 \left\{
   \ket{l}\equiv
   \ket{i,j}
   \equiv
  c^\dag_{i}c^\dag_{j}\ket{0}
  \,,\quad
  i = 1,\dots,N
  \,,\,
  j=i+1,\dots,N
  \right\}
  \,,
\end{equation}
\ie, double occupation of a site is not possible. Since we can further not
distinguish between the states $\ket{i,j}$ and $\ket{j,i}$, the dimension of
the electronic Hilbert space is reduced from $N^2$ (singlet case) to
$N(N-1)/2$. Consequently, for the same system size, simulations for the
triplet case will be much faster.

\subsubsection{Many-electron case}\label{sec:partition-function-many-electrons}

The one-electron QMC algorithm can easily be extended to the spinless
Holstein model with many electrons.  For the latter, assuming $V=0$, the interaction term
in equation~(\ref{eq:LF-H-local}) reduces to $\tilde{H}_\text{ee}=-\Ep\sum_i
\on_i$. Therefore, the grand-canonical Hamiltonian becomes
\begin{equation}\label{eq:many-electrons:Hspinless}
  \tilde{\H}
  =
  \tilde{H} - \mu \sum_i \on_i
  =
  \underbrace{%
  -t\sum_{\las ij\ras} c^\dag_i c^{\nag}_j \rme^{\rmi\gamma(\op_i - \op_j)}
  }_{\tilde{H}_\text{kin}}
  +
  \tilde{H}_\text{ph}
  -
  \underbrace{%
  (\Ep + \mu) \sum_{\phantom{\las}i\phantom{\ras}} \on_i
  }_{\tilde{H}'_\text{ee}}
  \,,
\end{equation}
where $\mu$ denotes the chemical potential. For half filling
$n=0.5$ [$N/2$ spinless fermions on $N$ sites,
cf equation~(\ref{eq:many-electrons:n})], the latter is
given by $\mu=-\Ep$, whereas for $n\neq 0.5$, it has to be adjusted to yield
the carrier density of interest.

The approximation to the partition function may again be cast into the form
of equation~(\ref{eq:polaron:Z}), with $\wb$ as defined by
equations~(\ref{eq:polaron:omega}) and~(\ref{eq:polaron:action}),
respectively. The fermionic weight is given by
\begin{equation}
  \wf
  =
  \tr_\text{f}(\oB_1\oB_{2}\cdots\oB_L)
  \,,\quad
  \oB_\tau
  =
  \rme^{-\dtau\tilde{H}_\text{kin}^{(\tau)}}
  \rme^{-\dtau \tilde{H}'_\text{ee}}
  \,.
\end{equation}
Following Blankenbecler \etal \cite{BlScSu81}, the fermion degrees of freedom
can be integrated out exactly leading to
\begin{equation}\label{eq:many-electrons:omega}
  \wf
   =
   \det (1 +  B_1\,\cdots\,B_L)
   \equiv
   \det (1 + \Omega)\,,
\end{equation}
where the $N^\rD\times N^\rD$ matrix $B_\tau$ is given by
\begin{equation}\label{eq:many-electrons:matprod}
  B_\tau
  =
  D_\tau\,\kappa\,D^\dag_\tau\,\V
  \,.
\end{equation}
Here $\kappa$ and $D_\tau$ are identical to
equation~(\ref{eq:polaron:matrices}), and
\begin{equation}
  (\V)_{ij}
  =
  \delta_{ij}\,\rme^{\dtau(\Ep+\mu)}
  \,.
\end{equation}

There is a close relation to the one-electron Green function
\begin{equation}
  G_{ij}
  =
  \underbrace{%
  \las \tilde{c}^{\nag}_i \tilde{c}^\dag_j\ras
  }_{G^{a}_{ij}}
  +
  \underbrace{%
  \las \tilde{c}^{\dag}_i \tilde{c}^{\nag}_j\ras
  }_{G^{b}_{ij}}
  \,.
\end{equation}
In real space and imaginary time, we have \cite{BlScSu81,Hi85}
\begin{equation}\label{eq:many-electrons:GA}
  G^{a}_{ij}
  =
  \las \tilde{c}^{\nag}_i \tilde{c}^\dag_j\ras
  =
  (1+\Omega)^{-1}_{ij}
  \,,\quad
  G^{b}_{ij}
  =
  \delta_{ij} - G^{a}_{ij}
  =
  (\Omega\,G^{a})_{ji}
  \,.
\end{equation}

At this stage, with the above results for the partition function, a QMC
simulation of the transformed Holstein model would proceed as follows. In
each MC step, a pair of indices $(i_0,\tau_0)$ on the $N^\rD\times L$ lattice of
phonon momenta $p_{i,\tau}$ is chosen at random. At this site, a change
$p_{i_0,\tau_0}\mapsto p_{i_0,\tau_0}+\Delta p$ of the phonon configuration
is proposed. To decide upon the acceptance of the new configuration using the
Metropolis algorithm \cite{wvl1992}, the corresponding
weights $\wb\wf$ and $\wb'\wf'$ have to be calculated. Due to the local
updating process, the computation of the change of the bosonic weight
$\Delta\wb=\wb'/\wb$ is very fast, which is not the case for the fermionic weight
$\Delta\wf=\wf'/\wf$. By varying $\tau_0$ sequentially from 1 to
$L$ instead of picking random values, the calculation of the ratio of the
fermionic weights can be reduced to only two matrix multiplications.

It turns out that a local updating as described above does not
permit efficient simulations for small phonon frequencies or low
temperatures. Therefore, we shall introduce an alternative global
updating in terms of principal components in section~\ref{sec:qmc-pc}.

\subsection{Observables}\label{sec:QMC_observables}

Using the transformed Hamiltonian~(\ref{eq:LF-H-local}),
the expectation value
\begin{equation}
  \las O\ras
  =
  \Z^{-1}\,\tr\, \hat{       O} \,\rme^{-\beta        H}
  =
  \Z^{-1}\,\tr\, \hat{\tilde{O}}\,\rme^{-\beta \tilde{H}}
\end{equation}
of an observables $O$ is computed according to
\begin{equation}
  \las O\ras
  = \Z^{-1}\,\tr_\text{f} \int\,\rmd p\,
  \bra{p}\hat{\tilde{O}}\,\rme^{-\beta\tilde{H}}\ket{p}
  \,.
\end{equation}

As a result of the analytic integration over the phonon coordinates $\ox$,
interesting observables such as the correlation function $\las\on_i
\ox_j\ras$ are difficult to measure accurately. Other quantities
such as the quasiparticle weight, and the closely related effective mass
\cite{WeRoFe96}, can be determined from the one-electron Green function at
long imaginary times \cite{BrCaAsMu01}, but results for one electron or two
electrons would not be as accurate
as in existing work (\eg, \cite{JeWh98,RoBrLi99III,KuTrBo02}).

The situation is strikingly different in the many-electron
case, for which many methods fail to produce results of high accuracy for
large systems and physically relevant parameters.
Moreover, other important observables, such as the one-electron Green
function, can be calculated with our approach.

\subsubsection{One electron}\label{sec:obs-one-electron}

The electronic kinetic energy is defined as
\begin{equation}\label{eq:polaron:Ekin}
  \Ekin
  =
  \las \tilde{H}_\text{kin} \ras
  =
  -t \Z^{-1}\,
  \sum_{\las ij\ras}\text{Tr}\;
  \big( \,c_i^\dag c^{\nag}_j\,
  \rme^{\rmi\gamma(\op_i-\op_j)} \,\rme^{-\beta \tilde{H}}\,
  \big)
  \,.
\end{equation}
Repeating the steps used to derive the partition function, and
noting that the additional phase factors in equation~(\ref{eq:polaron:Ekin})
again lead to the same matrix $\Omega$ as in
equation~(\ref{eq:polaron:omega2}), we find
\begin{eqnarray}\nonumber
  \Ekin
  &=&
  -t \Z^{-1}_L\,\sum_{\las ij\ras}\int\,\D p\,\wb\sum_n
  \bra{n}\Omega c^\dag_i c^{\nag}_j\ket{n}
  \\
  &=&
  -t \Z^{-1}_L\,\sum_{\las ij\ras}\int\,\D p\,\wb
  \bra{j}\Omega\ket{i}
\end{eqnarray}
with the one-electron states~(\ref{eq:polaron:oneelbasis}).
Introducing the matrix elements $(\Omega)_{ij} = \bra{i}\Omega\ket{j}$ and the
expectation value with respect to $\wb$,
\begin{equation}\label{eq:polaron:Ob}
  \las O \ras_\text{b}
  = \frac{\int\,\D p\,\wb\; O(p)}{ \int\,\D p\,\wb}
\end{equation}
we obtain
\begin{equation}\label{eq:polaron:Ek}
  \Ekin
  =
  -t\; \frac{\sum_{\las ij\ras}\;\las \Omega_{ji}\ras_\text{b}}
  {
    \sum_{i}\;\las \Omega_{ii}\ras_\text{b}
  }
  \,.
\end{equation}
Here we have anticipated the reweighting discussed in
section~\ref{sec:qmc-reweighting}.

The total energy can be obtained from
$E=-\partial(\ln\Z)/\partial\beta$ as
\begin{eqnarray}\label{eq:polaron:E0}\nonumber
  E
  &=&
  \Ekin + \frac{\om_0}{2}\sum_i\la p_i^2\ra
  + E'_\text{ph} - \Ep
  \,,
  \\
  E'_\text{ph}
  &=&
  \frac{N^\rD}{2\dtau}-\frac{1}{2\om_0\dtau^2L}
  \sum_{i,\tau}\la\left(p_{i,\tau}-p_{i,\tau+1}\right)^2\ra
  \,.
\end{eqnarray}
To compare with other work we subtract the ground-state energy of the phonons,
$E_{0,\text{ph}}=N^\rD\om_0/2$.

\subsubsection{Two electrons}\label{sec:obs-two-electrons}

For two electrons, exploiting spin symmetry, we have
\begin{equation}
  \Ekin
  =
  -t\sum_{\las ij\ras\si}\las\tilde{c}^\dag_{i\si}\tilde{c}^{\nag}_{j\si}\ras
  =
  -2t \sum_{\las ij\ras}
  \las
  c^\dag_{i\UP} c^{\nag}_{j\UP} \rme^{\rmi\gamma(\op_i-\op_j)}
  \ras
  \,.
\end{equation}
A simple calculation gives
\begin{equation}
  \las\tilde{c}^\dag_{i\UP}\tilde{c}^{\nag}_{j\UP}\ras
  =
  \Z_L^{-1}\int\,\D p\,w_\text{b}
   \rme^{\rmi\gamma(p_{i,1}-p_{j,1})}
  \tr_\text{f}
  (
  \Omega\, c^\dag_{i\UP}c^{\nag}_{j\UP}
  )
  \,.
\end{equation}
Writing out explicitly the fermionic trace we obtain
\begin{eqnarray}\nonumber
  \tr_\text{f}
  (\Omega\, c^\dag_{i\UP}c^{\nag}_{j\UP})
  &=&
  \sum_{i'j'}
  \bra{i',j'}
  \Omega c^\dag_{i\UP}c^{\nag}_{j\UP}
  \ket{i',j'}
  \\
  &=&
  \sum_{j'}
  \bra{j,j'}\Omega\ket{i,j'}
  \,,
\end{eqnarray}
and the kinetic energy finally becomes
\begin{equation}
  \Ekin
  =
  -2t\Z_L^{-1}\int\,\D p\,w_\text{b}
  \sum_{\las ij\ras}\sum_{j'}
  \rme^{\rmi\gamma(p_{i,1}-p_{j,1})}
  \bra{j,j'}\Omega\ket{i,j'}
  \,.
\end{equation}
In addition to $\Ekin$, we shall also consider the correlation function
\begin{equation}\label{eq:bipolaron:rho}
  \rho(\delta)
  =
  \sum_i \las \on_{i\UP} \on_{i+\delta\DO}\ras
  \,,
  \quad
  \delta = 0,1,\dots,N/2-1
\end{equation}
depending on the distance $\delta$. We find
\begin{equation}
  \rho(\delta)
  =
  \Z_L^{-1}\int\,\D p\,w_\text{b} \sum_i
  \bra{i,i+\delta}\Omega\ket{i,i+\delta}
  \,.
\end{equation}

\subsubsection{Many-electron case}\label{sec:obs-many-electrons}

The calculation of observables within the formalism presented here is similar
to the standard determinant QMC method \cite{BlScSu81,Hi85,LoGu92}.
For an equal-time (\ie, static) observable $O$ we have
\begin{equation}
  \las O \ras_\text{b}
  =
  \frac{\int\,\D p\,\wb \wf \tr_\text{f} (\hat{O} \oB_1\cdots\oB_L)}
  {\int\,\D p\,\wb}
  \,.
\end{equation}
The carrier density
\begin{equation}\label{eq:many-electrons:n}
  n
  =
  \frac{1}{N^{\rD}}\sum_i\las \on_i \ras
\end{equation}
may be calculated from $G^{b}$
[equation~(\ref{eq:many-electrons:GA})] using $\las\on_i\ras = \las
G^{b}_{ii}\ras$.

Similarly, the modulus of the kinetic energy per site is
given by
\begin{equation}\label{eq:many-electrons:Ek}
  \Ek
  =
  \frac{t}{N^{\rD}} \sum_{\las ij\ras}\las G^{b}_{ji}\ras\,.
\end{equation}
Equal-time two-particle correlation functions such as
\begin{equation}
  \rho(\delta)
  =
  \sum_i
  \las
  \on_i \on_{i+\delta}
  \ras
\end{equation}
may be calculated in the same way as in \cite{BlScSu81,Hi85}. For a given phonon configuration, Wick's
Theorem \cite{Ma90} yields
\begin{eqnarray}\nonumber
  \las \on_i \on_j\ras_p
  &=&
  \las c^\dag_i c^{\nag}_i c^\dag_j c^{\nag}_j\ras_p
  \\\nonumber
  &=&
  \las c^\dag_i c^{\nag}_i\ras_p \las c^\dag_j c^{\nag}_j\ras_p
  +
  \las c^\dag_i c^{\nag}_j\ras_p \las c^{\nag}_i c^{\dag}_j\ras_p
  \\
  &=&
  G^{b}_{ii} G^{b}_{jj} + G^{b}_{ij} G^{a}_{ij}
  \,,
\end{eqnarray}
and $\las \on_i \on_j\ras$ is then determined by averaging
over all phonon configurations.

The time-dependent one-particle Green function
\begin{equation}\label{eq:many-electrons:Gb_tau}
  G^{b}(\bm{k},\tau)
  =
  \las c^{\dag}_{\bm{k}} (\tau)c^{\nag}_{\bm{k}} \ras
  =
  \las \rme^{\tau\H} c^{\dag}_{\bm{k}} \rme^{-\tau\H} c^{\nag}_{\bm{k}} \ras
\end{equation}
is related to the momentum-- and energy-dependent spectral function
\begin{equation}\label{eq:many-electrons:akwqmc}
  A(\bk,\om-\mu)
  =
  -\frac{1}{\pi}
  \Im G^b(\bk,\om-\mu)
\end{equation}
through
\begin{equation}\label{eq:many-electrons:maxent}
  G^{b}(\bk,\tau)
  =
  \int_{-\infty}^\infty\,d \om\,\frac{\rme^{-\tau(\om-\mu)}A(\bk,\om-\mu)}
  {1 + \rme^{-\beta(\om-\mu)}}\,.
\end{equation}
The inversion of the above relation is ill-conditioned and requires the use
of the maximum entropy method \cite{HoNevdLWeLoFe04,wvl1992,JaGu96}. Fourier
transformation leads to
\begin{equation}
  G^{b}(\bk,\tau)
  =
  \frac{1}{N^\mathrm{D}}\sum_{ij}  \rme^{\rmi \bm{k}\cdot(\bm{r}_i-\bm{r}_j)} G^b_{ij}(\tau)\,.
\end{equation}
The allowed imaginary times are $\tau_l=l\dtau$, with non-negative
integers $0\leq l\leq L$. Within the QMC approach, we have
\cite{BlScSu81,Hi85}
\begin{equation}\label{eq:many-electrons:Gb_tau_QMC}
  G^{b}_{ij}(\tau_l)
  =
  (
   G^{a} B_1\cdots B_l
  )_{ji}
    \,.
\end{equation}
The one-electron density of states is given by
\begin{equation}\label{eq:many-electrons:dosgen}
  N(\om-\mu)
  =
  -\frac{1}{\pi} \Im G(\om-\mu)
  \,,
\end{equation}
where $G(\om-\mu)=(N^\mathrm{D})^{-1}\sum_{\bk} G(\bk,\om-\mu)$. It may be obtained numerically
via
\begin{equation}\label{eq:many-electrons:DOS}
  N(\tau) = G^{b}_{ii}(\tau)\,,
\end{equation}
and subsequent analytical continuation.

\subsubsection{Suzuki-Trotter error}\label{sec:obs-trotter}

The error associated with the approximation made in, \eg,
equation~(\ref{eq:polaron:suzuki-trotter}) can be systematically reduced by
using smaller values of $\dtau$. In practice, there are two strategies to
handle this so-called {\it Suzuki-Trotter error}. Owing to the usually large
numerical effort for QMC simulations, $\dtau$ is often simply chosen such
that the systematic error is smaller than the statistical errors for
observables. A second, more satisfactory, but also more costly method is to
run simulations at different values of $\dtau$, and to exploit the $\dtau^2$
dependence of the results to extrapolate to $\dtau=0$.

For the results in section~\ref{sec:results}, we have used a scaling toward
$\dtau=0$ based on typical values $\dtau=0.1$, 0.075 and 0.05 to obtain the
results for one and two electrons. In contrast, for the numerically more
demanding calculations of dynamic properties in the many-electron case,
$\dtau=0.1$ has been chosen. This is justified by the uncertainties in the
analytical continuation.

\subsection{Reweighting}\label{sec:qmc-reweighting}

As pointed out at the end of section~\ref{sec:partition-function}, the
calculation of the change of the fermionic weight $\wf$ represents the most
time-consuming part of the updating process. Consequently, it would be highly
desirable to avoid the evaluation of $\wf$. This may be achieved by using
only the bosonic weight $\wb$ in the updating, and treating $\wf$ as part of
the observables. For the expectation value of an observable $O$, such a
reweighting requires calculation of
\begin{equation}\label{eq:polaron:reweighting}
  \las O \ras
  =
  \frac{\las O\, \wf\ras_\text{b}}{\las \wf\ras_\text{b}}
  \,,
\end{equation}
where the subscript ``b'' indicates that the average is computed based on
$\wb$ only [cf equation~(\ref{eq:polaron:Ob})].

Reweighting of the probability distribution is frequently used in MC
simulations if a minus-sign problem occurs \cite{wvl1992}. Here, the
splitting into the configuration weight $\wb$ and the observable $O \wf$ is
practicable provided the variance of both $\wf$ and $O \wf$ is small, which is the
case after the LF transformation. Furthermore, we require a significant
overlap of the two distributions, which may be quantified using the
Kullback-Leibler number \cite{HoEvvdL03}, in order to avoid prohibitive
statistical noise.  In fact, our calculations show that, in general, for the
{\it untransformed} model the reweighting method cannot be applied. For a
detailed discussion of this point in the one-electron case see
\cite{HoEvvdL03}. Here we merely note that no problems arise when simulating
the transformed model.

Apart from the significant advantage that the fermionic weight $\wf$ only has
to be calculated when observables are measured, the reweighting method
becomes particularly effective in the present case when combined with the
principal component representation introduced in section~\ref{sec:qmc-pc}. In
this case, we will be able to perform an exact sampling of the phonons
without any autocorrelations. For a reliable error analysis for observables
calculated according to equation~(\ref{eq:polaron:reweighting}) the Jackknife procedure
\cite{DavHin} is applied.

\subsection{Principal components}\label{sec:qmc-pc}

The reweighting method allows us, in principle, to skip enough
sweeps between measurements to reduce autocorrelations to a minimum. However,
even though a single phonon update requires negligible computer time compared to
the evaluation of $\wf$, for critical parameters, an enormous number of such
steps will be necessary between successive measurements \cite{HoEvvdL03}. On top of that, reliable results require
knowledge of the longest autocorrelation times, which have to be determined
in separate simulations for each set of parameters.

Due to the structure of the bosonic action $S_\text{b}$ [see
equation~(\ref{eq:polaron:action})], even relatively small (local)
changes to the phonon momenta lead to large variations in $S_\mathrm{b}$ and
hence the weight $\wb$. As a consequence, only minor
changes may be proposed in order to reach a reasonable acceptance rate.
Unfortunately, this strategy is the very origin of autocorrelations.

The problem can be overcome by a
transformation to the normal modes of the phonons (along the imaginary time axis), so
that we can sample completely uncorrelated configurations. As the
fermion degrees of freedom are treated exactly, the resulting QMC method is
then indeed free of any autocorrelations.  

To find such a transformation, let us recall the form of
the bosonic action, given by equation~(\ref{eq:polaron:action-w-matrix}), which
we write as
\begin{equation}\label{eq:polaron:pc}
  S_\text{b}
  =
  \sum_i \bm{p}_i^\text{T} A \bm{p}_i
  =
  \sum_i \bm{p}_i^\text{T} A^{1/2} A^{1/2} \bm{p}_i
  \equiv
  \sum_i \vec{\xi}_i^\text{T}\cdot\vec{\xi}_i
\end{equation}
with the {\it principal components} $\vec{\xi}_i=A^{1/2}\bm{p}_i$, in terms of which the
bosonic weight takes the simple Gaussian form
\begin{equation}\label{eq:polaron:action_quad}
  \wb
  =
  \rme^{-\dtau\sum_i \vec{\xi}^\text{T}_i\cdot\vec{\xi}_i}
  \,.
\end{equation}
The sampling can now be performed directly in terms of the new variables
$\vec{\xi}$. To calculate observables we have to transform back to the
physical momenta $\vec{p}$ using $A^{-1/2}$. Comparison with
equation~(\ref{eq:polaron:action-w-matrix}) shows that instead of the
ill-conditioned matrix $A$ we now have the ideal case that we can easily
generate exact samples of a Gaussian distribution. With the new coordinates
$\bm{\xi}$, the probability distribution can be sampled exactly, \eg, by the
Box-M\"{u}ller method \cite{numrec_web}. In contrast to a standard Markov
chain MC simulation, every new configuration is accepted and measurements can
be made at each step, so that simulation times are significantly reduced.

From the definition of the principal components it is obvious that an update
of a single variable $\xi_{i,\tau}$, say, actually corresponds to a change of
all $p_{i,\tau'}$\,, $\tau'=1,\dots,L$.  Thus, in terms of the original
phonon momenta $\vec{p}$, the updating becomes non-local.

The principal component representation can be used for one, two and many
electrons, since the bosonic action [equation~(\ref{eq:polaron:action_quad})]
is identical. This even holds for models including, \eg, spin-spin
interactions, as long as the phonon operators enter in the same form as in
the Holstein model.

An important point is the combination of the principal components with the
reweighting method. Using the latter, the changes to the original momenta
$\bm{p}$, which are made in the simulation, do not depend in any way on the
electronic degrees of freedom. Thus we are actually sampling a set of
independent harmonic oscillators, as described by $S_\text{b}$. The crucial
requirement for the success of this method is the use of the LF transformed
model, in which the (bi-)polaron effects are separated from the zero-point
motion of the oscillators around their current equilibrium positions.

Finally, as there is no need for a warm-up phase, and owing to the
statistical independence of the configurations, the present algorithm is
perfectly suited for parallelization.

\subsection{Minus-sign problem}\label{sec:sign}

The motivation for our development of a novel QMC approach to Holstein models
was to improve on the performance of existing methods, especially in the
many-electron case. As pointed out in \cite{HoEvvdL05}, the LF transformation causes a sign
problem even for the pure Holstein model which, in general, may significantly
affect the applicability of the method. Therefore, we briefly discuss the
resulting limitations, focussing on the many-electron case.

We shall see that there is a fundamental difference between simulations for
one or two electrons---the carrier density being zero in the thermodynamic
limit---and grand-canonical calculations at finite density $n>0$. Whereas for
one or two carriers the sign problem turns out to be rather uncritical---the
average sign approaches unity upon increasing system size, in contrast to the
usual behaviour \cite{wvl1992}---restrictions are encountered in simulations
of the many-electron case.

\begin{figure}[t]
  \centering
  \includegraphics[width=0.45\textwidth]{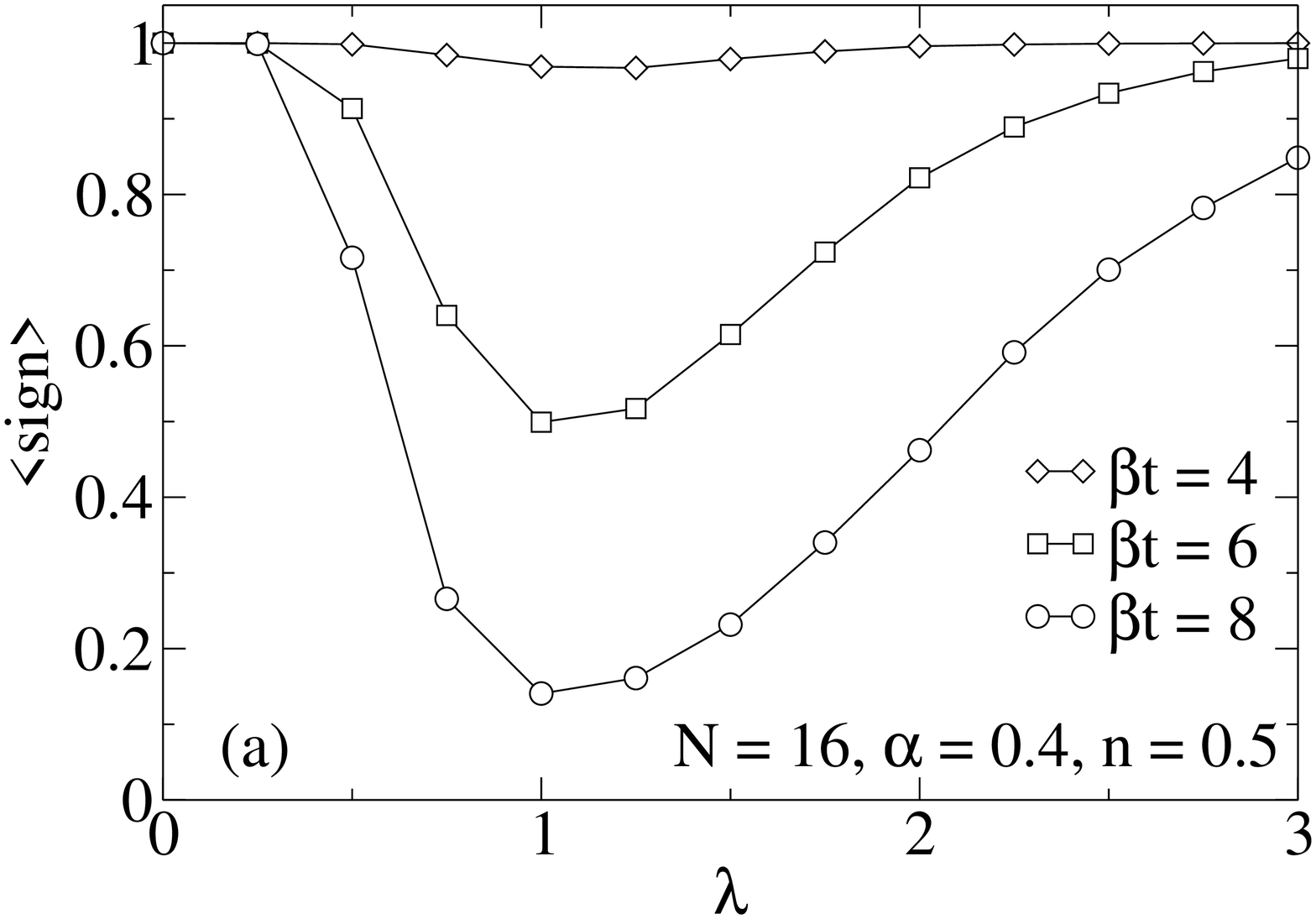}
  \includegraphics[width=0.45\textwidth]{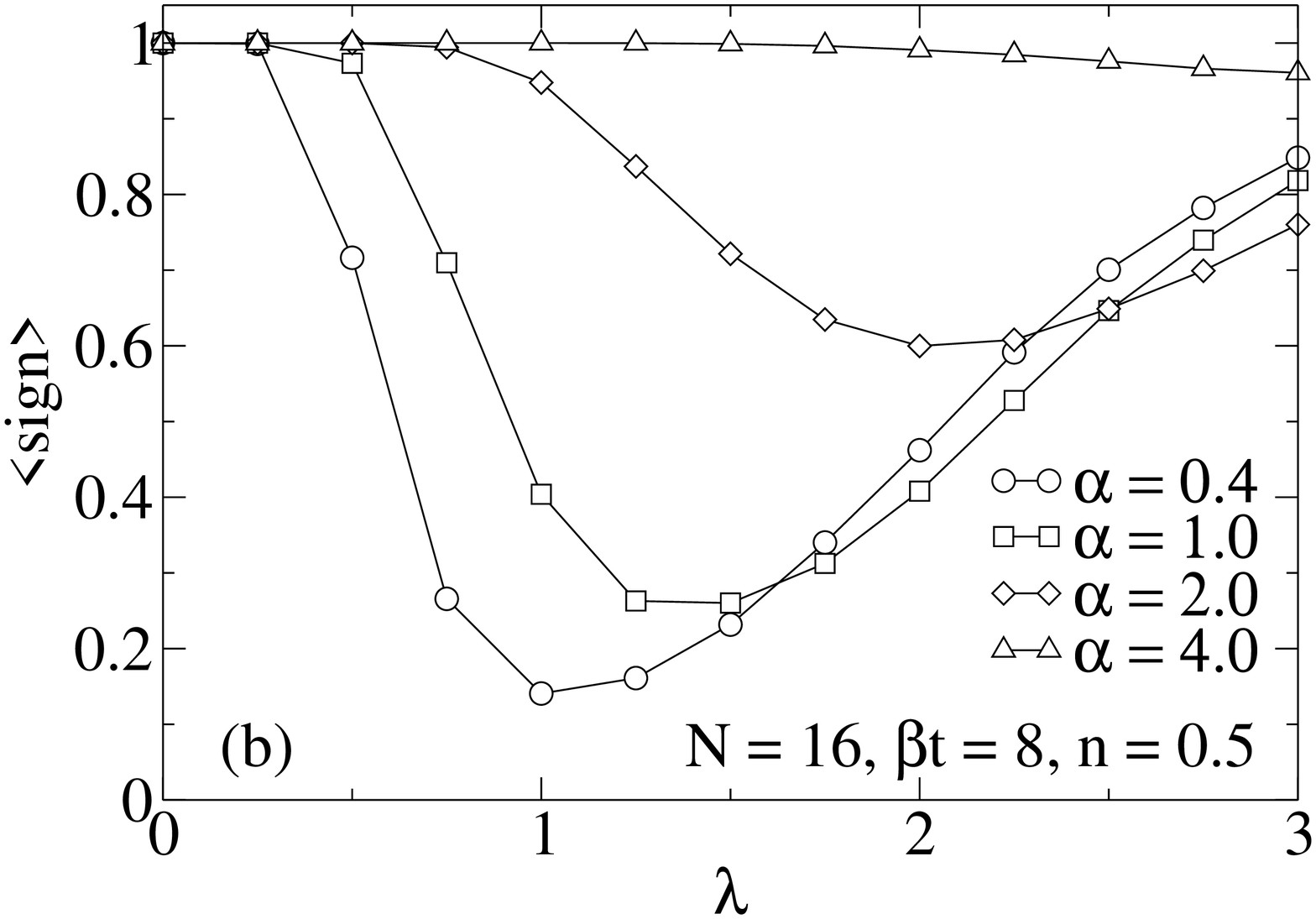}
  \caption{\label{fig:many-electrons:sign_beta_omega} Average sign
    $\sign$ in the many-electron case as a function of el-ph
    coupling $\lambda$ in $\rD=1$ (a) for different inverse
    temperatures $\beta$, and (b) for different values of the adiabaticity ratio
    $\alpha$.  Lines are guides to the eye, and errorbars are smaller than the
    symbols shown. The data presented in figures~\ref{fig:many-electrons:sign_beta_omega}
    and~\ref{fig:many-electrons:sign_n_N} are for $\dtau=0.05$. [Taken from \cite{HoNevdLWeLoFe04}.]}
\end{figure}

Since $\wb$ is strictly positive, we define the average sign as

\begin{equation}\label{eq:polaron:sign}
  \sign
  =
  \las\wf\ras_\text{b} / \las|\wf|\ras_\text{b}
  \,.
\end{equation}

For simplicity, we first show results for $n=0.5$, while the effect of
band filling will be discussed later. The choice $n=0.5$ is convenient since
we know the chemical potential, and we shall see below that the sign problem
is most pronounced for a half-filled band. Moreover, most existing QMC
results for the spinless Holstein model are for half filling (see references
in \cite{HoEvvdL03}).

Figure~\ref{fig:many-electrons:sign_beta_omega}(a) shows the dependence of
$\sign$ on the el-ph coupling strength. It takes on a minimum near
$\lambda=1$ (for $\alpha<1$) that becomes more
pronounced with decreasing temperature. At weak coupling (WC) and SC,
$\sign\approx1$, so that accurate simulations can be carried out.
These results are quite similar to the cases of one or two electrons
\cite{Hohenadler04}.

The dependence on phonon frequency
[figure~\ref{fig:many-electrons:sign_beta_omega}(b)] also bears a close
resemblance to the polaron problem \cite{Hohenadler04}.  Whereas $\sign$ becomes very small
for $\alpha\ll1$, it increases noticeably in the non-adiabatic regime $\alpha>1$,
permitting efficient and accurate simulations.

\begin{figure}[t]
  \centering
  \includegraphics[width=0.45\textwidth]{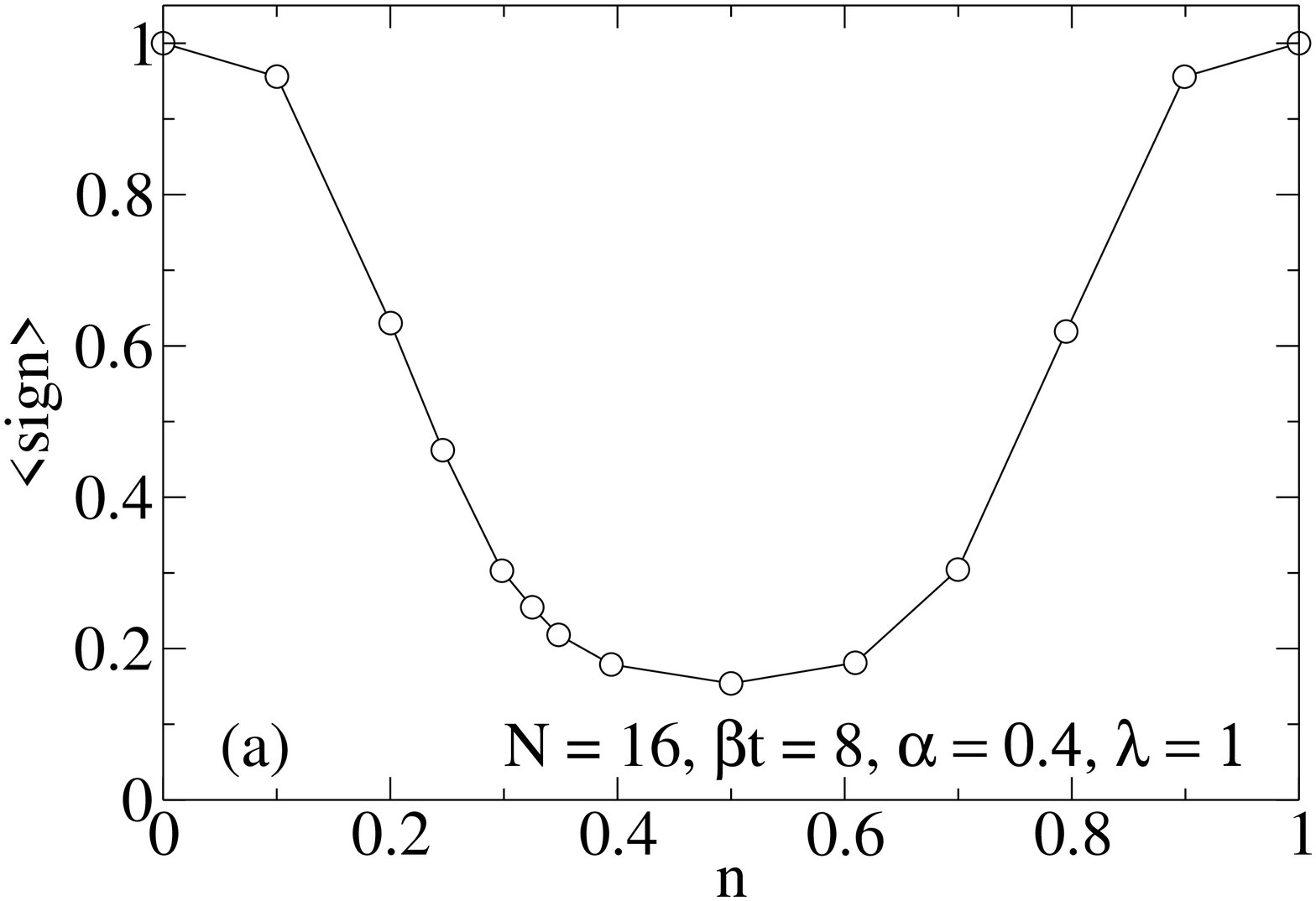}
  \includegraphics[width=0.45\textwidth]{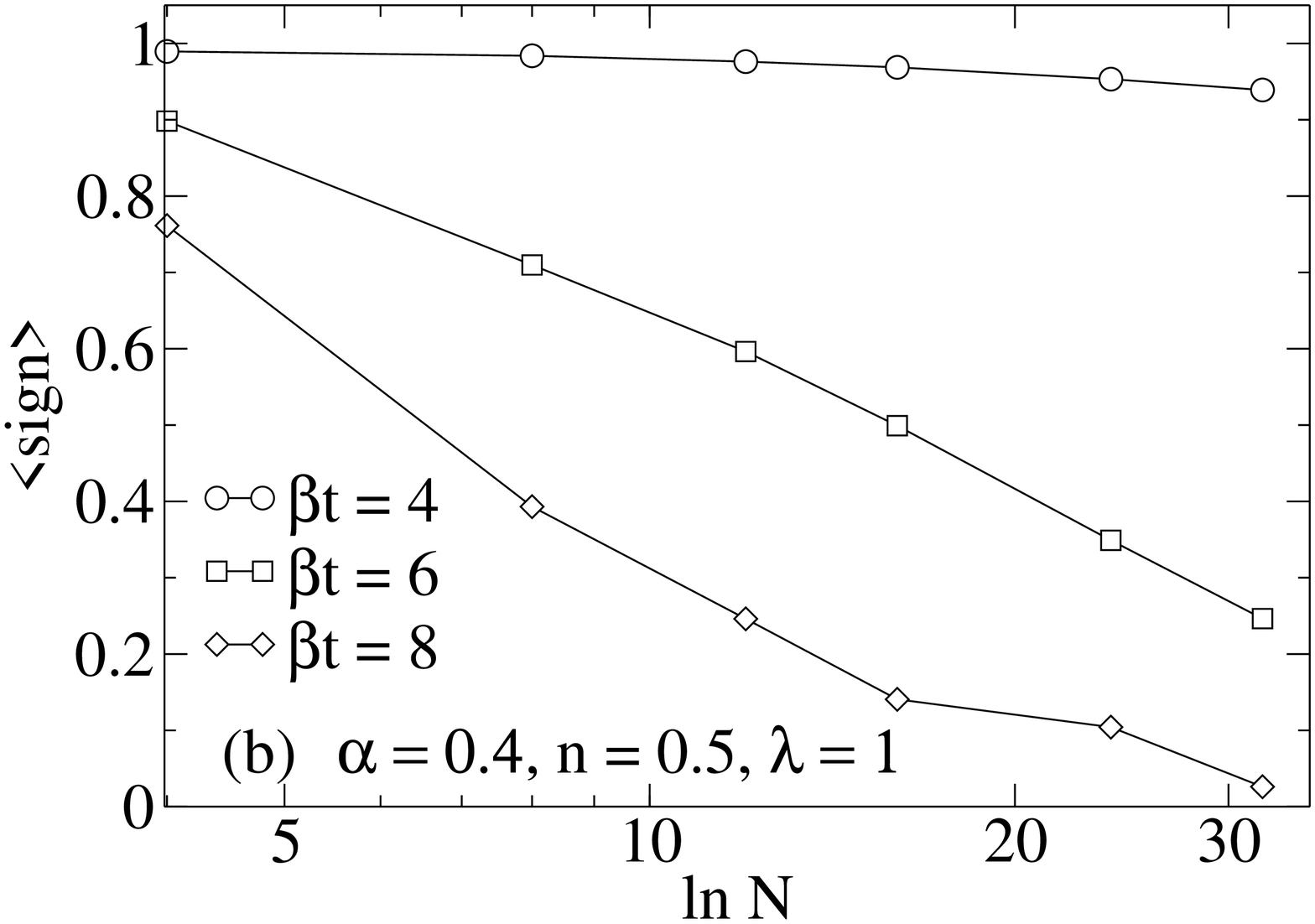}
\caption{\label{fig:many-electrons:sign_n_N}
  Average sign in the many-electron case as a function of (a) band filling
  $n$, and (b) system size $N$.}
\end{figure}

As illustrated in figure~\ref{fig:many-electrons:sign_n_N}(a), the average
sign depends strongly on the band filling $n$. While it is close to one in
the vicinity of $n=0$ or $n=1$ (equivalent to one or two electrons), a significant reduction is
visible near half filling $n=0.5$. The minimum occurs at $n=0.5$, and the
results display particle-hole symmetry as expected. Here we have chosen
$\beta t=8$, $\alpha=0.4$ and $\lambda=1$, for which the sign problem is most
noticeable according to figure~\ref{fig:many-electrons:sign_beta_omega}.

In figure~\ref{fig:many-electrons:sign_n_N}(b), we report the
average sign as a function of system size, again for $n=0.5$. The dependence is
strikingly different from the one-electron case. While in the latter
$\sign\rightarrow1$ as $N\rightarrow\infty$ \cite{HoEvvdL03,Hohenadler04},
here the average sign decreases nearly exponentially with increasing system
size, a behaviour well-known from QMC simulations of Hubbard models \cite{wvl1992}. Obviously,
this limits the applicability of our method. However, we
shall see below that we can nevertheless obtain accurate results at low
temperatures, small phonon frequencies, and over a large range of the
el-ph coupling strength. Moreover, we would like to point out that for
such parameters, other methods suffer strongly from autocorrelations,
rendering simulations extremely difficult.

The dependence of the sign problem on the dimension of the system is again
similar to the single-electron case \cite{Hohenadler04}. The minimum  at
intermediate $\lambda$ becomes more pronounced for the same
parameters $N$, $\alpha$, $\beta t$ and $\lambda$ as one increases the
dimension of the cluster.

To conclude with, we would like to point out that, in principle, the sign
problem can be compensated by performing sufficiently long QMC runs, but
we have to keep in mind that the statistical errors increase proportional to
$\sign^{-2}$ \cite{wvl1992}, setting a practical limit to the accuracy.

\subsection{Comparison with other approaches}\label{sec:qmc_comparison}

The QMC method presented above seems to be most advantageous---as compared to
other approaches---in the case of the spinless Holstein model with many
electrons. For the latter, other methods are severely restricted by
autocorrelations, rendering accurate simulations in the physically important
adiabatic, IC regime virtually impossible even at
moderately low temperatures. In contrast, the present method enables us to
study the single-particle spectrum on rather large clusters and for a wide
range of model parameters and band filling (see
section~\ref{sec:res-manypol}). Unfortunately, the generalization to the
spinful Hubbard-Holstein model suffers severely from the sign problem.

For the polaron and the bipolaron problem, our method requires more computer
time than other QMC algorithms \cite{dRLa82,deRaLa86,Ko98,Mac04}. However, we
are able to consider practically all parameter regimes on reasonably large
clusters in one (polaron and bipolaron problem) and two dimensions (polaron
problem).

Finally, a discussion of the scaling of computer time with the system
parameters can be found in \cite{HoEvvdL03,HoEvvdL05,HovdL05}.

\section{Selected results}
\label{sec:results}

We now come to a selection of results obtained with the methods discussed so
far, most of which have been published before
\cite{HoEvvdL03,HoEvvdL05,HovdL05,HoNevdLWeLoFe04}. 
Note that errorbars will be suppressed in the figures
if smaller than the symbolsize. Moreover, lines connecting data points
are guides to the eye only.

\subsection{Small-polaron cross-over}\label{sec:res-polaron}

The Holstein model with a single electron (for a review see
\cite{FeAlHoWe06}) exhibits a cross-over from a large
polaron ($\rD=1$) or a quasi-free electron ($\rD>1$) to a small polaron with
increasing el-ph coupling strength.

\subsubsection{Quantum Monte Carlo}

To investigate the small-polaron cross-over, following previous work
\cite{dRLa82,dRLa83,Ko97,WeRoFe96,JeWh98,dMeRa97,RoBrLi99,RoBrLi99III,KuTrBo02},
we calculate the electronic kinetic energy $\Ekin$ given by
equation~(\ref{eq:polaron:Ek}).  As we shall compare results for different
dimensions, we define the normalized quantity
\begin{equation}\label{eq:Ek}
  \Ek
  =
  \Ekin/(-2t\mathrm{D})
\end{equation}
with $\Ek=1$ for $T=0$ and $\lambda=0$.

The inverse temperature will be fixed to $\beta t = 10$, low enough to
identify the cross-over. Calculations at even lower
temperatures can easily be done for $\alpha>1$, but $\alpha<1$ requires very large
numbers of measurements to ensure satisfactorily small statistical
errors. System sizes were 32 sites in
1D, a $12\times12$ cluster in 2D, and a $6\times6\times6$ lattice in 3D. In
contrast to $\rD=1$, 2, where results are well converged with
respect to system size, non-negligible finite-size effects (maximal relative
changes of up to 20 \% between $N=5$ and $N=6$ for $\alpha\ll1$; much smaller
changes otherwise) are observed in three dimensions. Moreover, for small $N$,
effects due to thermal population of states with non-zero momentum
$\bm{k}$---absent in ground-state calculations---are visible, as discussed
below. Nevertheless, the main characteristics are well visible already for
$N=6$. For a detailed study of finite-size and finite-temperature effects see
\cite{Hohenadler04}.

\begin{figure}[t]
  \centering
  \includegraphics[width=0.45\textwidth]{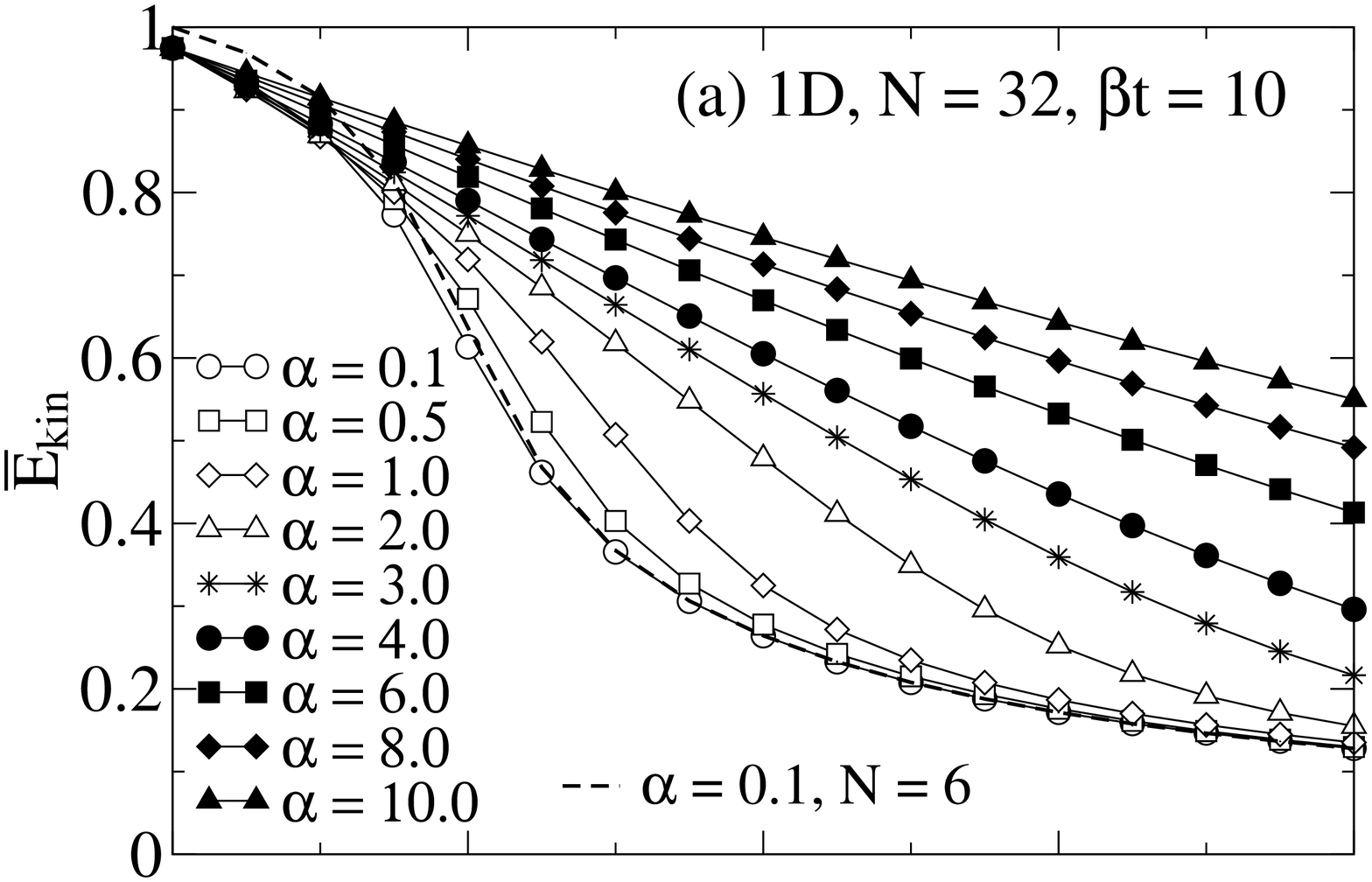}\\
  \includegraphics[width=0.45\textwidth]{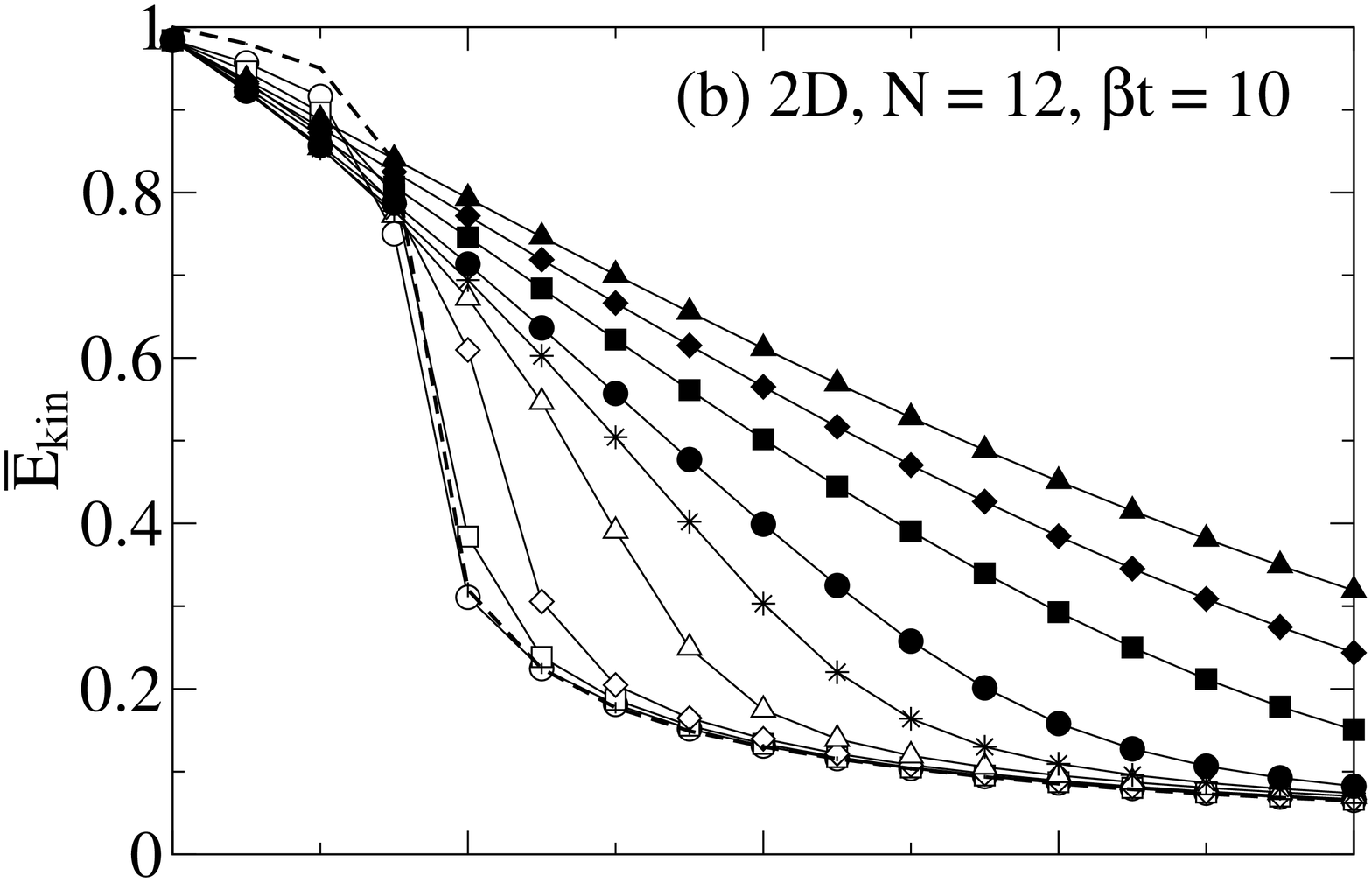}\\
  \includegraphics[width=0.45\textwidth]{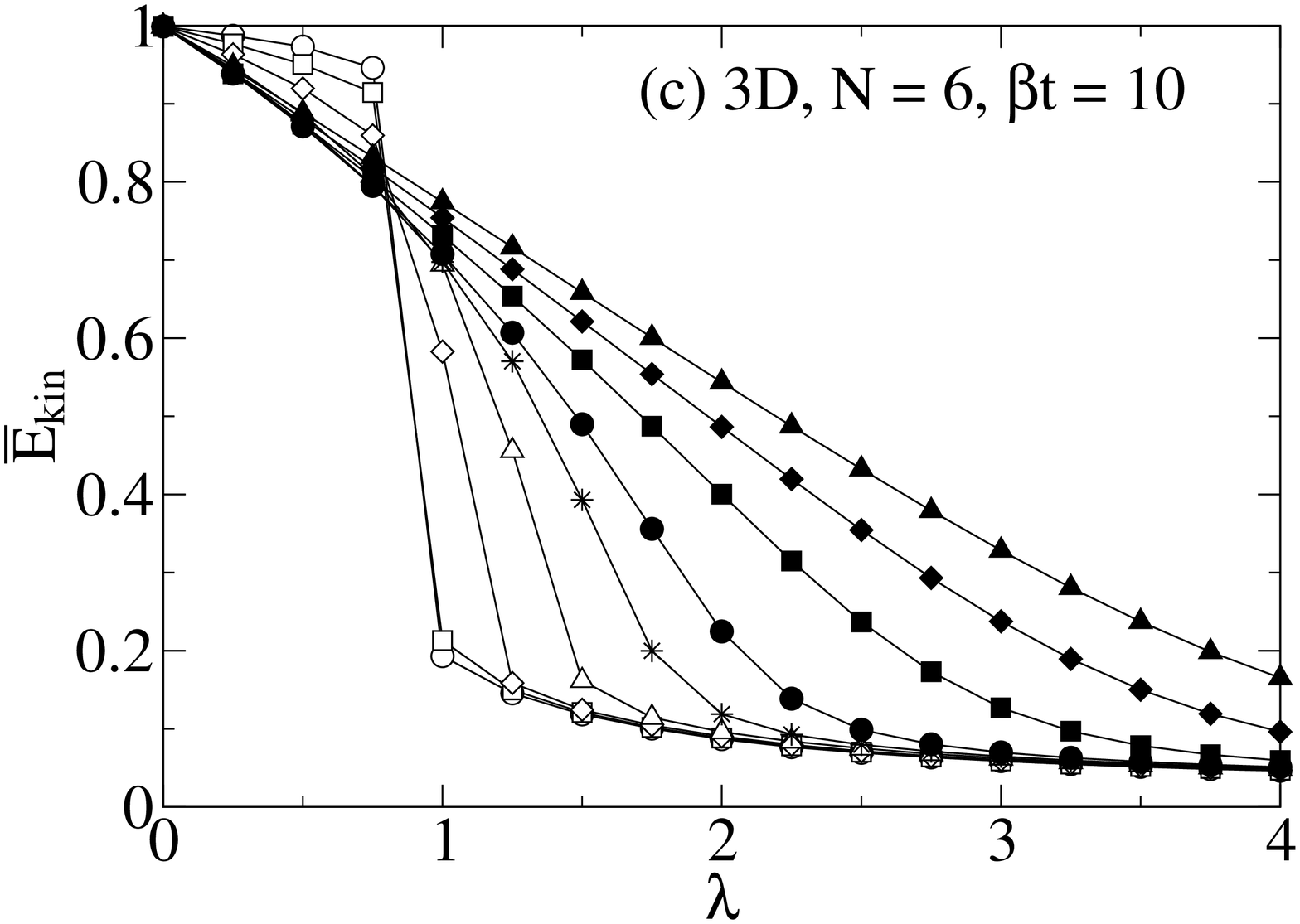}
  \caption{\label{fig:Ek} Normalized kinetic energy $\Ek$
    [equation~(\ref{eq:Ek})] of the Holstein model with one electron from QMC
    as a
    function of el-ph coupling $\lambda$ for different adiabaticity ratios
    $\alpha$ and different dimensions D of the lattice ($N$ denotes the
    linear cluster size). Here and in subsequent figures, QMC data have been
    extrapolated to $\dtau=0$ (see section~\ref{sec:obs-trotter}).  [Taken
    from \cite{HoEvvdL05}.]}
\end{figure}

Figure~\ref{fig:Ek} shows $\Ek$ as a function of the el-ph coupling
$\lambda$ for different phonon frequencies varying over two orders of
magnitude, in one to three dimensions. Generally, the kinetic energy is large
at WC, where the ground state consists of a weakly dressed
electron ($\rD>1$) or a large polaron ($\rD=1$). It reduces more or less
strongly---depending on $\alpha$---in the SC regime, where a
small, heavy polaron exists, defined as an electron surrounded by a lattice
distortion essentially localized at the same site. The finite values of $\Ek$
even for large $\lambda$ are a result of undirected motion of the electron
inside the surrounding phonon cloud. In contrast, the quasiparticle weight is
exponentially reduced in the SC regime (see, \eg,
\cite{KuTrBo02}), whereas the effective mass becomes exponentially large.

In all dimensions, the phonon frequency has a crucial influence on the
behaviour of the kinetic energy. While in the adiabatic regime $\alpha<1$ the
small-polaron cross-over is determined by the condition
$\lambda=\Ep/2t\rD>1$, the corresponding criterion for
$\alpha>1$ is $g^2=\Ep/\om_0>1$. The former condition reflects the fact that
the loss in kinetic energy of the electron has to be outweighed by a gain in
potential energy in order to make small-polaron formation favourable. The
latter condition expresses the increasing importance of the lattice energy
for $\alpha>1$, since the formation of a ``localized'' state requires a sizable
lattice distortion. As a consequence, for large phonon frequencies, the
critical coupling shifts to $\lambda_\mathrm{c}>1$, whereas for $\alpha<1$ we
have $\lambda_\mathrm{c}= 1$.  Additionally, the decrease of $\Ek$ at
$\lambda_\mathrm{c}$ becomes significantly sharper with decreasing phonon
frequency.

Concerning the effect of dimensionality, figure~\ref{fig:Ek} reveals that, for
fixed $\alpha$, the small-polaron cross-over becomes more abrupt in higher
dimensions, with a very sharp decrease in 3D. Nevertheless, there is no
real phase transition \cite{Loe88}. Figure~\ref{fig:Ek} also contains results
for $N=6$ in one and two dimensions, \ie, for the same linear cluster size as
in 3D (dashed lines). Clearly, for such small clusters, the spacing between
the discrete allowed momenta $\bm{k}$ is too large to permit substantial
thermal population, so that results are closer to the ground state [\eg,
$\Ek(\lambda=0)\approx 1$], and exhibit a slightly more pronounced decrease
near the critical coupling. However, the sharpening of the latter with
increasing dimensionality is still well visible.

\subsubsection{Variational approach}

\begin{figure}[t]
  \centering
  \includegraphics[width=0.45\textwidth]{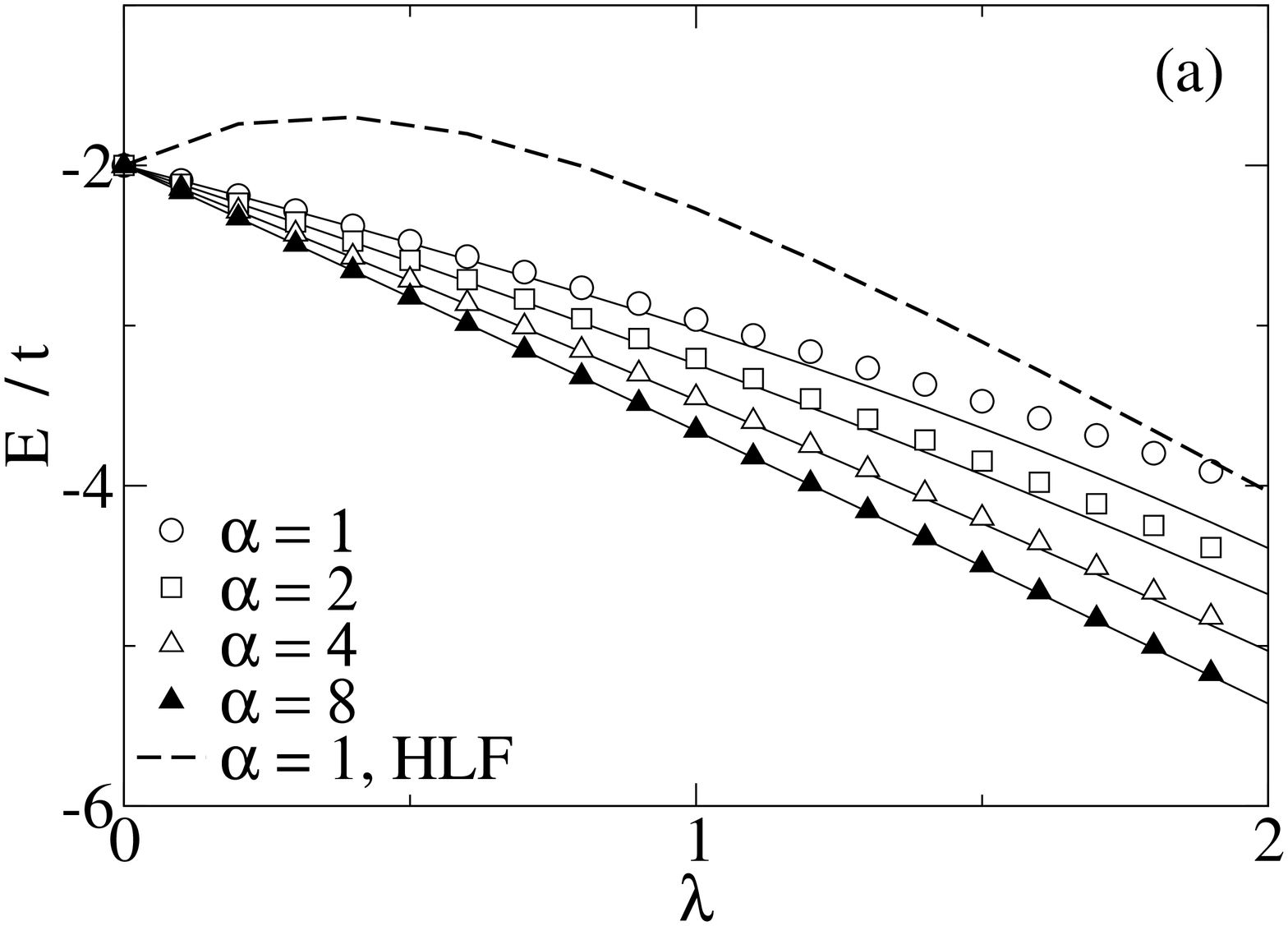}
  \includegraphics[width=0.45\textwidth]{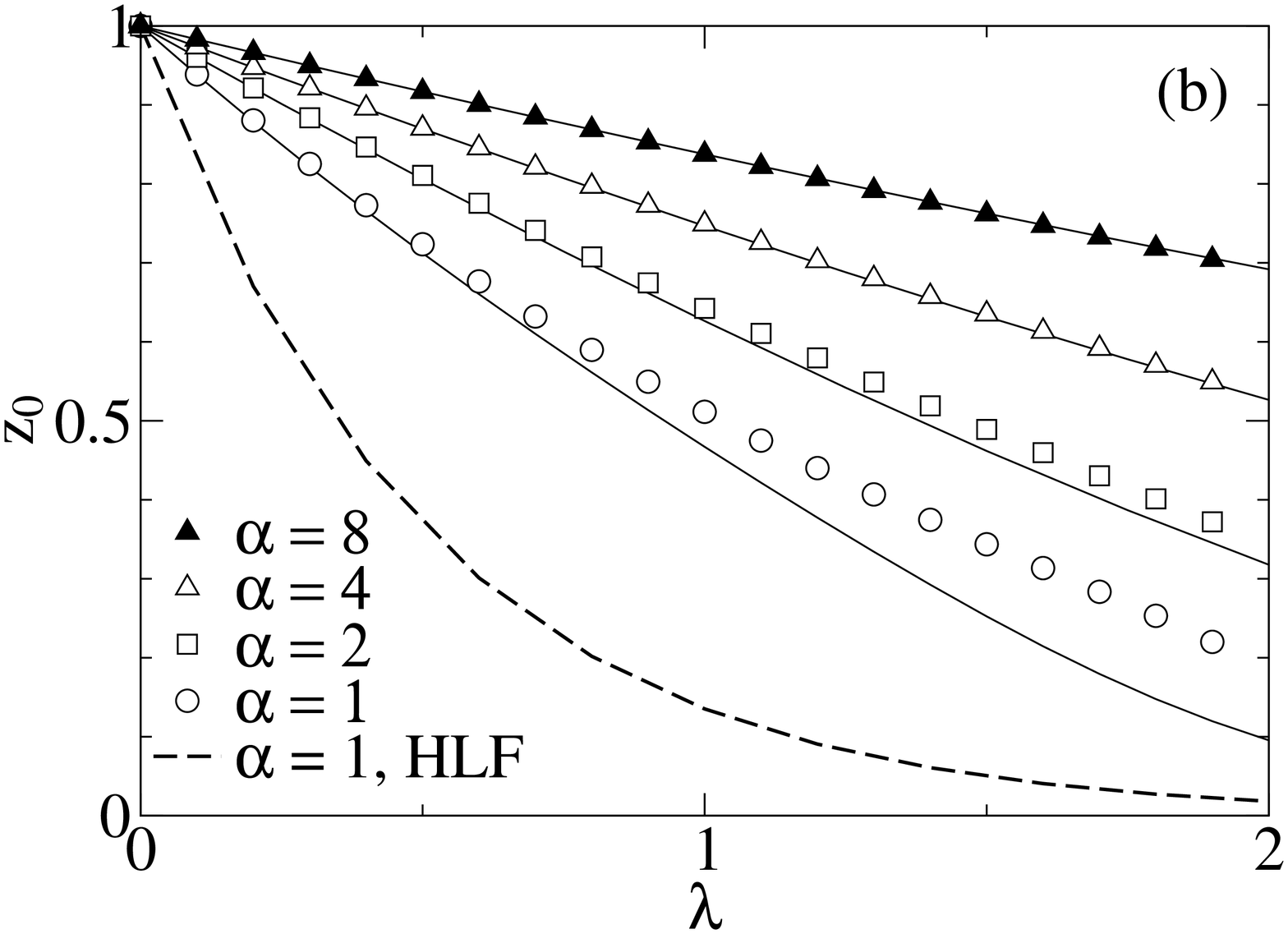}
  \caption{\label{fig:polaron:E0z0vpa} Total energy $E$ (a) and quasiparticle
    weight $z_0$ (b) for $N=4$ as functions of the el-ph coupling $\lambda$
    for different values of the adiabaticity ratio $\alpha$.  Symbols
    correspond to variational results and full lines represent exact $T=0$
    data obtained with the Lanczos method \cite{Mars95}. Dashed lines are
    results of the HLF approximation.  [Taken from \cite{HoEvvdL03}.]}
\end{figure}

To test the validity of the variational approach of section~\ref{sec:vpa} we
have calculated the total energy [equation~(\ref{eq:polaron:evs})] and the
quasiparticle weight [equation~(\ref{eq:polaron:z0})] on a cluster with $N=4$
for various values of $\alpha$.  A comparison with exact diagonalization
results \cite{Mars95} is depicted in figure~\ref{fig:polaron:E0z0vpa}. We
only consider the regime $\alpha\geq1$ where the zero-phonon approximation is
expected to be justified.  The overall agreement is strikingly good.  Minor
deviations from the exact results increase with decreasing $\alpha$.  For the
smallest frequency shown, $\alpha=1$, the result of the HLF approximation is
also reported. Clearly, the variational method represents a significant
improvement over the HLF approximation, underlining the importance of taking
into account non-local distortions. Similar conclusions can be drawn for larger system
sizes (see figure~3 in \cite{HoEvvdL03}).

\begin{figure}[h!]
  \centering
  \includegraphics[height=0.33\textwidth]{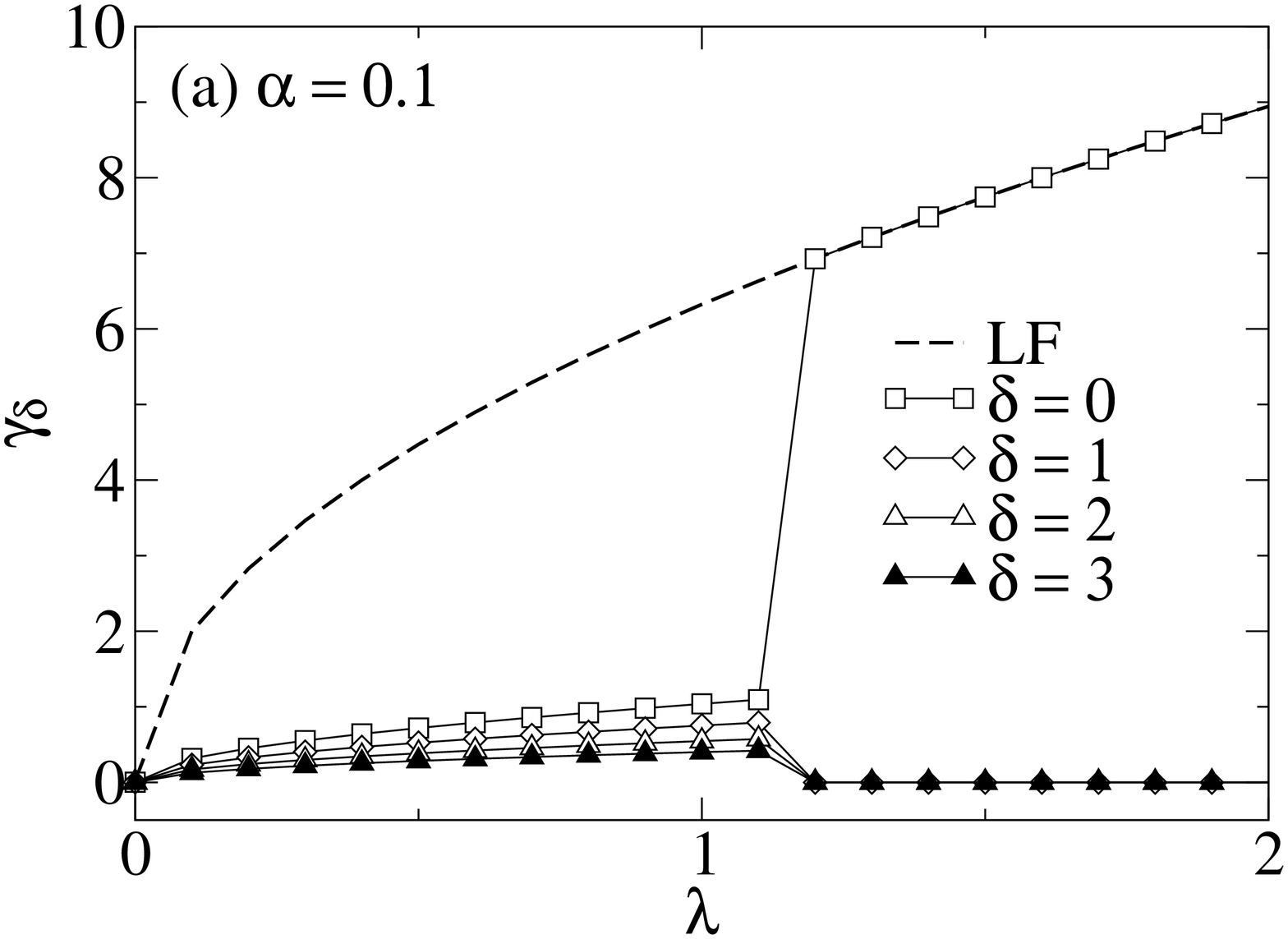}
  \includegraphics[height=0.33\textwidth]{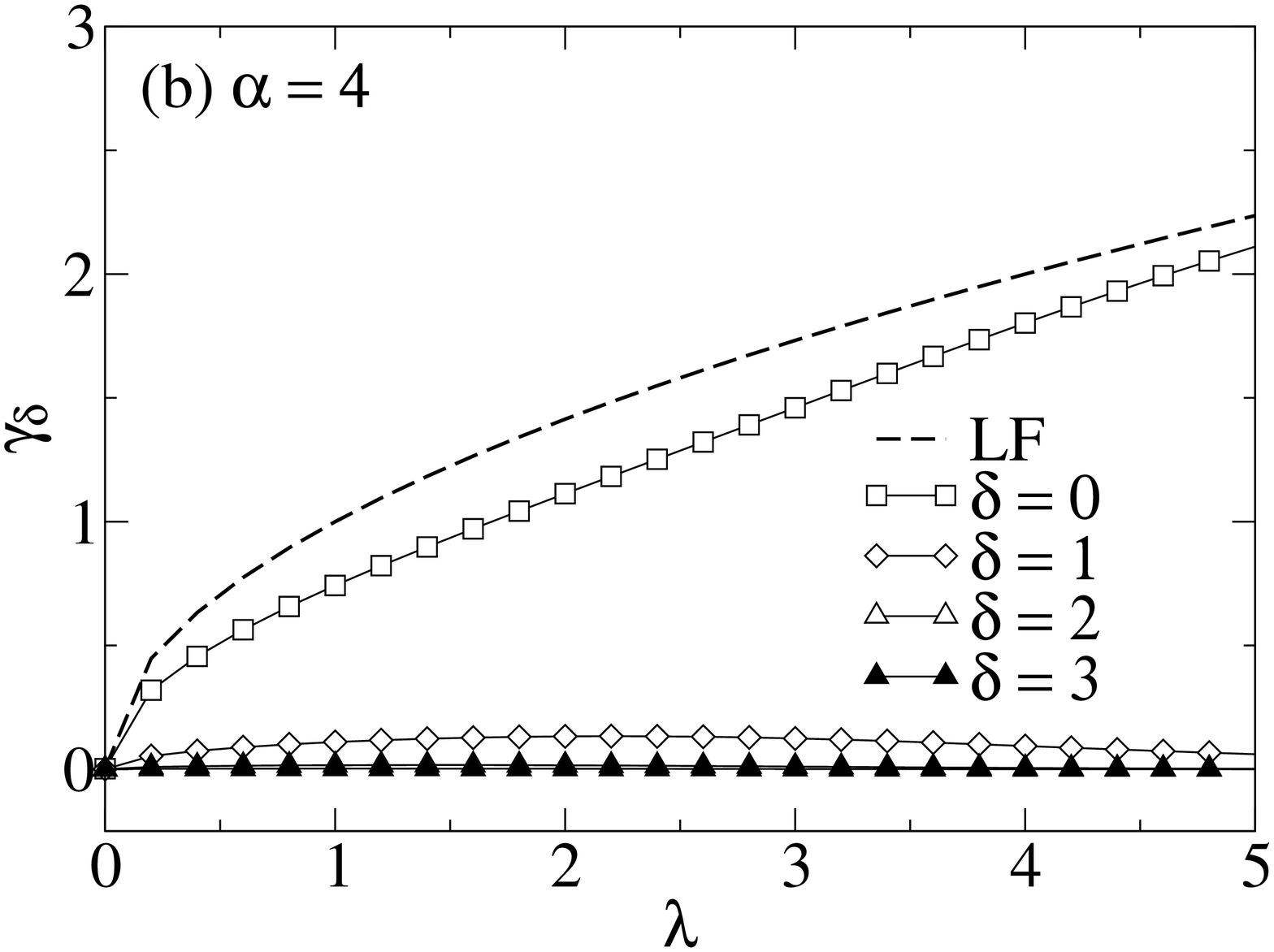}
\caption{\label{fig:polaron:gammas}
  Polaron-size parameter $\gamma_\delta$ for $N=16$ as a function of the el-ph
  coupling $\lambda$ for various distances $\delta$ in the (a) adiabatic and
  (b) anti-adiabatic regime. Also shown is the
  LF parameter $\gamma$ [equation~(\ref{eq:polaron:gamma})]. [Taken from \cite{HoEvvdL03}.]}
\end{figure}

In figure~\ref{fig:polaron:gammas} we present results for the variational
displacement fields $\gamma_\delta$, which provide a measure for the polaron
size. For $\alpha=0.1$ we see an abrupt cross-over from a large to a small
polaron at $\lambda\approx 1.2$. For smaller $\lambda$, the electron induces
lattice distortions at neighboring sites even at a distance of more than
three lattice constants. Above $\lambda\approx1.2$ we have a mobile small
polaron extending over a single site only. In contrast, for the
anti-adiabatic case $\alpha=4$, the cross-over is much more gradual, and
$\gamma_1>0$ even for $\lambda\gg1$. The same behaviour has been found by
Marsiglio \cite{Marsiglio95} who determined the correlation function $\las
\on_i \ox_{i+\delta}\ras$ by exact diagonalization; within the variational
approach $\las \on_i \ox_{i+\delta}\ras= \gamma_\delta$. Although in Marsiglio's
results the cross-over to a small polaron for $\alpha=0.1$ occurs at a
smaller value of the coupling $\lambda\approx 1$, the simple variational
approach reproduces the main characteristics.

\subsection{Bipolaron formation in the extended Holstein-Hubbard model}\label{sec:res-bipolaron}

In contrast to Cooper pairing of electrons with opposite momentum, two
electrons may also form a bound state by travelling sufficiently close in
real space. Bipolaron formation
may be studied in the framework of the 1D extended Holstein-Hubbard model, and
a brief review of previous work has been given in
\cite{HoAivdL04,HovdL05}. Here we merely note that depending on the choice of
parameters, the ground state of the model may either consist of two polarons,
a large bipolaron, an inter-site bipolaron or a small bipolaron (in the
singlet case). A summary of the conditions on the model parameters is given
in table~\ref{tab:bipolaron:bipolaronconditions}. Whereas existing work is
almost exclusively concerned with the singlet case, here we shall also
consider two electrons of the same spin. Triplet bipolarons are expected to
play a role, \eg, in the ferromagnetic state of the manganites
\cite{AlBr99,AlBr99_2,David_AiP}. Furthermore, we are not aware of any
previous work for $V>0$.

\begin{table}[t]
  \centering
  \caption{\label{tab:bipolaron:bipolaronconditions}Conditions for the
    existence of different singlet bipolaron states in the
    one-dimensional Holstein-Hubbard model \cite{HovdL05}.}
  \begin{tabular}{cc|cc||c|c|c}\hline\hline
    \multicolumn{4}{c}{$U=0$}  & \multicolumn{3}{c}{$U>0$} \\\hline
    Large bipolaron &&& Small bipolaron & Two     & Inter-site & Small    \\
                    &&&                 & polarons& bipolaron  & bipolaron\\\hline
    $\lambda<0.5$   &&& $\lambda>0.5$ & $U>2\Ep$ (WC) &  $U<2\Ep$ (WC) & \\
    or              &&& and           &  &   & $U\ll2\Ep$\\
    $g<0.5$ &&& $g>0.5$  & $U>4\Ep$ (SC)    &
    $U<4\Ep$ (SC) & \\\hline
    \hline
  \end{tabular}
\end{table}

\subsubsection{Quantum Monte Carlo}\label{sec:res_qmc}

Owing to the increased numerical effort compared to the one-electron
case, we shall only present results for $N\leq12$ in one
dimension. However, finite-size effects are small even for the most critical
parameters \cite{HovdL05}.

\begin{figure}[t]
  \centering
  \includegraphics[height=0.34\textwidth]{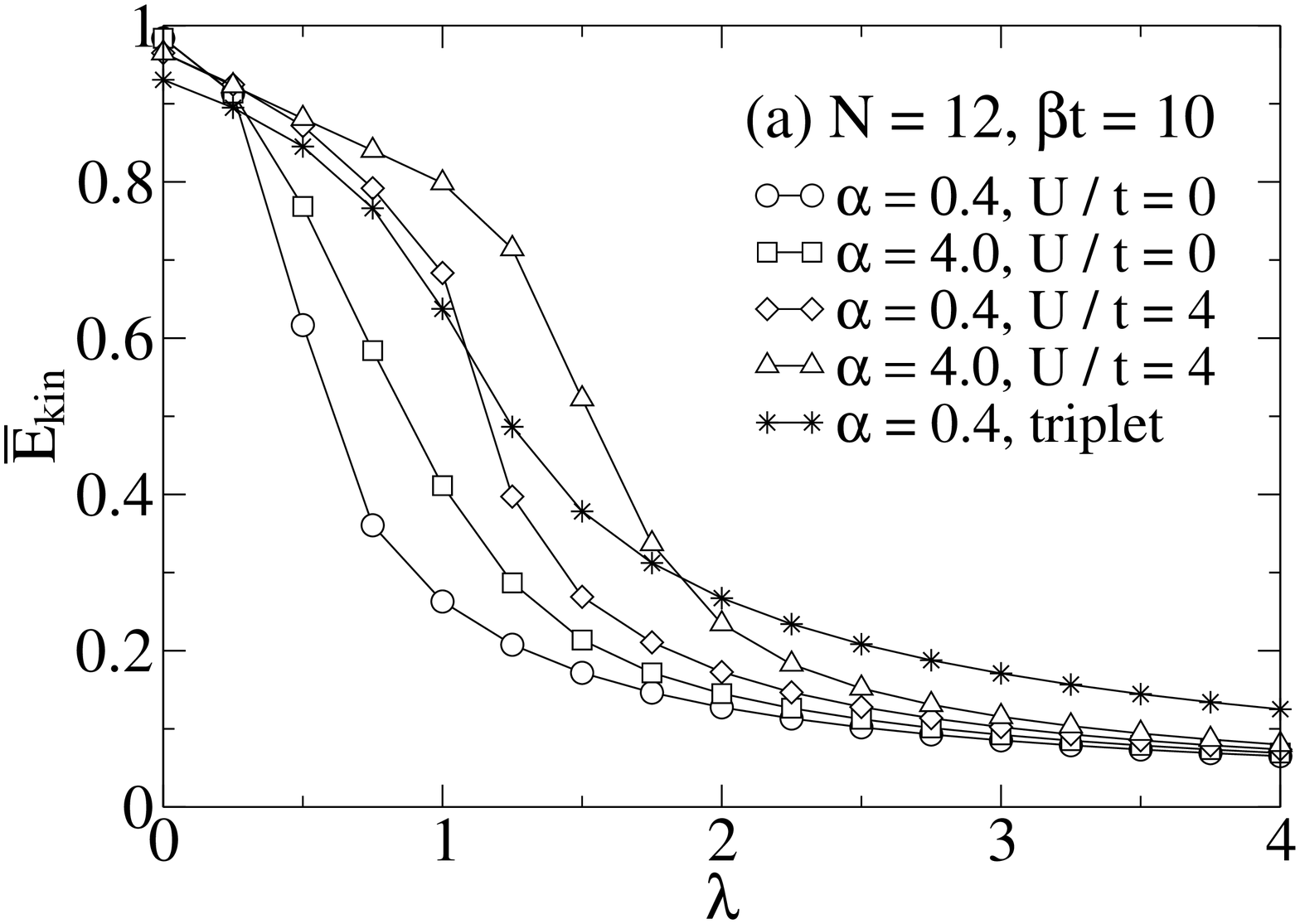}
  \includegraphics[height=0.34\textwidth]{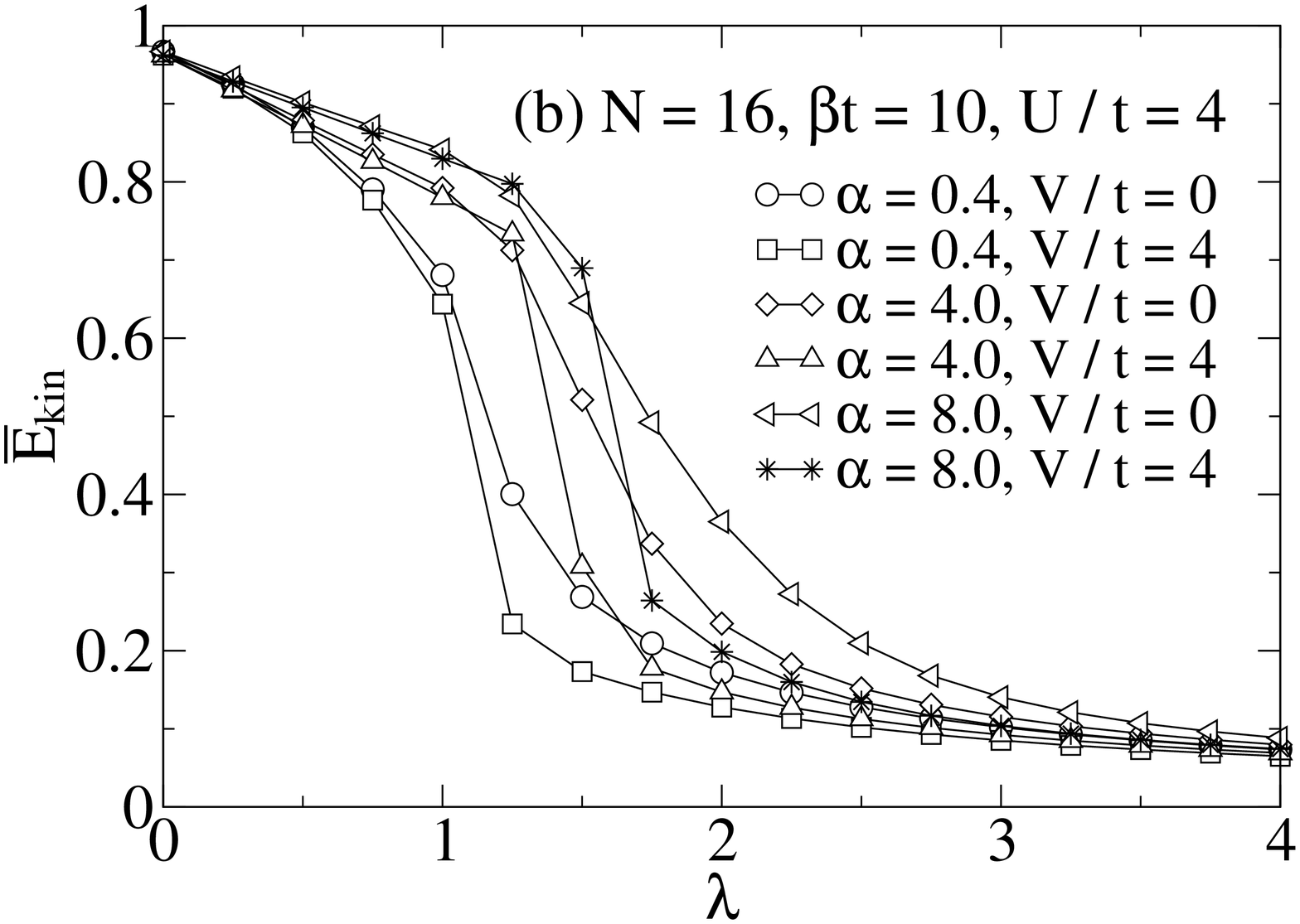}
 \caption{\label{fig:Ek_lambda_omega}
   Normalized kinetic energy $\Ek$ [equation~(\ref{eq:ekeff})] from QMC as a
   function of the el-ph coupling $\lambda$ for different values of the adiabaticity ratio
   $\alpha$, the on-site repulsion $U$ and the
   nearest-neighbour repulsion $V$. [(a) taken from \cite{HovdL05}.]}
\end{figure}

We define the effective kinetic energy of the two electrons as
\begin{equation}\label{eq:ekeff}
  \Ek
  =
  \Ekin/(-4t)
  \,.
\end{equation}
In figure~\ref{fig:Ek_lambda_omega}(a) we depict $\Ek$ as a function of the
el-ph coupling for different values of $\alpha$ and $\Ub$, at $\beta t=10$,
\ie, much closer to the ground state than in some previous work \cite{deRaLa86}.

Figure~\ref{fig:Ek_lambda_omega}(a) reveals a strong decrease of $\Ek$ near
$\lambda=0.5$ for $\alpha=0.4$ and $\Ub=0$. With increasing $\alpha$, the
cross-over becomes less pronounced, and shifts to larger values of $\lambda$.
For the same value of $\alpha$, the cross-over to a small bipolaron is
sharper than the small-polaron cross-over [cf figure~\ref{fig:Ek}(a)]. For
finite on-site repulsion $\Ub=4$, $\Ek$ remains fairly large up to
$\lambda\approx1$ (for $\alpha=0.4$), in agreement with the SC result
$\lambda_\text{c}=1$ for $\Ub=4$ (see discussion in \cite{HoAivdL04}).  At
even stronger coupling, the Hubbard repulsion is overcome, and a small
bipolaron is formed. Again, the critical coupling increases with phonon
frequency. Finally, the kinetic energy in the triplet case (corresponding to
$U/t=\infty$) is comparable to the results for $U/t=4$ up to
$\lambda\approx1$, but significantly larger in the SC regime since on-site
bipolaron formation is not possible.

The influence of nearest-neighour repulsion $V$ is revealed in
figure~\ref{fig:Ek_lambda_omega}(b), again for $\Ub=4$. For all values of
$\alpha$ shown, the cross-over sharpens noticeably for $V>0$. The reason is
that $V>0$ suppresses the (more mobile) inter-site bipolaron state, leading
to a direct cross-over from a large to a small bipolaron. 

The nature of the bipolaron state is revealed by the correlation function
$\rho(\delta)$ [equation~(\ref{eq:bipolaron:rho})], which gives the
probability for the two electrons to be separated by a distance
$\delta\geq0$, and provides a measure of the bipolaron size. The phonon
frequency determines the degree of retardation of the el-ph interaction, and
thereby limits the distance between the two electrons in a bound state. In
the sequel, we shall focus on the most interesting case of small phonon
frequencies, which has often been avoided in previous work for reasons
outlined in section~\ref{sec:qmc}.

\begin{figure}[t]
  \centering
  \includegraphics[width=0.45\textwidth]{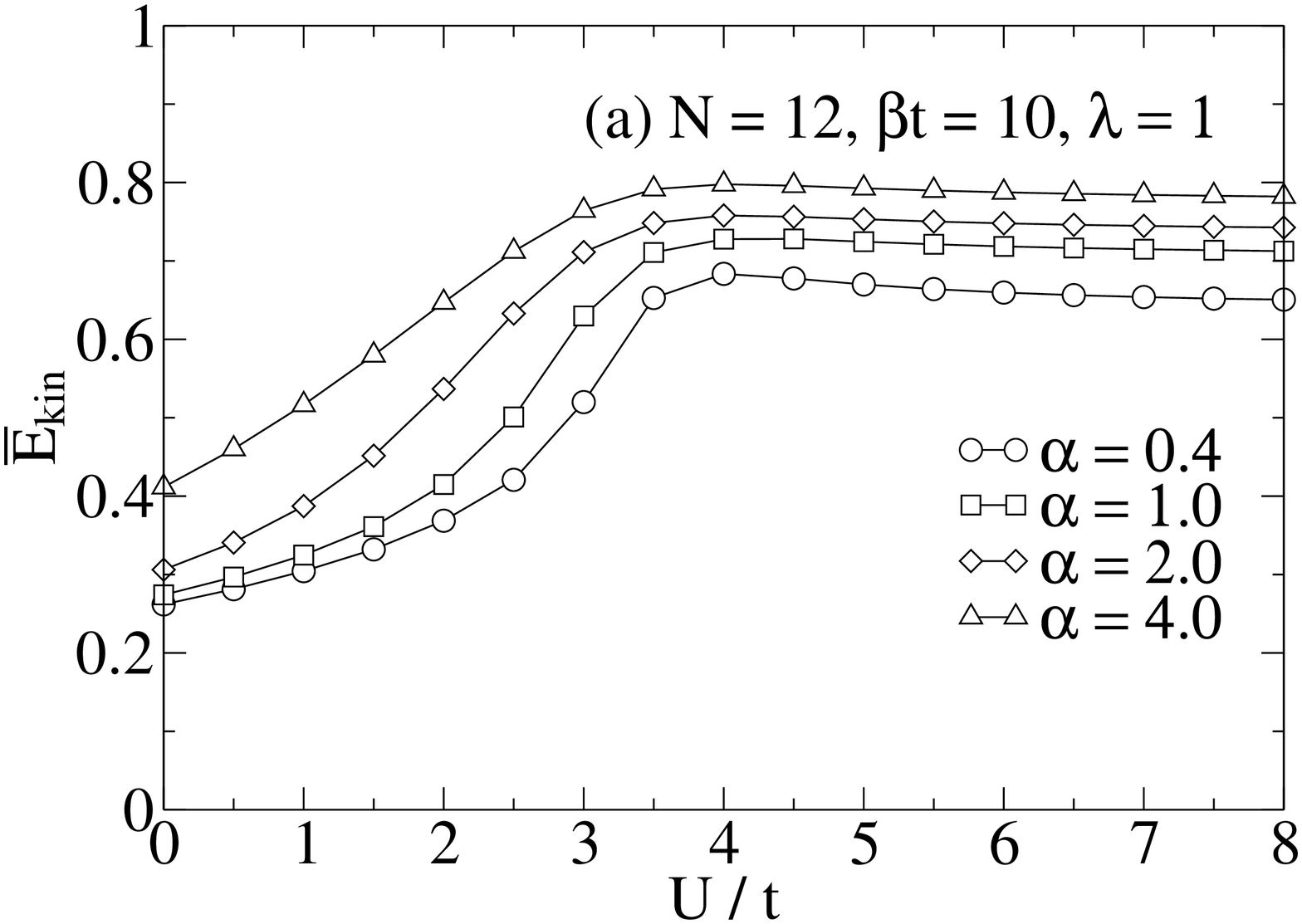}
  \includegraphics[width=0.45\textwidth]{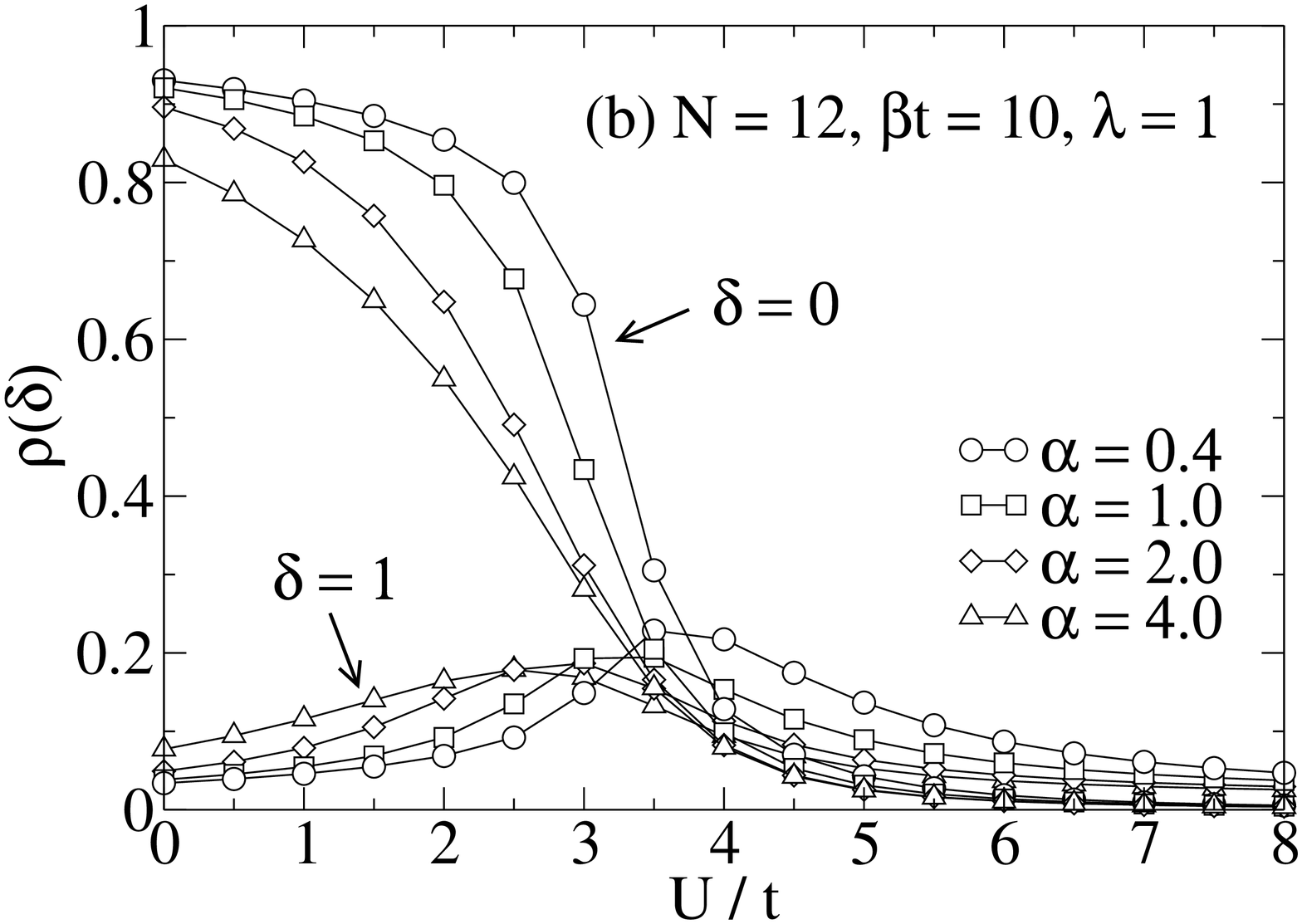}
 \caption{\label{fig:S0S1}
   (a) Normalized kinetic energy $\Ek$ and (b) correlation functions
   $\rho(0)$, $\rho(1)$ from QMC as a function of the Hubbard repulsion $\Ub$ for
   different values of the adiabaticity ratio $\alpha$. [Taken from \cite{HovdL05}.]}
\end{figure}

Starting with $U\ll\Ep$, a cross-over from a small to an inter-site bipolaron to two
weakly bound polarons takes place upon increasing the Hubbard interaction
\cite{BoKaTr00}. Since the latter competes with the retarded el-ph interaction,
the phonon frequency is expected to be an important parameter. In
figure~\ref{fig:S0S1}, we show the kinetic energy and the correlation function
$\rho(\delta)$ as a function of $\Ub$ for IC $\lambda=1$. Starting from a
small bipolaron for $\Ub=0$, the kinetic energy increases with increasing
Hubbard repulsion, equivalent to a reduction of the effective bipolaron
mass \cite{BoKaTr00,ElShBoKuTr03}.
Although the cross-over is slightly washed out by the finite
temperature in our simulations, there is a well-conceivable increase in
$\Ek$ up to $\Ub\approx4$, above which the kinetic energy begins to decrease
slowly. The increase of $\Ek$ originates from the breakup of the small
bipolaron, as indicated by the decrease of $\rho(0)$ in
figure~\ref{fig:S0S1}(b). Close to $\Ub=4$, the curves for $\rho(0)$
and $\rho(1)$ cross, and it becomes more favourable for the two electrons to
reside on neighboring sites.  The inter-site
bipolaron only exists below a critical Hubbard repulsion $U_\text{c}$. The latter is given by
$U_\text{c}=2\Ep$ (\ie, here $U_\text{c}/t=4$) at weak el-ph
coupling, and by $U_\text{c}=4\Ep$ at SC. For an intermediate
value $\lambda=1$ as in figure~\ref{fig:S0S1}, the cross-over from the
inter-site state to two weakly bound polarons is expected to occur somewhere
in between, but is difficult to locate exactly from the QMC results.

Figure~\ref{fig:S0S1} further illustrates that the cross-over becomes steeper with
decreasing phonon frequency. In the adiabatic limit $\alpha=0$, it has been
shown to be a first-order phase transition \cite{PrAu99}, whereas for $\alpha>0$
retardation effects suppress any non-analytic behaviour. At the same
$\Ub$, $\Ek$ increases with $\alpha$ since for a fixed $\lambda$, the
bipolaron becomes more weakly bound. For the same reason, the cross-over to an inter-site
bipolaron---showing up in figure~\ref{fig:S0S1} as a crossing of $\rho(0)$ and
$\rho(1)$---shifts to smaller values of $\Ub$.

\begin{figure}[t]
  \centering
  \includegraphics[width=0.45\textwidth]{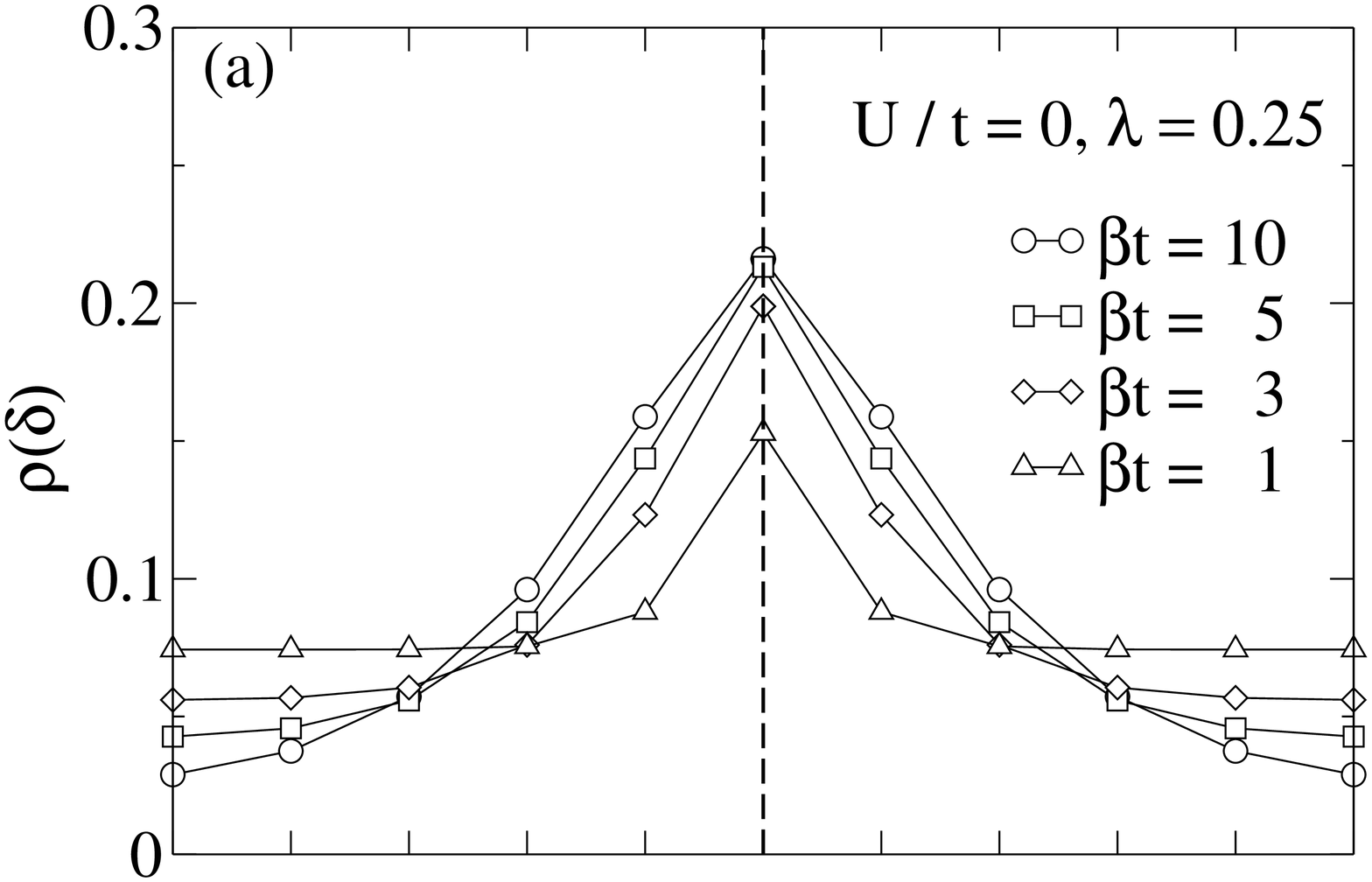}\\
  \includegraphics[width=0.45\textwidth]{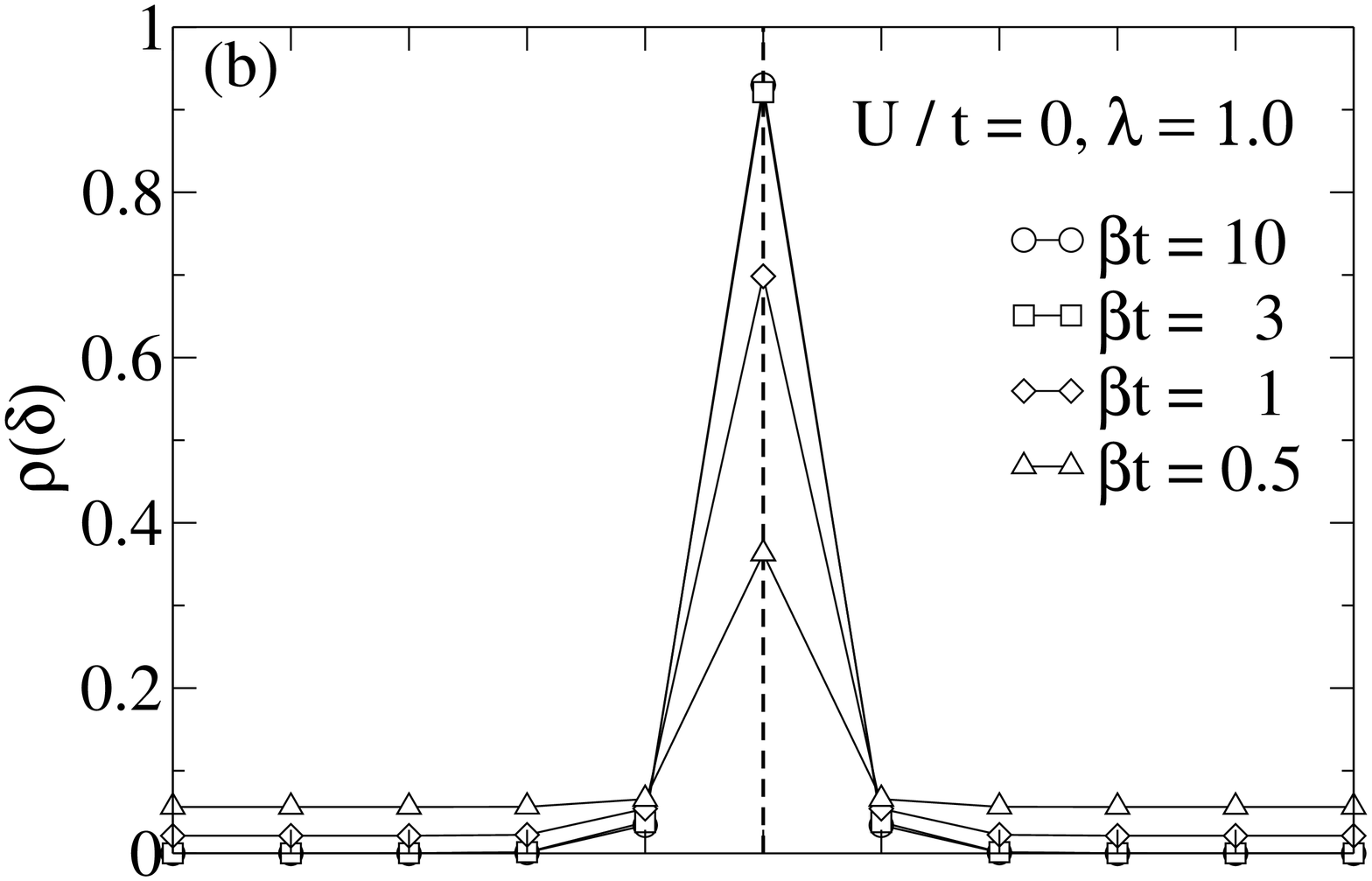}\\
  \includegraphics[width=0.45\textwidth]{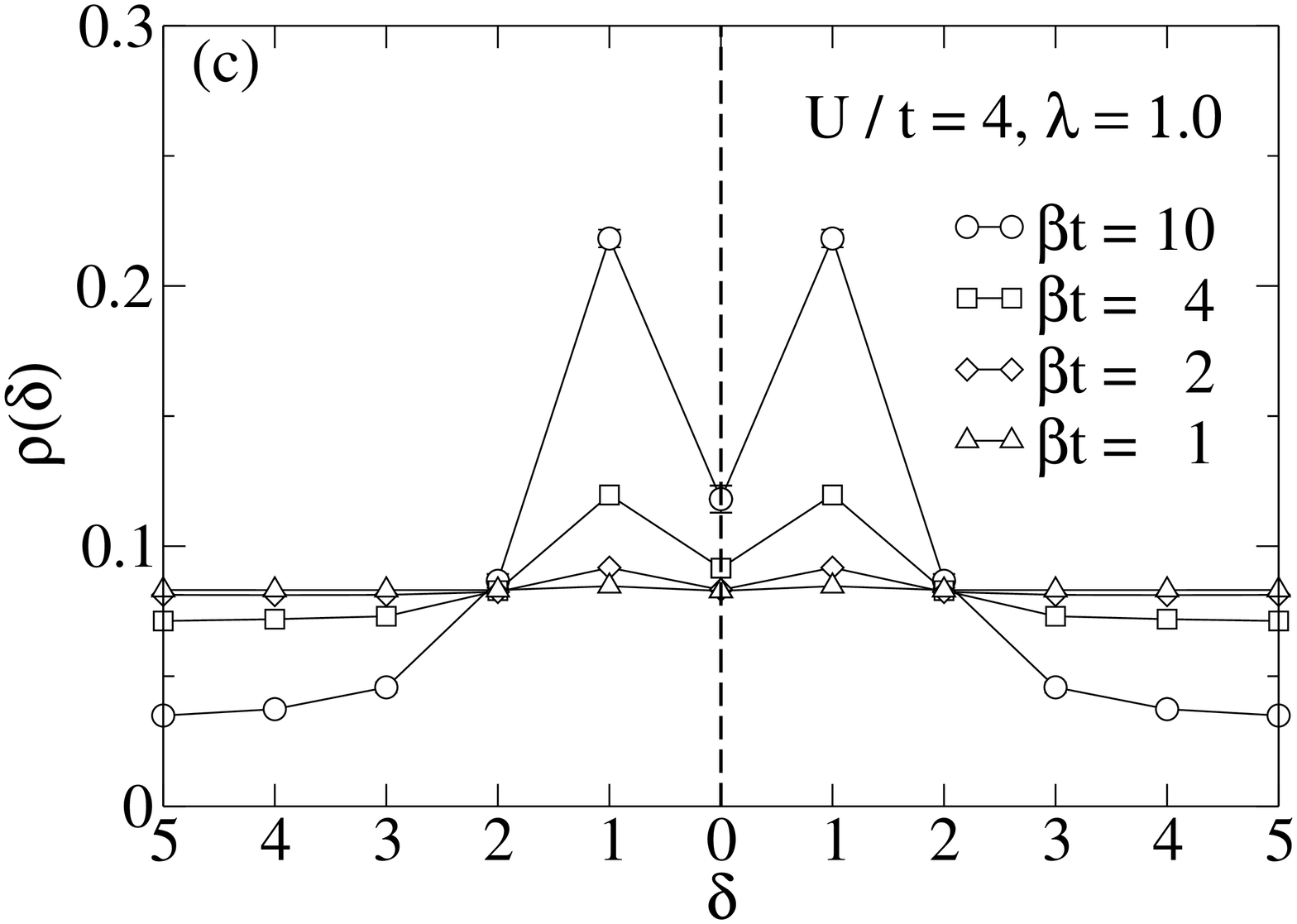}
  \caption{\label{fig:rho_temp}
    Correlation function $\rho(\delta)$ from QMC as a function of $\delta$ for
    different inverse temperatures $\beta$, $N=12$ and $\alpha=0.4$.
    [Taken from \cite{HovdL05}.]}
\end{figure}

Let us now consider the effect of temperature on $\rho(\delta)$. To this end,
we plot in figures~\ref{fig:rho_temp}(a)\,--\,(c) $\rho(\delta)$ at different
temperatures, for parameters corresponding to the three regimes of a large,
small and inter-site bipolaron, respectively.

For the parameters in figure~\ref{fig:rho_temp}(a) ($\Ub=0$, $\lambda=0.25$), the two electrons are
most likely to occupy the same site, but
the bipolaron extends over a distance of several lattice constants. Clearly, in this
regime, the cluster size $N=12$ used here is not completely satisfactory, but
still provides a fairly accurate description as can be deduced from
calculations for $N=14$ (not shown). Nevertheless, on such a small cluster,
no clear distinction between an extended bipolaron and two weakly bound
polarons can be made.  As the temperature increases from $\beta t=10$ to
$\beta t = 1$, the probability distribution broadens noticeably, \ie, it
becomes more likely for the two electrons to be further apart. In particular,
for the highest temperature shown, $\rho(0)$ has reduced by about 30 \%
compared to $\beta t=10$.

A different behaviour is observed for the small bipolaron, which exist at
stronger el-ph coupling $\lambda=1$. Figure~\ref{fig:rho_temp}(b) reveals
that $\rho(\delta)$ peaks strongly at $\delta=0$, but is very small for
$\delta>0$ at low temperatures. Increasing temperature, $\rho(\delta)$
remains virtually unchanged up to $\beta t = 3$. Only at very
high temperatures there occurs a noticeable transfer of probability from
$\delta=0$ to $\delta>0$. At the highest temperature shown, $\beta t=0.5$, the two
electrons have a non-negligible probability for traveling a finite distance
$\delta>0$ apart, although most of the probability is still contained in the
peak located at $\delta=0$.

Finally, we consider in figure~\ref{fig:rho_temp}(c) the inter-site bipolaron,
taking $\Ub=4$ and $\lambda=1$ (cf figure~2 in \cite{HovdL05}). At
low temperatures, $\rho(\delta)$ takes on a
maximum for $\delta=1$. For smaller values of $\beta t$, the latter
diminishes, until at $\beta t=1$, the distribution is completely flat, so
that all $\delta$ are equally likely.

The different sensitivity of the bipolaron states to changes in temperature
found above can be explained by their different binding energies. The latter
is given by $\Delta E_0=E_0^{(2)}-2 E_0^{(1)}$, where $E_0^{(1)}$
and $E_0^{(2)}$ denote the ground-state energy of the model with one and two
electrons, respectively.

\begin{figure}[t]
  \centering
  \includegraphics[width=0.45\textwidth]{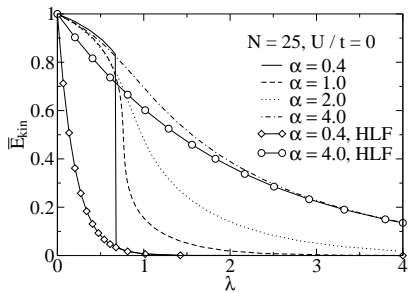}
 \caption{\label{fig:vpa_Ek}
   Variational results for the normalized kinetic energy $\Ek$ as a function
   of the el-ph coupling $\lambda$, and for different adiabaticity ratios
   $\alpha$. Also shown are results of the HLF approximation. [Taken from \cite{HovdL05}.]}
\end{figure}

Generally, the thermal dissociation is expected to occur at a temperature
such that the thermal energy $k_\text{B} T = (\beta T)^{-1}$ becomes
comparable to $\Delta E_0$, in accordance with our numerical data.  The large
and the inter-site bipolaron are relatively weakly bound as a result of the
rather small effective interaction $U_\text{eff}\approx U-2\Ep$
\cite{HoAivdL04}.  The binding energies are $\Delta
E_0\approx-(0.32\pm0.08) t$ and $-(0.28\pm0.08)t$, respectively, so that we
expect a critical inverse temperature $\beta t\approx 2.5$\,--\,5, in
agreement with figures~\ref{fig:rho_temp}(a) and (c).  In contrast, the small
bipolaron in figure~\ref{fig:rho_temp}(b) has a significantly larger binding
energy $\Delta E\approx-(3.43\pm0.09)t$, and therefore remains stable up to
$\beta t\approx 0.3$. Thermal dissociation of bipolarons occurs at even
lower temperatures for $V>0$, especially in the triplet
case, owing to the reduced binding energy.

\subsubsection{Variational approach}\label{sec:res_vpa}

Whereas the QMC approach is limited to finite temperatures and relatively
small clusters, the variational method of section~\ref{sec:vpa} yields
ground-state results on much larger systems. To scrutinize the quality of the
variational method, we compare the ground-state energy for $\Ub=0$ to the
most accurate approach currently available in one dimension, namely the
variational diagonalization \cite{BoKaTr00}. We find a good agreement over
the whole range of $\lambda$. As expected from the nature of the
approximation, slight deviations occur for $\alpha\lesssim1$, similar to the
one-electron case.

Despite the success in calculating the total energy---being the quantity that
is optimized---one has to be careful not to overestimate the validity of any
variational method.  To reveal the shortcomings of the current approach, we
show in figure~\ref{fig:vpa_Ek} the normalized kinetic energy
$\Ek=t_\text{eff}$ [see equations~(\ref{eq:bipolaron:teff})
and~(\ref{eq:ekeff})] as a function of el-ph coupling, and for different
$\alpha$. We have chosen $N=25$ to ensure negligible finite-size effects. In
principle, figure~\ref{fig:vpa_Ek} displays a behaviour similar to the QMC
data in figure~\ref{fig:Ek_lambda_omega}(a). There is a jump-like decrease of
$\Ek$ near $\lambda=0.5$ for $\alpha=0.4$, which becomes washed out and moves
to larger $\lambda$ with increasing phonon frequency. For $\alpha=0.4$, the
cross-over in the variational results is much too steep, regardless of the
fact that the latter are for $T=0$, a common defect of variational methods.
Moreover, for $\alpha=0.4$\,--\,2, the variational kinetic energy is too
small above the bipolaron cross-over compared to the QMC data, whereas for
$\alpha=4$, the decay of $\Ek$ with increasing $\lambda$ is too slow.

The reason for the failure is the absence
of retardation effects, which play a dominant role in the formation of
bipolaron states. The increased importance of the phonon dynamics---not
included in the variational method---for the two-electron problem leads to a
less good agreement with exact results than in the one-electron
case. In particular, our variational results overestimate the
position of the cross-over (figure~\ref{fig:vpa_Ek}) compared to the
value $\lambda_\text{c}=0.5$ expected in the adiabatic regime.  Nevertheless,
the method represents a significant improvement over the simple HLF
approximation, due to the variational determination of the parameters
$\gamma_{ij}$. This is illustrated in figure~\ref{fig:vpa_Ek}, where we also
show the HLF result $\Ek = \rme^{-g^2}$ for $\alpha=0.4$ and 4.0. In contrast to the
variational approach, the HLF approximation yields an exponentially decreasing
kinetic energy for all values of the phonon frequency.  Whereas such behaviour
actually occurs in the anti-adiabatic limit $\alpha\to\infty$, the
situation is different for small $\alpha$ [see figures~\ref{fig:Ek_lambda_omega}(a)
and~\ref{fig:vpa_Ek}]. The
variational method presented here accounts qualitatively for the influence of
the phonon frequency on bipolaron formation.

\begin{figure}[t]
  \centering
  \includegraphics[width=0.45\textwidth]{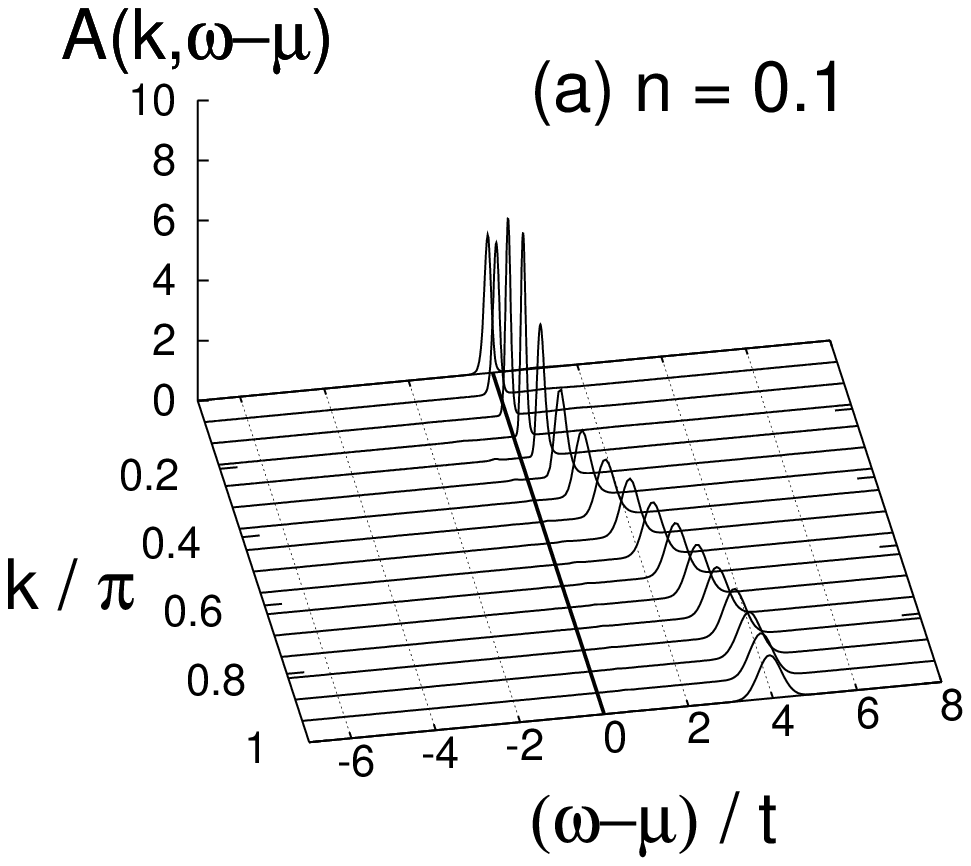}
  \includegraphics[width=0.45\textwidth]{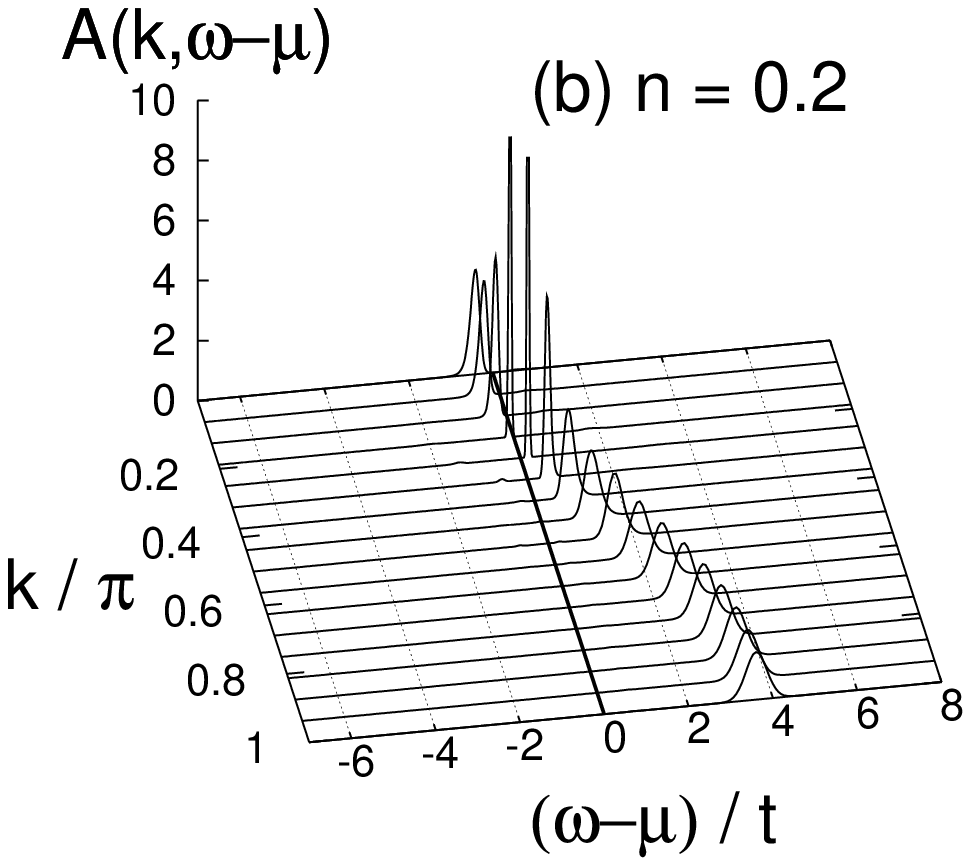}\\
  \includegraphics[width=0.45\textwidth]{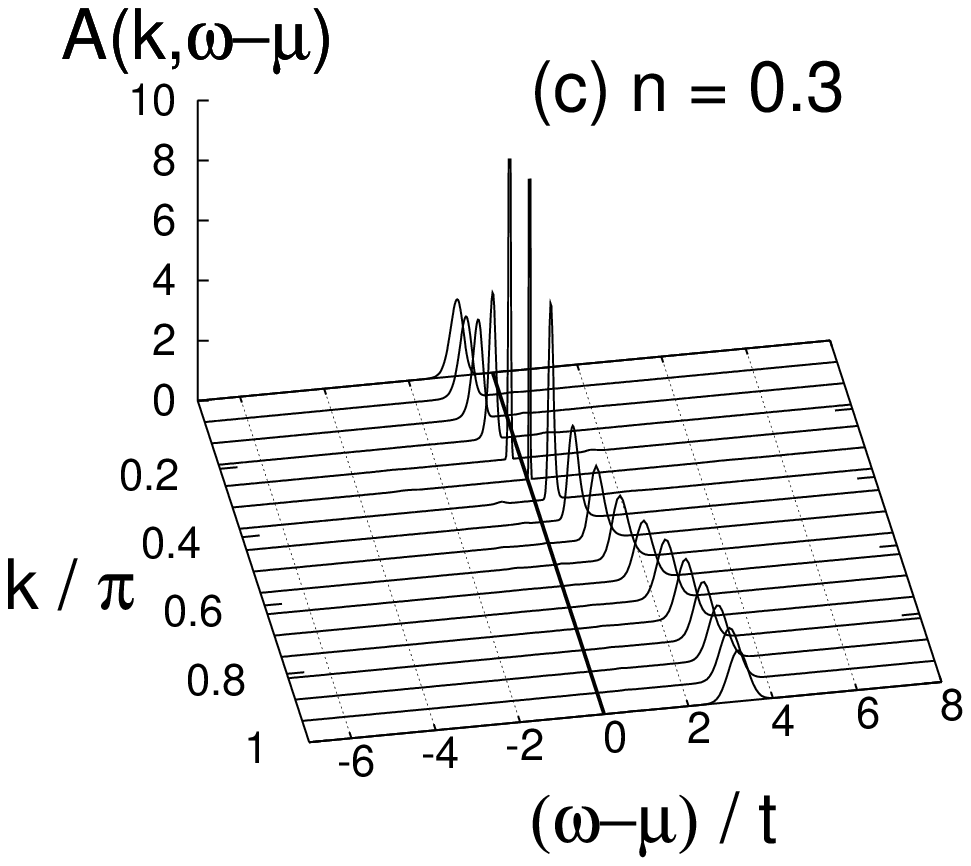}
  \includegraphics[width=0.45\textwidth]{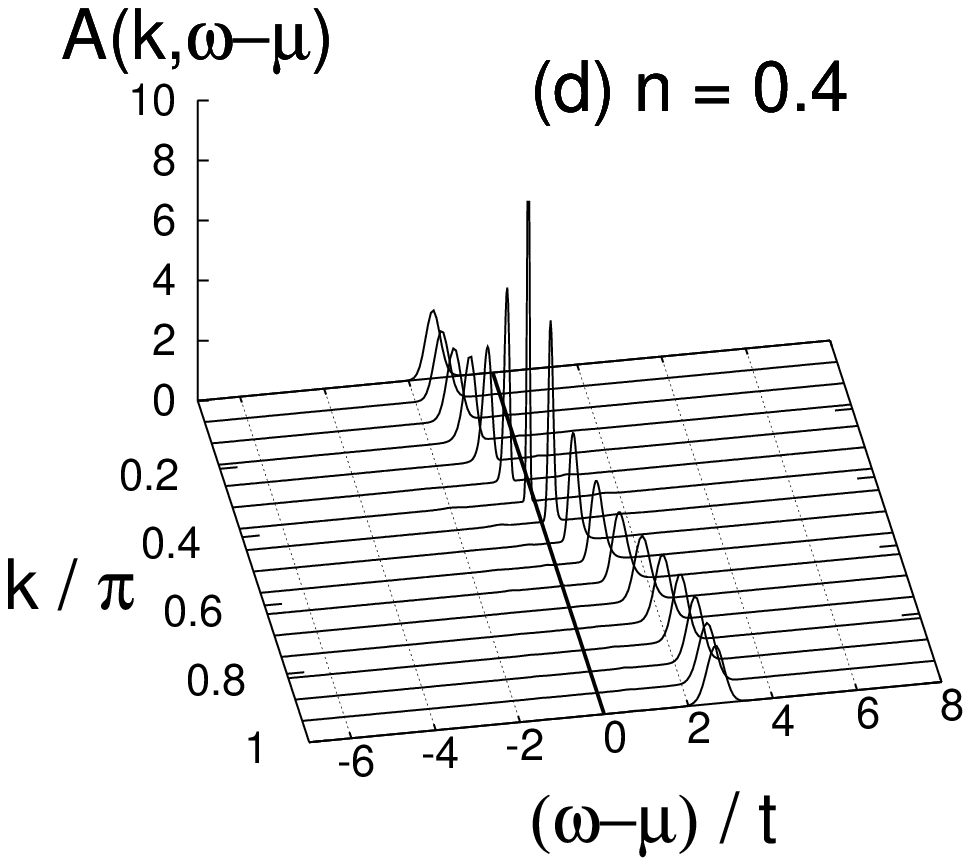}
  \caption{\label{fig:results:QMC_lambda0.1}
    One-electron spectral function $A(k,\om-\mu)$ from QMC for different band
    fillings $n$, $N=32$, $\beta t=8$, $\alpha=0.4$, and $\lambda=0.1$. Here
    and in subsequent figures $\dtau=0.1$. [Taken from \cite{HoNevdLWeLoFe04}.]}
\end{figure}

\subsection{Many-polaron problem}\label{sec:res-manypol}

We review recent results on the carrier-density dependence of photoemission
spectra of many-polaron systems in the framework of the spinless Holstein
model~(\ref{eq:many-electrons:Hspinless}) in one dimension. We shall see that the sensitivity
to changes in $n$ strongly depends on the phonon frequency and el-ph coupling
strength, with the most interesting physics being observed in the adiabatic,
IC regime often realized experimentally. This regime is characterized by the
existence of large polarons at low carrier density. At larger densities, a
substantial overlap of the single-particle wavefunctions occurs, leading to a
dissociation of the individual polarons and finally to a restructuring of the
whole many-particle ground state. Note that the many-polaron problem has
since been studied also by means of other methods
\cite{LoHoFe06,HoWeAlFe05,WeBiHoScFe05}, confirming the original findings of
\cite{HoNevdLWeLoFe04}. 

\subsubsection{Weak coupling}

For WC $\lambda=0.1$, the sign problem is not severe
(section~\ref{sec:sign}) so that simulations can easily be
performed for large lattices with $N=32$, making the dispersion of
quasiparticle features well visible.

\begin{figure}[t]
  \centering
  \includegraphics[width=0.6\textwidth]{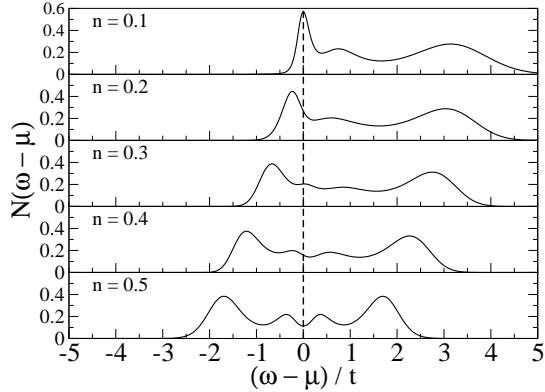}
\caption{\label{fig:results:dos_wc}
  One-electron density of states $N(\om-\mu)$ from QMC for different band
  fillings $n$, $N=32$, $\beta t=8$, $\alpha=0.4$ and $\lambda=0.1$. [Taken from \cite{HoNevdLWeLoFe04}.]}
\end{figure}

Figure~\ref{fig:results:QMC_lambda0.1} shows the evolution of the
one-electron spectral function $A(k,\om-\mu)$ with increasing electron density
$n$. At first sight, we see that the spectra bear a close resemblance to the
free-electron case, \ie, there is a strongly dispersive band running from
$-2t$ to $2t$ which can be attributed to weakly dressed
electrons.  As
expected, the height (width) of the peaks increases (decreases) significantly
in the vicinity of the Fermi momentum $k_\text{F}$, determined by
the crossing of the band with the chemical potential.  However, in contrast to
the case of a rigid tight-binding band, we shall see below
(figure~\ref{fig:results:dos_wc}) that a significant
redistribution of spectral weight occurs with increasing $n$.

We would like to point out that the apparent absence of any phonon
signatures in figure~\ref{fig:results:QMC_lambda0.1} is not a defect of the
maximum entropy method, but results from the large scale of the $z$-axis chosen. As a
consequence, the peaks running close to the bare band dominate the spectra
and suppress any small phonon peaks present. At higher resolution,
for all densities $n=0.1$\,--\,0.4, we observe the band
flattening \cite{Stephan,WeFe97,HoAivdL03} at large wavevectors which
originates from the intersection of the approximately free-electron
dispersion with the bare phonon energy at $\om-\mu=\om_0$.

To complete our discussion of the WC regime, we show in
figure~\ref{fig:results:dos_wc} the one-electron density of states (DOS)
$N(\om-\mu)$ given by equation~(\ref{eq:many-electrons:DOS}). Clearly, for
small $n$, there is a peak with large spectral weight at the Fermi level. In
contrast, for large $n$, the tendency toward formation of a Peierls--
(band--) insulating state at $n=0.5$ suppresses the DOS at the Fermi level,
although we are well below the critical value of $\lambda$ at which the
cross-over to the insulating state takes place at $T=0$
\cite{BuMKHa98,HoWeBiAlFe06}.  The additional small features separated from
$\mu$ by the bare phonon energy $\om_0$ will be discussed below.

\subsubsection{Strong coupling}
\begin{figure}[t]
  \centering
  \includegraphics[width=0.45\textwidth]{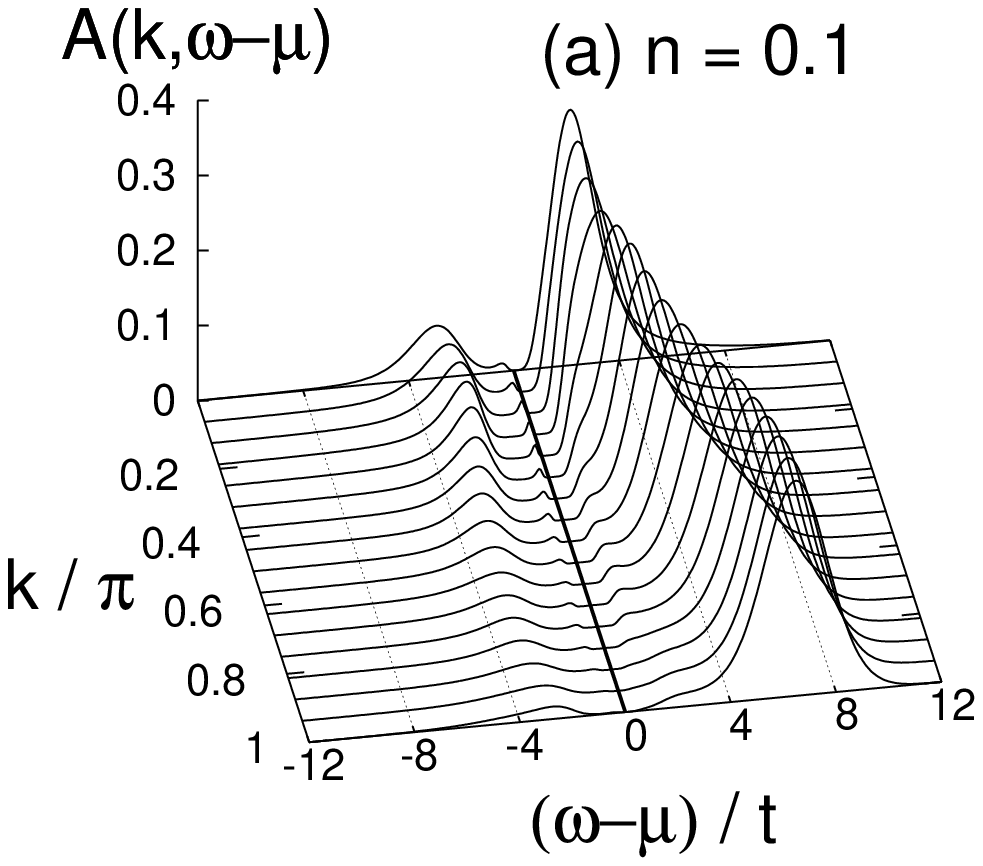}
  \includegraphics[width=0.45\textwidth]{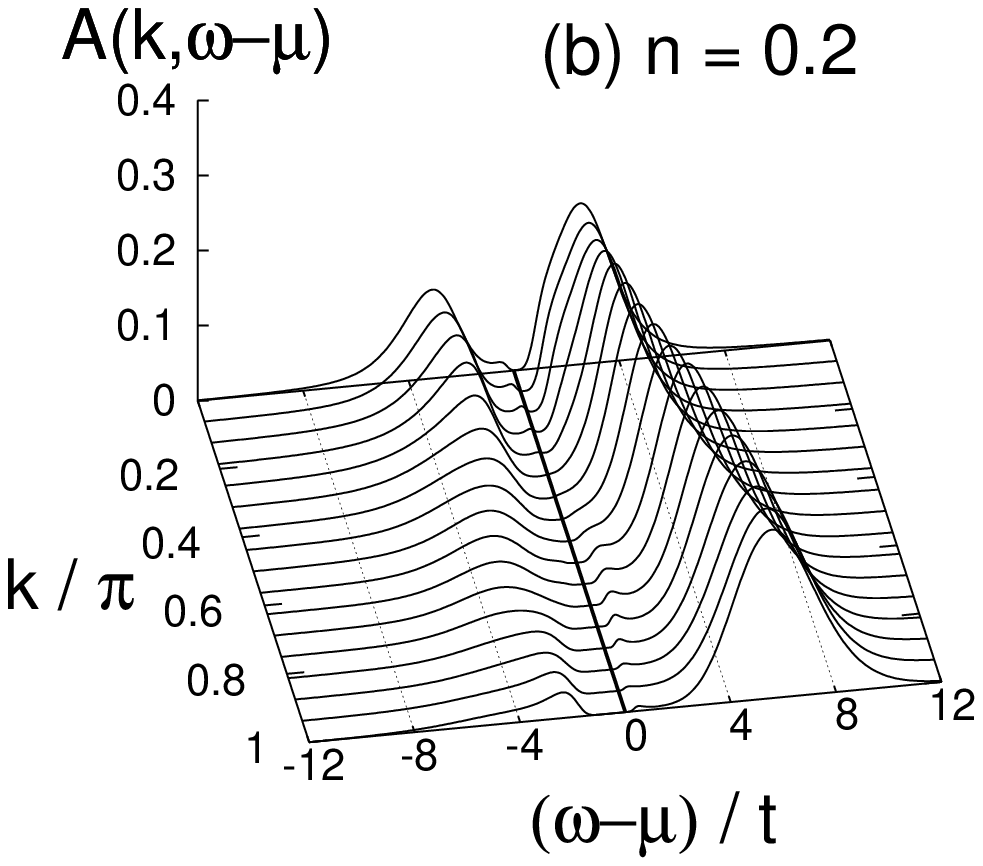}\\
  \includegraphics[width=0.45\textwidth]{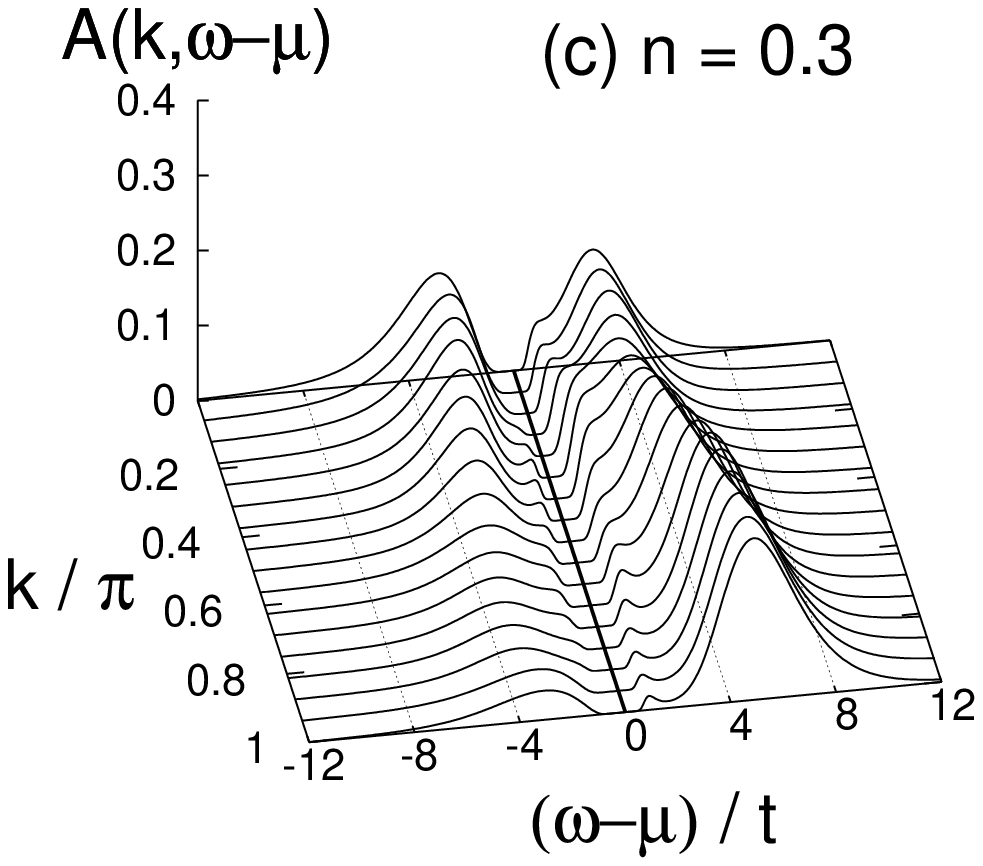}
  \includegraphics[width=0.45\textwidth]{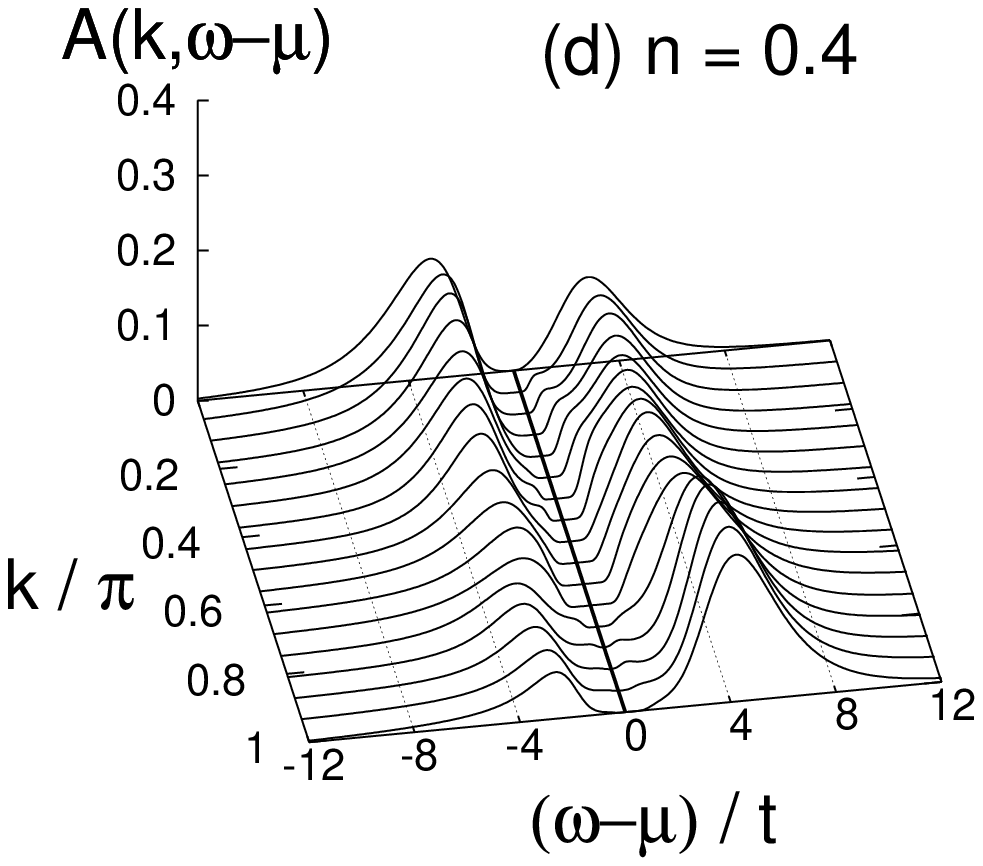}
  \caption{\label{fig:results:QMC_lambda2.0}
    One-electron spectral function $A(k,\om-\mu)$ from QMC for different band
    fillings $n$, $N=32$, $\beta t=8$, $\alpha=0.4$, and $\lambda=2$.
  [Taken from \cite{HoNevdLWeLoFe04}.]}
\end{figure}

We now turn to the SC limit taking $\lambda=2$.
At low density $n=0.1$ [figure~\ref{fig:results:QMC_lambda2.0}(a)], we expect
the well-known, almost flat polaron band having exponentially reduced
spectral weight (given by $e^{-g^2}$ in the single-electron, SC limit) which,
nevertheless, can give rise to coherent transport at $T=0$. As discussed in
\cite{HoNevdLWeLoFe04}, such weak signatures are difficult to determine
accurately using the maximum entropy method. Generally, it is known that the
reliability of dynamic properties obtained by means of the maximum entropy
method crucially depends on the size of statistical errors and the
general structure of the spectra. A detailed discussion of this point has
been given in \cite{HoNevdLWeLoFe04}.

Besides, the spectrum consists of two incoherent features located above and
below the chemical potential, which reflect the phonon-mediated transitions
to high-energy electron states. Here, the maximum of the photoemission
spectra ($\om-\mu>0$) follows a tight-binding cosine dispersion.  The
incoherent part of the spectra is broadened according to the phonon
distribution.

For all band fillings, the chemical potential is expected to be located in a
narrow polaron band with little spectral weight. There exists a finite gap to the
photoemission (inverse photoemission) parts of the spectrum, so
that the system typifies as a polaronic metal. We shall see below that a
completely different behaviour is observed at IC. Notice that the incoherent
inverse photoemission (photoemission) signatures are more pronounced at small
(large) wavevectors.

Finally, for $n=0.4$ [figure~\ref{fig:results:QMC_lambda2.0}(d)], the
incoherent features lie rather close to the Fermi level, thus being
accessible by low-energy excitations. Now, the photoemission spectrum for
$k<\pi/2$ is almost symmetric to the inverse photoemission spectrum for
$k>\pi/2$ and already reveals the gapped structure which occurs at $n=0.5$
due to charge-density-wave formation accompanied by a Peierls distortion
\cite{HoWeBiAlFe06}.

\begin{figure}[t]
  \centering
  \includegraphics[width=0.6\textwidth]{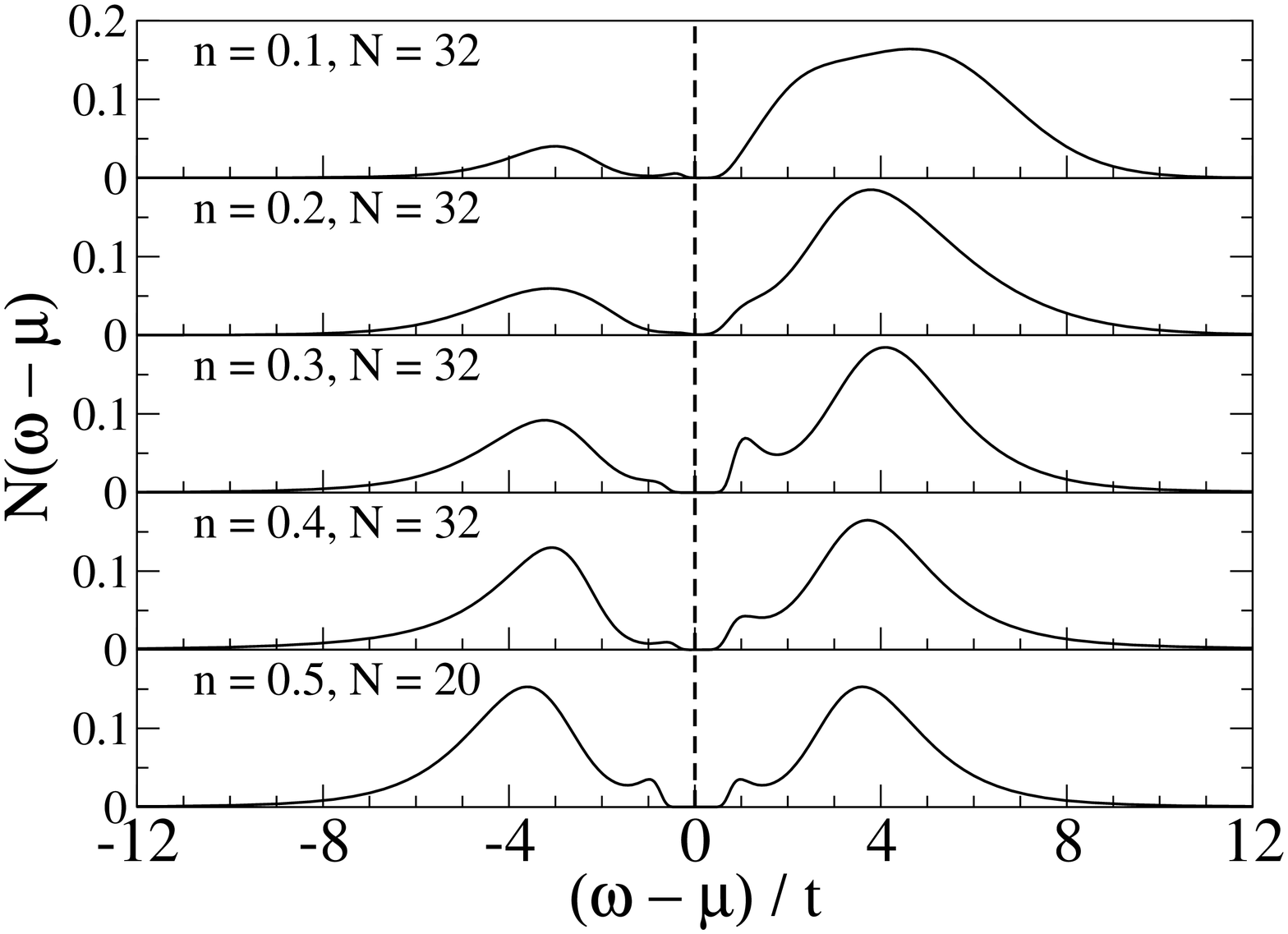}
\caption{\label{fig:results:dos_sc}
  One-electron density of states $N(\om-\mu)$ from QMC for different band
  fillings $n$ and cluster sizes $N$, $\beta t=8$, $\alpha=0.4$ and
  $\lambda=2$. [Taken from \cite{HoNevdLWeLoFe04}.]}
\end{figure}

As in the WC case discussed above, the properties of the system also manifest
itself in the DOS, shown in figure~\ref{fig:results:dos_sc}. Owing to the
strong el-ph interaction, the spectral weight at the chemical potential is
exponentially small for all fillings $n$. At half filling, the DOS exhibits
particle-hole symmetry, and the system can be described as a Peierls
insulator, consisting of a polaronic superlattice.  In contrast to the WC
case, the ground state is characterized as a polaronic insulator rather than
as a band insulator.

\subsubsection{Intermediate coupling}

As discussed in the introduction, a cross-over from a polaronic state to a
system with weakly dressed electrons can be expected in the IC regime. Here
we choose $\lambda=1$, which corresponds to the
critical value for the small-polaron cross-over in the one-electron problem
[cf figure~\ref{fig:Ek}(a)]. Owing to the sign problem, which is
particularly noticeable for $\lambda=1$ (see
figure~\ref{fig:many-electrons:sign_beta_omega}), we have to decrease the
system size as we increase the electron density $n$.

\begin{figure}[t]
  \centering
  \includegraphics[width=0.45\textwidth]{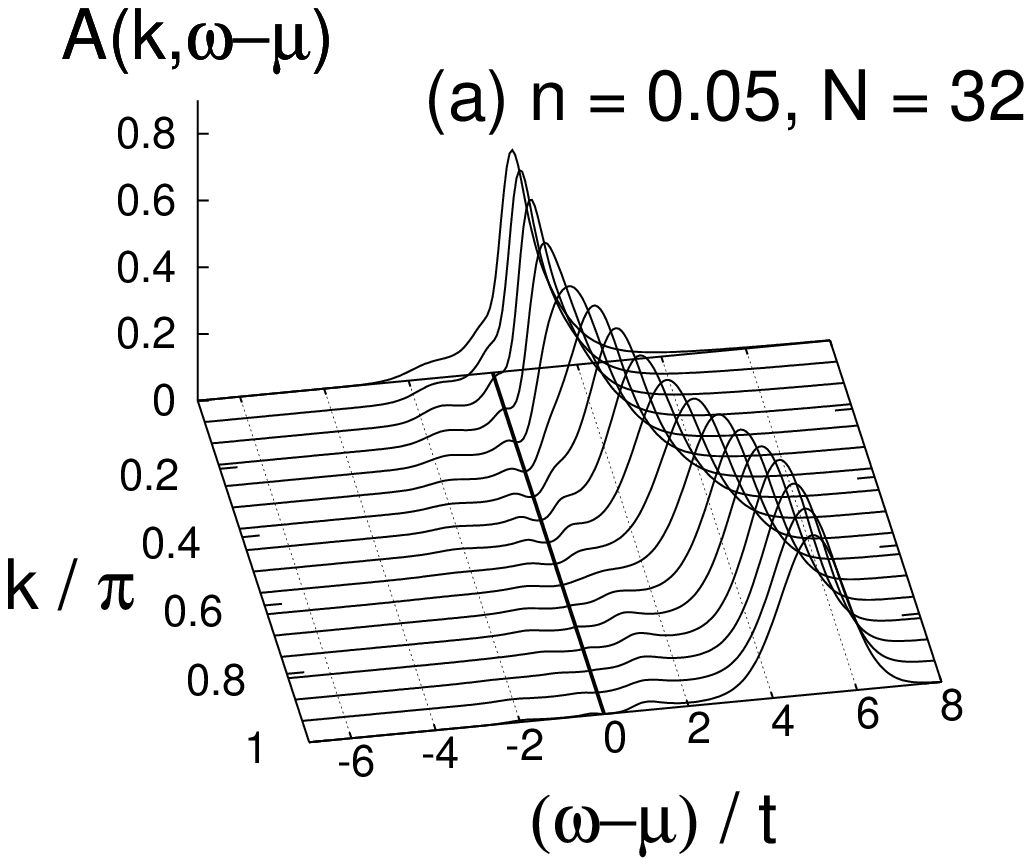}
  \includegraphics[width=0.45\textwidth]{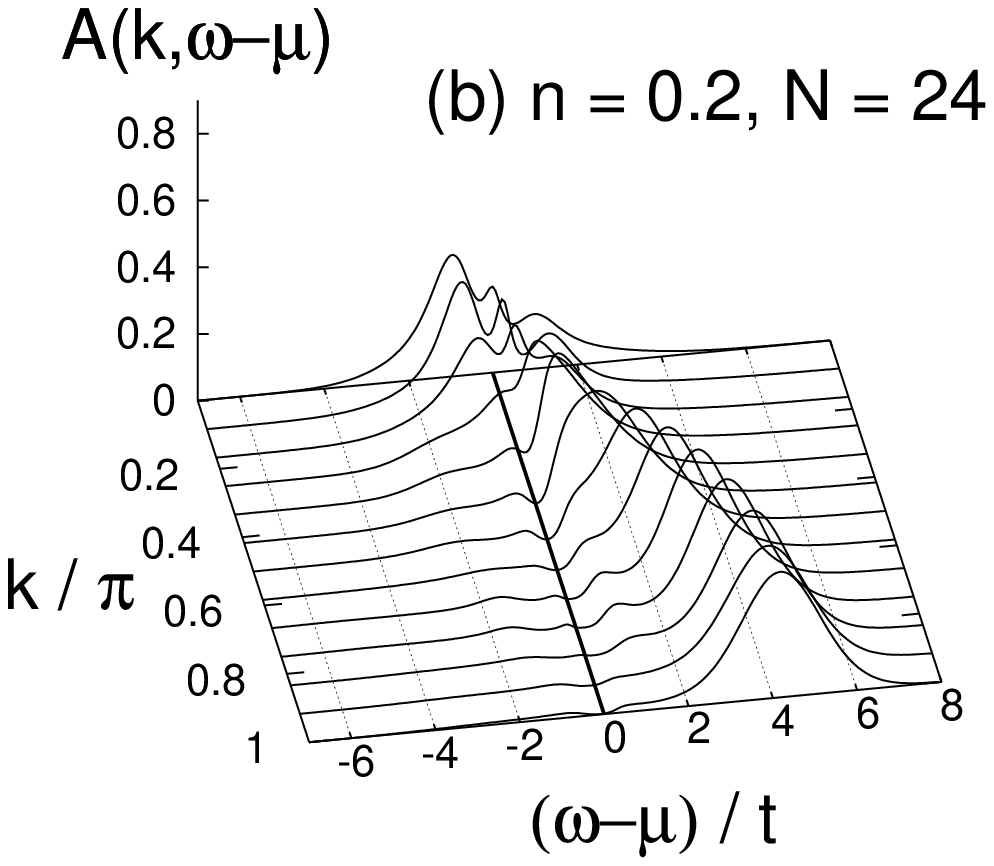}\\
  \includegraphics[width=0.45\textwidth]{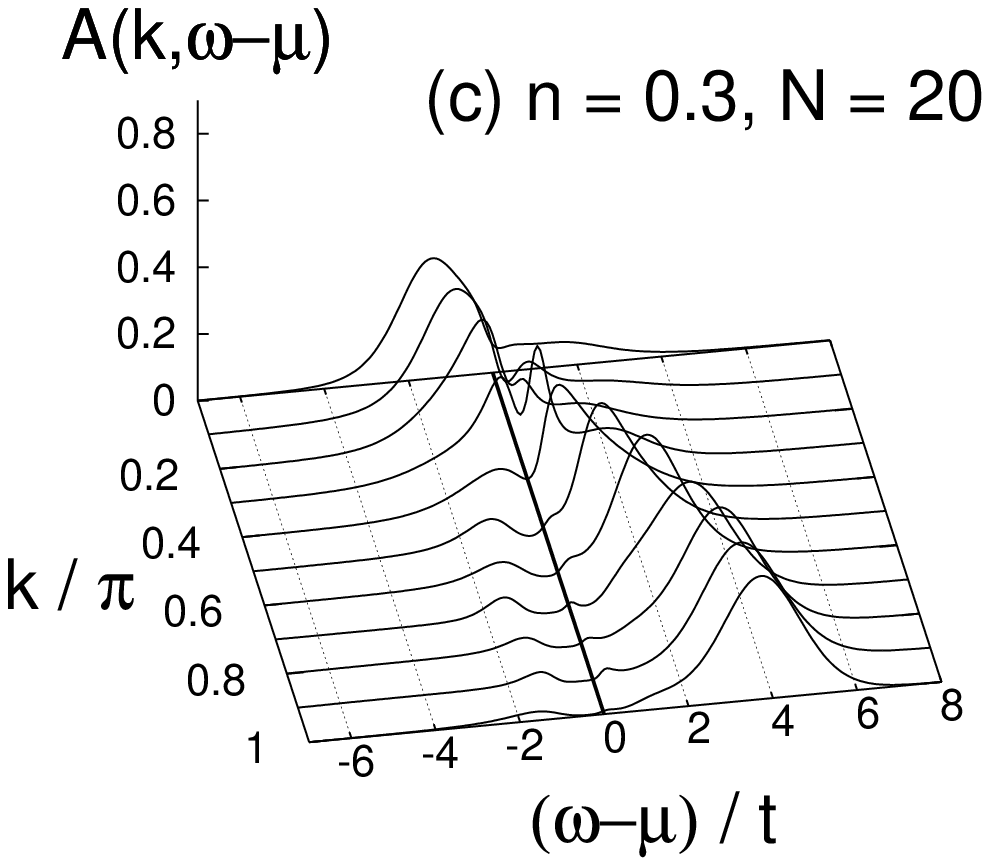}
  \includegraphics[width=0.45\textwidth]{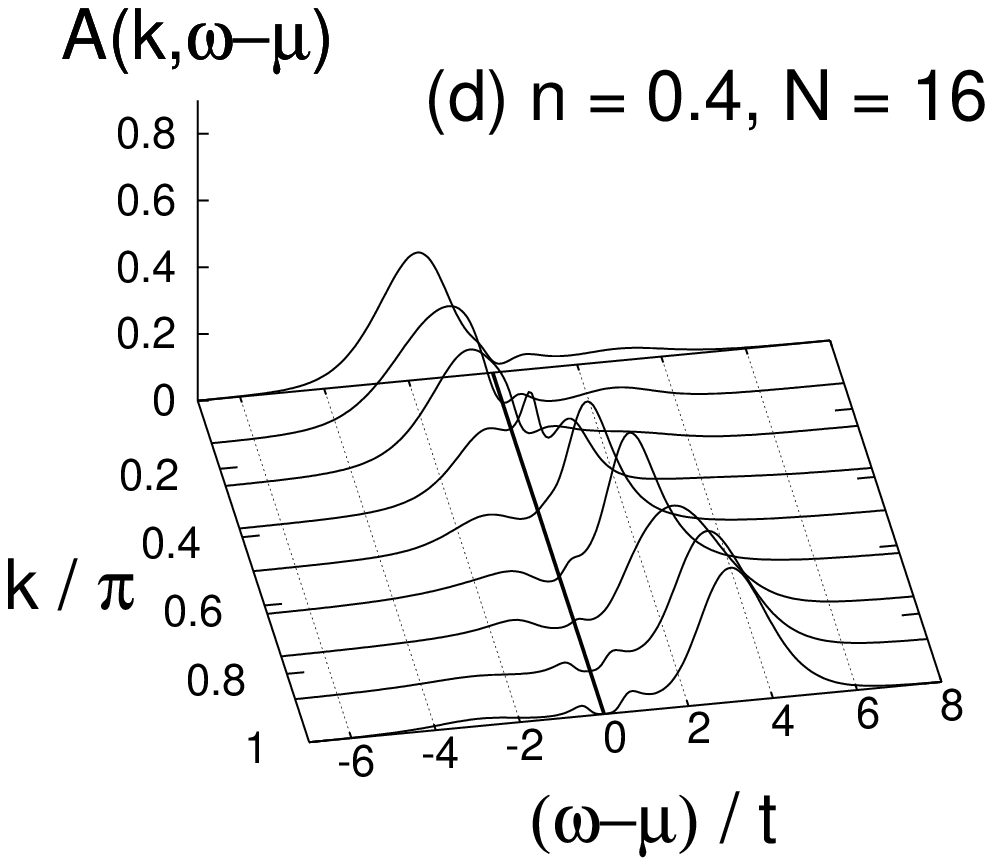}
  \caption{\label{fig:results:QMC_lambda1.0}
    One-electron spectral function $A(k,\om-\mu)$ from QMC for different band
    fillings $n$ and cluster sizes $N$, $\beta t=8$, $\alpha=0.4$, and
    $\lambda=1$. [Taken from \cite{HoNevdLWeLoFe04}.]}
\end{figure}

We shall see that the cross-over is rather difficult to detect from the QMC
results only. However, the data presented here are perfectly consistent with
more recent studies employing other methods such as exact diagonalization
\cite{HoNevdLWeLoFe04}, cluster perturbation theory \cite{HoWeAlFe05} or
self-energy calculations \cite{LoHoFe06}.

Figure~\ref{fig:results:QMC_lambda1.0} shows the spectral function for
$\lambda=1$ and increasing band filling.  Owing to the overlap of large
polarons in the IC regime, we start with a very low
density $n=0.05$ [figure~\ref{fig:results:QMC_lambda1.0}(a)]. Compared to
the behaviour for $\lambda=2$ [figure~\ref{fig:results:QMC_lambda2.0}(a)], we
notice that the polaron band now lies much closer to the incoherent features,
and that there is a mixing of these two parts of the spectrum at small values
of $k$. Nevertheless, the almost flat polaron band is well visible for large
$k$.

With increasing density, the polaron band merges with the incoherent peaks at
higher energies, signaling the above-anticipated density-driven cross-over
from a polaronic to a (diffusive) metallic state, with the broad main band
crossing the Fermi level. 

\begin{figure}[t]
  \centering
  \includegraphics[width=0.6\textwidth]{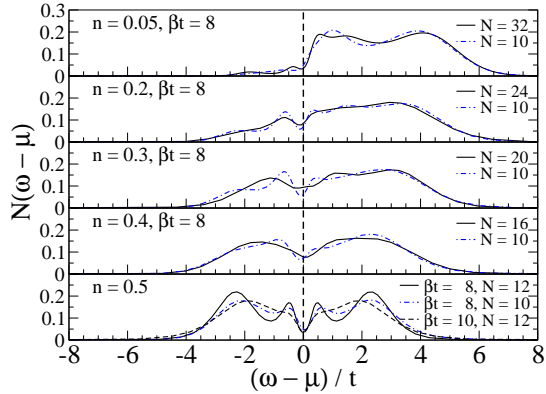}
\caption{\label{fig:results:dos_ic}
  One-electron density of states $N(\om-\mu)$ from QMC for different band
  fillings $n$, cluster sizes $N$ and inverse temperatures $\beta$. Here
  $\alpha=0.4$ and $\lambda=1$. [Taken from \cite{HoNevdLWeLoFe04}.]}
\end{figure}

Further information about the density dependence can be obtained from the
one-electron DOS. The latter is presented in figure~\ref{fig:results:dos_ic}
for different fillings $n=0.05$\,--\,0.5.  As in
figure~\ref{fig:results:QMC_lambda1.0}, the cluster size is reduced with
increasing $n$ in order to cope with the sign problem. To illustrate the
rather small influence of finite-size effects,
figure~\ref{fig:results:dos_ic} also contains results for $N=10$.

For low density $n=0.05$, the DOS in figure~\ref{fig:results:dos_ic} lies in
between the results for WC and SC discussed above. Although the spectral
weight at the chemical potential is strongly reduced compared to
$\lambda=0.1$, $N(0)$ is still significantly larger than for $\lambda=2$.

When the density is increased to $n=0.2$, the DOS at the chemical potential
increases, as a result of the dissociation of polarons. Increasing $n$
further, a pseudogap begins to form at $\mu$, which is a precursor of the
charge-density-wave gap at half filling and zero temperature.

In the case of half filling $n=0.5$, the DOS has become symmetric with
respect to $\mu$.  There are broad features located either side of the
chemical potential, which take on maxima close to $\om-\mu=\pm\Ep$. However,
apart from the SC case, where the single-polaron binding energy is still a
relevant energy scale, the position of these peaks is rather determined by
the energy of the upper and lower bands, split by the formation of a Peierls
state. The gap of size $\sim\lambda$ expected for the insulating
charge-ordered state at $T=0$ is partially filled in due to the finite
temperature considered here.

Furthermore, we find additional, much smaller features roughly separated from
$\mu$ by the bare phonon frequency $\om_0$, whose height decreases with
decreasing temperature, as revealed by the results for $\beta t=10$
(figure~\ref{fig:results:dos_ic}). These peaks---not present at $T=0$
\cite{SyHuBeWeFe04,HoWeBiAlFe06}---arise from thermally activated transitions
to states with additional phonons excited, and are also visible in
figures~\ref{fig:results:dos_wc} and~\ref{fig:results:dos_sc}.  While for WC
($\lambda=0.1$, figure~\ref{fig:results:dos_wc}), the maximum of these
features is almost exactly located at $|\om-\mu|=\om_0$, it moves to
$|\om-\mu|\approx1.25\om_0$ for IC ($\lambda=1$,
figure~\ref{fig:results:dos_ic}), and finally to $|\om-\mu|\approx2.5\om_0$
for SC ($\lambda=2$, figure~\ref{fig:results:dos_sc}). Although the exact
positions of the peaks are subject to uncertainties due to the maximum
entropy method, this evolution reflects the shift of the maximum in the phonon
distribution function with increasing coupling. The maximum entropy method
yields an envelope of the multiple peaks separated by $\om_0$.

\subsubsection{Anti-adiabatic regime}

The comparison of the spectral functions for $n=0.1$ and $n=0.3$ in
figure~10 of \cite{HoNevdLWeLoFe04} reveals that there is no density-driven
cross-over of the system as observed in the adiabatic case even for the
critical value $g^2=1$. In particular, owing to
the large phonon energy, there are no low-energy excitations close to the
polaron band, so that the latter remains well separated from the incoherent
features even for $n=0.3$. Furthermore, the spectral weight of the polaron
band also remains almost unchanged as we increase the density from $n=0.1$
to $n=0.3$. Consequently, almost independent small
polarons are formed also at finite electron densities, in accordance with
previous findings for small systems \cite{CaGrSt99}.

\section{Summary}
\label{sec:summary}

We have reviewed quantum Monte Carlo and variational approaches to Holstein
models based on Lang-Firsov transformations of the Hamiltonian. The methods
have been applied to investigate single polarons and bipolarons,
respectively, as well as a many-polaron system.

The variational methods include displacements of the lattice at all lattice
sites, which enables them to quite accurately describe large polaron or
bipolaron states.

Using the transformed Hamiltonian, we have shown that quantum Monte Carlo
simulation can be based on exact sampling without autocorrelations, which
proves to be an enormous advantage for small phonon frequencies or low
temperatures. Indeed, we have used a grand-canonical algorithm to obtain
dynamical properties of many-polaron systems in all interesting parameter
regimes. Such simulations are currently not possible with other Monte Carlo
methods.

\section*{Acknowledgements}

We thank A.~R.~Bishop, H.~G.~Evertz, H.~Fehske, J.~Loos, D.~Neuber, W.~von
der Linden, G.~Wellein for fruitful discussion.


\begin{thebibliography}{10}

\bibitem{LangFirsov}
I.~G. Lang and Y.~A. Firsov, Zh. Eksp. Teor. Fiz. {\bf 43},  1843  (1962),
  [Sov. Phys. JETP {\bf 16}, 1301 (1962)].

\bibitem{Ho59a}
T. Holstein, Ann. Phys. (N.Y.) {\bf 8},  325; {\bf 8}, 343  (1959).

\bibitem{Ma90}
G.~D. Mahan, {\em Many-Particle Physics}, 2nd  ed. (Plenum Press, New York,
  1990).

\bibitem{dMeRa97}
E.~V.~L. {de Mello} and J. Ranninger, Phys. Rev. B {\bf 55},  14 872  (1997).

\bibitem{Robin97}
J.~M. Robin, Phys. Rev. B {\bf 56},  13 634  (1997).

\bibitem{FeLoWe00}
H. Fehske, J. Loos, and G. Wellein, Phys. Rev. B {\bf 61},  8016  (2000).

\bibitem{HoEvvdL03}
M. Hohenadler, H.~G. Evertz, and W. {von der Linden}, Phys. Rev. B {\bf 69},
  024301  (2004).

\bibitem{ZhJeWh99}
C. Zhang, E. Jeckelmann, and S.~R. White, Phys. Rev. B {\bf 60},  14 092
  (1999).

\bibitem{HovdL05}
M. Hohenadler and W. von~der Linden, Phys. Rev. B {\bf 71},  184309  (2005).

\bibitem{HoEvvdL05}
M. Hohenadler, H.~G. Evertz, and W. {von der Linden}, phys. stat. sol. (b) {\bf
  242},  1406  (2005).

\bibitem{HoNevdLWeLoFe04}
M. Hohenadler {\it et~al.}, Phys. Rev. B {\bf 71},  245111  (2005).

\bibitem{wvl1992}
W. {von der Linden}, Phys. Rep. {\bf 220},  53  (1992).

\bibitem{LoGu92}
E.~Y. Loh and J.~E. Gubernatis,  in {\em Electronic Phase Transitions}, edited
  by W. Hanke and Y.~V. Kopaev (Elsevier Science Publishers, North-Holland,
  Amsterdam, 1992), Chap.~4.

\bibitem{BlScSu81}
R. Blankenbecler, D.~J. Scalapino, and R.~L. Sugar, Phys. Rev. D {\bf 24},
  2278  (1981).

\bibitem{Hi85}
J.~E. Hirsch, Phys. Rev. B {\bf 31},  4403  (1985).

\bibitem{WeRoFe96}
G. Wellein, H. R\"oder, and H. Fehske, Phys. Rev. B {\bf 53},  9666  (1996).

\bibitem{BrCaAsMu01}
M. Brunner, S. Capponi, F.~F. Assaad, and A. Muramatsu, Phys. Rev. B {\bf 63},
  R180511  (2001).

\bibitem{JeWh98}
E. Jeckelmann and S.~R. White, Phys. Rev. B {\bf 57},  6376  (1998).

\bibitem{RoBrLi99III}
A.~H. Romero, D.~W. Brown, and K. Lindenberg, Phys. Rev. B {\bf 60},  14080
  (1999).

\bibitem{KuTrBo02}
L.~C. Ku, S.~A. Trugman, and J. Bon\v{c}a, Phys. Rev. B {\bf 65},  174306
  (2002).

\bibitem{JaGu96}
M. Jarrell and J.~E. Gubernatis, Phys. Rep. {\bf 269},  133  (1996).

\bibitem{DavHin}
A.~C. Davison and D.~V. Hinkley, {\em Bootstrap Methods and their Application}
  (Cambridge University Press, Cambridge, UK, 1997).

\bibitem{numrec_web}
W.~H. Press, Numerical Recipes in {F}ortran 77, {\ttfamily
  http://www.numrec.com}.

\bibitem{Hohenadler04}
M. Hohenadler, Ph.D. thesis, TU Graz, 2004.

\bibitem{dRLa82}
H. {De Raedt} and A. Lagendijk, Phys. Rev. Lett. {\bf 49},  1522  (1982).

\bibitem{deRaLa86}
H. {De Raedt} and A. Lagendijk, Z. Phys. B: Condens. Matter {\bf 65},  43
  (1986).

\bibitem{Ko98}
P.~E. Kornilovitch, Phys. Rev. Lett. {\bf 81},  5382  (1998).

\bibitem{Mac04}
A. Macridin, G.~A. Sawatzky, and M. Jarrell, Phys. Rev. B {\bf 69},  245111
  (2004).

\bibitem{FeAlHoWe06}
H. Fehske, A. Alvermann, M. Hohenadler, and G. Wellein,  in {\em Polarons in
  Bulk Materials and Systems with Reduced Dimensionality}, {\em Proc. Int.
  School of Physics ``Enrico Fermi'', Course CLXI}, edited by G. Iadonisi, J.
  Ranninger, and G. {De Filippis} (IOS Press, Amsterdam, Oxford, Tokio,
  Washington DC, 2006), pp.\ 285--296.

\bibitem{dRLa83}
H. {De Raedt} and A. Lagendijk, Phys. Rev. B {\bf 27},  6097  (1983).

\bibitem{Ko97}
P.~E. Kornilovitch, J. Phys.: Condens. Matter {\bf 9},  10675  (1997).

\bibitem{RoBrLi99}
A.~H. Romero, D.~W. Brown, and K. Lindenberg, Phys. Lett. A {\bf 254},  287
  (1999).

\bibitem{Loe88}
H. Lowen, Phys. Rev. B {\bf 37},  8661  (1988).

\bibitem{Mars95}
F. Marsiglio,  in {\em Recent Progress in Many-Body Theories}, edited by E.
  Schachinger, H. Mitter, and H. Sormann (Plenum Press, New York, 1995),
  Vol.~4.

\bibitem{Marsiglio95}
F. Marsiglio, Physica C {\bf 244},  21  (1995).

\bibitem{HoAivdL04}
M. Hohenadler, M. Aichhorn, and W. {von der Linden}, Phys. Rev. B {\bf 71},
  014302  (2005).

\bibitem{AlBr99}
A.~S. Alexandrov and A.~M. Bratkovsky, Phys. Rev. Lett. {\bf 82},  141  (1999).

\bibitem{AlBr99_2}
A.~S. Alexandrov and A.~M. Bratkovsky, J. Phys.: Condens. Matter {\bf 11},
  L531  (1999).

\bibitem{David_AiP}
D.~M. Edwards, Adv. Phys. {\bf 51},  1259  (2002).

\bibitem{BoKaTr00}
J. Bon\v{c}a, T. Katra\v{s}nik, and S.~A. Trugman, Phys. Rev. Lett. {\bf 84},
  3153  (2000).

\bibitem{ElShBoKuTr03}
S. {El Shawish}, J. Bon\v{c}a, L.~C. Ku, and S.~A. Trugman, Phys. Rev. B {\bf
  67},  014301  (2003).

\bibitem{PrAu99}
L. Proville and S. Aubry, Eur. Phys. J. B {\bf 11},  41  (1999).

\bibitem{LoHoFe06}
J. Loos, M. Hohenadler, and H. Fehske, J. Phys.: Condens. Matter {\bf 18},
  2453  (2006).

\bibitem{HoWeAlFe05}
M. Hohenadler, G. Wellein, A. Alvermann, and H. Fehske, Physica B {\bf
  378-380},  64  (2006).

\bibitem{WeBiHoScFe05}
G. Wellein {\it et~al.}, Physica B {\bf 378-380},  281  (2006).

\bibitem{Stephan}
W. Stephan, Phys. Rev. B {\bf 54},  8981  (1996).

\bibitem{WeFe97}
G. Wellein and H. Fehske, Phys. Rev. B {\bf 56},  4513  (1997).

\bibitem{HoAivdL03}
M. Hohenadler, M. Aichhorn, and W. {von der Linden}, Phys. Rev. B {\bf 68},
  184304  (2003).

\bibitem{BuMKHa98}
R.~J. Bursill, R.~H. McKenzie, and C.~J. Hamer, Phys. Rev. Lett. {\bf 80},
  5607  (1998).

\bibitem{HoWeBiAlFe06}
M. Hohenadler {\it et~al.}, Phys. Rev. B {\bf 73},  245120  (2006).

\bibitem{SyHuBeWeFe04}
S. Sykora {\it et~al.}, Phys. Rev. B {\bf 71},  045112  (2005).

\bibitem{CaGrSt99}
M. Capone, M. Grilli, and W. Stephan, Eur. Phys. J. B {\bf 11},  551  (1999).

\end{thebibliography}



\end{document}